\documentclass[a4paper,11pt]{article}
\usepackage{etex}      
\pdfoutput=1 

\usepackage{jheppub} 
\usepackage{amsmath,amssymb,amsthm,amscd,graphicx}
\input epsf.sty

\newcommand{\CF}{{\cal F}}

\newcommand{\CN}{{\cal N}}
\newcommand{\CO}{{\cal O}}
\newcommand{\CP}{{\cal P}}

\def\IZ{{\mathbb Z}}
\def\IR{{\mathbb R}}
\def\IC{{\mathbb C}}

\def\IP{{\mathbb P}}

\def\IF{{\mathbb F}}
\def\mO{\mathsf{O}}
\def\mW{\mathsf{W}}

\newcommand{\tr}{{\rm Tr}}
\newcommand{\re}{{\rm e}}
\newcommand{\ri}{\mathsf{i}}
\newcommand{\rd}{{\rm d}}

\newcommand{\mA}{\mathsf{A}}

\newcommand{\mP}{\mathsf{P}}

\newcommand{\mJ}{\mathsf{J}}
\newcommand{\mx}{\mathsf{x}}
\newcommand{\my}{\mathsf{y}}
\newcommand{\mm}{\mathsf{p}}
\newcommand{\im}{\mathsf{i}}
\newcommand{\mH}{\mathsf{H}}

\newcommand{\mq}{\mathsf{q}}
\newcommand{\mg}{\mathsf{g}}

\newcommand{\fad}{\operatorname{\Phi}_{\mathsf{b}}}
\newcommand{\mb}{{\mathsf{b}}}

\newcommand{\be}{\begin{equation}}
\newcommand{\ee}{\end{equation}}
\newcommand{\ba}{\begin{aligned}}
\newcommand{\ea}{\end{aligned}}
\newcommand{\ben}{\begin{eqnarray}\displaystyle}
\newcommand{\een}{\end{eqnarray}}

\newcommand{\figref}[1]{Fig.~\protect\ref{#1}}
\newcommand{\fadi}{\fad^*}

\newcommand{\mypsi}[2]{\operatorname{\Psi}_{#1,#2}}

\DeclareMathOperator{\Li}{Li}
\DeclareMathOperator{\Ai}{Ai}

\def\({\left(}
\def\){\right)}
\DeclareMathOperator{\real}{Re}

\usepackage[table]{xcolor}				
\usepackage{colortbl}
\usepackage{booktabs}					
\usepackage{multirow,bigdelim}			
\usepackage{subfig}						

\DeclareMathOperator{\imag}{Im}

\def\bC{\mathbb{C}}
\def\bF{\mathbb{F}}

\def\bR{\mathbb{R}}
\def\bZ{\mathbb{Z}}

\def\cC{\mathcal{C}}

\def\cF{\mathcal{F}}

\def\cN{\mathcal{N}}
\def\cO{\mathcal{O}}

\def\sb{\mathsf{b}}
\def\sp{\mathsf{p}}
\def\sq{\mathsf{q}}

\def\su{\mathsf{u}}

\def\sx{\mathsf{x}}
\def\sy{\mathsf{y}}
\def\sA{\mathsf{A}}
\def\sH{\mathsf{H}}
\def\sO{\mathsf{O}}
\def\sP{\mathsf{P}}

\def\sW{\mathsf{W}}

\title{\boldmath Operators and higher genus mirror curves}

\author{Santiago Codesido$^a$, Jie Gu$^b$, and Marcos Mari\~no$^a$}

\affiliation{$^a$D\'epartement de Physique Th\'eorique et Section de Math\'ematiques\\
Universit\'e de Gen\`eve, Gen\`eve, CH-1211 Switzerland\\
\\
$^b$Laboratoire de Physique Th\'eorique de l'\'{E}cole Normale Sup\'erieure\\
CNRS, PSL Research University, Sorbonne Universit\'{e}s, UPMC, 75005 Paris, France\\}

\emailAdd{santiago.codesido@unige.ch}
\emailAdd{jie.gu@lpt.ens.fr}
\emailAdd{marcos.marino@unige.ch}

\abstract{We perform further tests of the correspondence between spectral theory and topological strings, focusing on mirror curves of genus greater than one with nontrivial mass parameters. In particular, we analyze the 
geometry relevant to the $SU(3)$ relativistic Toda lattice, and the resolved $\IC^3/\IZ_6$ orbifold. Furthermore, we give evidence 
that the correspondence holds for arbitrary values of the mass parameters, where the quantization 
problem leads to resonant states. We also explore the relation between this correspondence and cluster integrable systems.}    

\begin{document}
\maketitle
\flushbottom

\section{Introduction}

Recently, a detailed, conjectural correspondence has been proposed between topological strings on toric Calabi--Yau (CY) manifolds, and the spectral theory of certain 
quantum-mechanical operators on the real line \cite{ghm} (see \cite{mmrev} for a review). The operators arise by the quantization of the mirror curve to the toric CY, 
as suggested originally in \cite{adkmv}. This correspondence builds upon previous work on quantization and mirror symmetry \cite{ns,mirmor,acdkv,km,hw}, 
and on the exact solution of the ABJM matrix model \cite{mptop,dmp,mp,hmo,hmo1, hmo2, hmmo} (reviewed in \cite{hmorev,mmabjmrev}.)  
It leads to exact formulae for the spectral traces and the Fredholm determinants of these operators, in terms 
of BPS invariants of the CY. Conversely, the genus expansion of the topological string free energy arises as an asymptotic expansion of their spectral traces, in a certain 
't Hooft-like regime. In this way, the correspondence provides a non-perturbative completion of the topological string free energy. Although the general correspondence 
of \cite{ghm} is still conjectural, it has passed many tests. In the last two years, techniques have been developed to calculate the corresponding quantities in spectral theory and shown to be 
in perfect agreement with the predictions of the conjecture \cite{kasmar,mz,kmz,oz,gkmr}. Other aspects of the correspondence have been discussed in \cite{lst,bgt,grassi,sugimoto}. 

The conjectural correspondence of \cite{ghm} was formulated for mirror curves of genus one. The generalization to 
curves of higher genus $g_\Sigma>1$ was proposed in \cite{cgm}. In this case, the quantization of the curve leads naturally to $g_\Sigma$ different operators, and one can define a 
generalized Fredholm determinant which encodes their spectral properties. In \cite{cgm}, the higher genus version of the correspondence 
was verified in detail in the example of the resolved $\IC^3/\IZ_5$ orbifold, arguably the simplest toric CY with a genus 
two mirror curve. It would be desirable to have more examples and tests of the conjecture in the higher genus case, which has been comparatively less 
explored. The first goal of this paper is to start filling this gap by analyzing in detail two important genus two geometries. After briefly reviewing the spectral theory of the 
operators associated to toric CY's in Sec.\ 2, we study in Sec.\ 3 the CY geometry that leads to the $SU(3)$ relativistic periodic Toda lattice 
(recently studied in \cite{hm}), and the resolved $\IC^3/\IZ_6$ orbifold. Both geometries have a mass parameter, 
which was absent in the example analyzed in \cite{cgm}. Geometrically, 
they both engineer $SU(3)$ super Yang--Mills theory \cite{kkv}, and they correspond to 5d, $SU(3)$ gauge theories on $S^1$ with different 
values of the 5d Chern--Simons coupling (see for example \cite{ikp}). 
Our focus is on the spectral traces of the operators obtained by quantization of their
mirror curves. In all cases, the conjectural formulae of \cite{cgm} pass the tests with flying colors. Along the way, we 
present in Sec.\ 2.3 the exact integral kernel for the inverse 
of a four-term operator which is relevant for the relativistic Toda case. This generalizes the results in \cite{kasmar,kmz}.

The operators arising from the quantization of mirror curves depend on the so-called 
mass parameters of the geometry \cite{hkp,hkrs}. They are only compact and of trace class when these parameters satisfy some 
positivity conditions. This is similar to what happens in 
ordinary quantum mechanics. A simple example is provided by the quartic oscillator, defined by the following Schr\"odinger operator on $L^2(\IR)$: 
\be
\label{qosc}
\mH= \frac{\mm^2}{ 2} + \frac{\mq^2}{ 2} + g \mq^4.
\ee
 The inverse operator $\mH^{-1}$ is trace class provided $g>0$ (see for example \cite{voros}). For $g<0$, the potential becomes unstable and there are no longer 
bound states. However, there are {\it resonant states} with a discrete set of {\it complex} 
eigenvalues that can be calculated by using complex dilatation techniques (see for example \cite{kpbook,reed-simon}). In section 5 of this paper, 
we point out that the spectral theory of the 
operators arising from mirror curves displays a very similar phenomenon. 
Namely, for general values of the mass parameters, these operators are no longer compact (this is easily seen by using semiclassical estimates). 
However, their spectral traces admit an analytic continuation to the complex planes of the mass parameters, with branch cuts, and they can be still exactly computed 
from topological string data by a natural extension of the conjecture in \cite{ghm,cgm}. 
In particular, we can construct an analytic continuation of the Fredholm determinant 
for arbitrary, complex values of the mass parameters. The vanishing locus of this analytically-continued function leads to discrete, complex values of the energy, which we interpret as resonances.

As explained in \cite{cgm}, since the $g_\Sigma$ operators arising from the mirror curve are closely related among themselves, 
the conjectural correspondence of \cite{cgm} leads to a single quantization condition, 
given by the vanishing of the generalized Fredholm determinant. However, by a construction of Goncharov and Kenyon \cite{gk}, 
it is possible to construct a quantum integrable system, called the cluster integrable system, for any toric CY, leading to $g_\Sigma$ commuting Hamiltonians. 
An exact quantization condition for this integrable system was proposed in \cite{hm,fhm}, based on the general philosophy of \cite{ns}. 
The result of \cite{fhm} generalizes the quantization condition of \cite{ghm} in the form presented in \cite{wzh}. Therefore, there are two different quantum 
problems arising from toric CY manifolds: on one hand we have the problem associated to the quantization of the mirror curve, 
which can be formulated in terms of 
non-commuting operators on $\IR$. On the other hand, we have the cluster integrable system of Goncharov and Kenyon, 
which leads to commuting Hamiltonians on $\IR^{g_\Sigma}$. In the genus one case, the two problems coincide, and in the higher 
genus case they should be closely related: as in the case of 
the standard Toda lattice, we expect that the quantization of the mirror curve leads to the quantum Baxter equation of 
the cluster integrable system. By requiring additional constraints on the solution 
of this equation, one should recover the $g_\Sigma$ quantization conditions. This program has not been pursued in detail. 
However, it was noticed in \cite{fhm}, in a genus two example, that an appropriate rotation of the 
variables in the generalized spectral determinant leads to two different functions on moduli space. 
The intersection of the vanishing loci of these functions turns out 
to coincide with the spectrum of the cluster integrable system. Recently, this observation has been generalized in 
\cite{swh}, and there is now empirical evidence that the generalized 
spectral determinant of \cite{cgm}, after appropriate rotations of the phases of the variables, leads to at least $g_\Sigma$ different functions 
whose zero loci intersect precisely on the spectrum of the cluster integrable system\footnote{As is obvious from this discussion, 
the quantization condition of \cite{cgm} is more general than the one in \cite{fhm}, as all existing results indicate that one can recover the latter from the former.}. 
Another goal of this paper is to further clarify the relation between the quantization conditions of \cite{cgm} and of \cite{fhm}. In particular, 
we will show in Sec.\ 5 that in some genus two geometries, reality and positivity conditions allow us 
to deduce the spectrum of the cluster integrable system from the generalized Fredholm determinant.

\section{Spectral theory and topological strings}

In this section, we review the construction of $g_\Sigma$ 
operators from the mirror curve of a toric CY, their spectral theory, and the connection to the topological string theory compactified on the toric CY. We mainly follow \cite{ghm, mz, kmz, cgm}. The material 
on toric CYs and mirror symmetry is standard, see for example \cite{hkt,ckyz,ck,horibook}.  

\subsection{Mirror curves and spectral theory}
\label{sc:mc}

Toric CY threefolds can be specified by a matrix of charges $Q_i^\alpha$, $i=0, \cdots, k+2$, $\alpha=1, \cdots, k$ satisfying the condition, 
\begin{equation}  
\sum_{i=0}^{k+2} Q_i^\alpha=0, \qquad \alpha = 1, \ldots, k .
\label{anomaly}
\end{equation}  
Their mirrors can be written in terms of
$3+k$ complex coordinates $Y^i\in \IC^*$, $i=0, \cdots, k+2$, which satisfy the constraint
\be
\label{dterm-Y}
\sum_{i=0}^{k+2} Q_i^\alpha Y^i=0,  \qquad \alpha = 1, \ldots, k.  
\ee
The mirror CY manifold $\widehat X$ is then given by 
\be
w^+w^-= W_X ,
\ee
where
\be
\label{wx}
W_X= \sum_{i=0}^{k+2} x_i \re^{Y_i}. 
\ee
It is possible to solve the constraints (\ref{dterm-Y}), modulo a global translation, in terms of two variables which we will denote by 
$x$, $y$, and we then obtain a function $W_X (\re^x, \re^y)$ from (\ref{wx}). The equation
\be
\label{riemann}
W_X (\re^x, \re^y)=0  
\ee
defines a Riemann surface $\Sigma$ embedded in $\IC^* \times \IC^*$, which we will call the {\it mirror curve} to the toric CY threefold $X$. We note that there is a group of reparametrization 
symmetries of the mirror curve given by \cite{akv}, 
\be
\label{can-t}
\begin{pmatrix} x \\ y \end{pmatrix}\rightarrow G \begin{pmatrix} x \\ y \end{pmatrix}, \qquad G\in {\rm SL}(2, \IZ). 
\ee
The moduli space of the mirror curve can be parametrized by the $k+3$ coefficients $x_i$ of its equation \eqref{wx}, among 
which three can be set to 1 by the $\IC^*$ scaling acting on $\re^x, \re^y$ and an overall $\IC^*$ scaling. Equivalently, one can use the Batyrev coordinates
\be
\label{z-moduli}
z_\alpha=\prod_{i=0}^{k+2} x_i^{Q_i^\alpha}, \qquad  \alpha = 1, \ldots, k,
\ee
which are invariant under the $\IC^*$ actions.
In order to write down the mirror curves, it is also useful to introduce a two-dimensional Newton polygon $\CN$. The points of this polygon are given by 
\be
\nu^{(i)}=(\nu^{(i)}_1, \nu^{(i)}_2),
\ee
in such a way that the extended vectors 
\be
\overline \nu^{(i)}=\left(1, \nu^{(i)}_1, \nu^{(i)}_2\right),\qquad i=0, \cdots, k+2,
\ee
satisfy the relations
\be
\label{qvecs}
\sum_{i=0}^{k+2} Q^\alpha_i \overline \nu^{(i)}=0. 
\ee
This Newton polygon is nothing else but the support of the 3d toric fan of the toric Calab-Yau threefold on a hyperplane located at $(1,*,*)$. The function on the l.h.s. of (\ref{riemann}) is then given by the Newton polynomial of the polygon $\CN$, 
\be
\label{coxp}
W_X(\re^x, \re^y)=\sum_{i=0}^{k+2}x_i  \exp\left( \nu^{(i)}_1 x+  \nu^{(i)}_2 y\right). 
 \ee
Clearly, there are many sets of vectors satisfying the relations (\ref{qvecs}), but they lead to curves differing in a reparametrization or a global translation, 
which are therefore equivalent. It can be seen that the genus of $\Sigma$, $g_\Sigma$, is given by the number of inner points of $\CN$. We also notice that among the coefficients $x_i$ of the mirror curve, $g_\Sigma$ of them are ``true" moduli of the geometry, corresponding to the inner points of $\CN$, while $r_\Sigma$ of them are the so-called ``mass parameters", corresponding to the points on the boundary of the polygon
(this distinction has been emphasized in \cite{hkp,hkrs}). In order to distinguish them, we will denote the former by $\kappa_i$, $i=1,\ldots,g_\Sigma$ and the latter by $\xi_j$, $j=1,\ldots, r_\Sigma$ . It is obvious 
that we can translate the Newton polygon in such a way that the inner point associated to a given $\kappa_i$ is the origin. In this way we obtain what we will call the {\it canonical} forms of the mirror curve
\be
\CO_i (x,y) + \kappa_i=0, \qquad i=1, \cdots, g_\Sigma, 
\ee
where $\CO_i(x, y)$ is a polynomial in $\re^x$, $\re^y$. Note that, for $g_\Sigma=1$, there is a single canonical form. Different canonical forms 
are related by reparametrizations of the form (\ref{can-t}) and by overall translations, which lead to overall 
monomials, so we will write 
\be
\label{o-rel}
\CO_i +\kappa_i  = \CP_{ij} \left( \CO_j +\kappa_j\right), \qquad i,j=1, \cdots, g_\Sigma, 
\ee
where $\CP_{ij}$ is of the form $\re^{ \lambda x + \mu y}$, $\lambda,\mu\in\bZ$. Equivalently, we can write 
\be
\label{oi-exp}
\CO_i = \CO_i^{(0)}+ \sum_{j \not=i} \kappa_j \CP_{ij}. 
\ee

The functions $\CO_i(x,y)$ appearing in the canonical forms of the mirror curves can be quantized \cite{ghm,cgm}. To do this, we 
simply promote the variables $x$, $y$ to 
Heisenberg operators $\mx$, $\my$ satisfying
\be
[\mx, \my]= \im \hbar, 
\ee
and we use the Weyl quantization prescription. In this way we obtain 
$g_\Sigma$ different operators, which we will denote by $\mO_i$, $i=1, \cdots, g_\Sigma$. These operators are self-adjoint. The equation of the mirror curve itself is promoted to an operator
\be
\mW_{X,i} \equiv \mO_i + \kappa_i \ ,
\ee
which we call the \emph{quantum mirror curve}. Different canonical forms are related by the quantum version of (\ref{o-rel}), 
\be
\label{op-rel}
\mO_i+ \kappa_i =\mP_{ij}^{1/2} \left( \mO_j  + \kappa_j \right) \mP^{1/2}_{ij}, \qquad i,j=1, \cdots, g_\Sigma, 
\ee
where $\mP_{ij}$ is the operator corresponding to the monomial $\CP_{ij}$. We will also denote by $\mO_i^{(0)}$ the operator associated to the function $\CO_i^{(0)}$ in (\ref{oi-exp}). 
This can be regarded as an ``unperturbed" operator, while the moduli $\kappa_j$ encode different 
perturbations of it. We also define the inverse operators, 
\be
\label{unp-inv}
\rho_i =\mO_i^{-1}, \quad \rho_i^{(0)}=\left( \mO_i^{(0)}\right)^{-1}, \qquad i=1, \cdots, g_\Sigma.
\ee
It is easy to see that \cite{cgm}
\be
\mP_{ij}= \mP_{ji}^{-1}, \qquad i \not=j, 
\ee
and
\be
\label{p-rels}
\mP_{ik}=\mP_{ij}^{1/2} \mP_{jk} \mP_{ij}^{1/2}, \qquad i\not= k. 
\ee

We want to study now the spectral theory of the operators $\mO_i$, $i=1, \cdots, g_\Sigma$. The appropriate object to consider turns out to be the {\it generalized spectral determinant} 
introduced in \cite{cgm}. Let us consider the following operators,  
\be
\label{ajl}
\mA_{jl}= \rho_j^{(0)} \mP_{jl}, \quad j, l=1, \cdots, g_{\Sigma}.  
\ee
Let us suppose that these operators are of trace class (this turns out to be the case in all known examples, provided some positivity conditions on the mass parameters are satisfied). 
Then, the generalized spectral determinant associated to the CY $X$ is given by 
\be
\label{gsd}
\Xi_X ( {\boldsymbol \kappa}; \hbar)= {\rm det} \left( 1+\kappa_1 \mA_{j1} +\cdots+ \kappa_{g_\Sigma} \mA_{j g_\Sigma}  \right). 
\ee
Due to the trace class property of the operators $\mA_{jl}$, this quantity is well-defined, and its definition does not 
depend on the choice of the index $j$, due to the similarity transformation
\be
\mA_{il}= \mP_{ij}^{-1/2} \mA_{jl} \mP_{ij}^{1/2}. 
\ee
As shown in \cite{simon-paper}, (\ref{gsd}) is 
an entire function of the moduli $\kappa_1, \cdots, \kappa_{g_{\Sigma}}$. In particular, it can be expanded 
around the origin ${\boldsymbol \kappa}=0$, as follows, 
\be
\label{or-exp}
\Xi_X ({\boldsymbol \kappa};\hbar)= \sum_{N_1\ge 0} \cdots \sum_{N_{g_\Sigma}\ge 0} Z_X(\boldsymbol{N}; \hbar) \kappa_1^{N_1} \cdots \kappa_{g_\Sigma}^{N_{g_\Sigma}},  
\ee
with the convention that 
\be
\label{zzeros}
Z_X(0, \cdots, 0;\hbar)=1. 
\ee
This expansion defines the (generalized) fermionic spectral traces $Z_X(\boldsymbol{N}; \hbar)$ of the toric CY $X$. Both $\Xi_X ({\boldsymbol \kappa};\hbar)$ and 
$Z_X(\boldsymbol{N}; \hbar)$ depend in addition on the mass parameters, gathered in a vector $\boldsymbol{\xi}$. When needed, we will indicate this dependence explicitly
and write $\Xi_X ({\boldsymbol \kappa};\boldsymbol{\xi},\hbar)$, $Z_X(\boldsymbol{N}; \boldsymbol{\xi}, \hbar)$. As shown in \cite{cgm}, one can use classical results in 
Fredholm theory to obtain determinant expressions for these fermionic traces. Let us consider the kernels $A_{jl}(x_m, x_n)$ of the operators defined in (\ref{ajl}), and let us construct the 
following matrix:
\be
R_j (x_m, x_n) = A_{jl}(x_m , x_n) \,\, \,\, \,\, \,\,  \text{if} \, \,  \,\, \,\, \,\,\sum_{s=1}^{l-1} N_s< m \le \sum_{s=1}^l N_s. 
\ee
Then, we have that \footnote{The determinant of the matrix $R_{j(x_m, x_n)}$ is independent of the label $j$, just like the Fredholm determinant.}
\be
\label{gen-fred-th}
Z_X(\boldsymbol{N}; \hbar)= \frac{1}{ N_1! \cdots N_{g_\Sigma}!} \int  {\rm det}_{m,n} \left(R_j(x_m, x_n)\right)\rd^N x, 
\ee
where
\be
N=\sum_{s=1}^{g_\Sigma} N_s. 
\ee
In the case $g_\Sigma=2$, this formula becomes
\be
\label{gtwo-matrix}
Z_X(N_1, N_2;\hbar)=\frac{1}{ N_1! N_2!} \int  {\rm det} \begin{pmatrix} A_{j1}(x_1, x_1)&  \cdots  & A_{j1}(x_1, x_N) \\
\vdots &  & \vdots\\
A_{j1}(x_{N_1}, x_1) & \cdots & A_{j1}( x_{N_1}, x_N) \\
A_{j2} (x_{N_1+1}, x_1) & \cdots & A_{j2}( x_{N_1+1}, x_N) \\
\vdots &  & \vdots \\
A_{j2} (x_{N}, x_1) & \cdots &A_{j2}( x_{N}, x_N) \end{pmatrix}\rd x_1 \cdots \rd x_{N}. 
\ee
One finds, for example
\be
\label{11-mixed}
\ba
Z_X(1, 1; \hbar)&= \tr \, \mA_{j1} \,  \tr  \, \mA_{j2} - \tr \left( \mA_{j1} \mA_{j2}\right)\\
&= \int \rd x_1 \rd x_2 \left( A_{j1}(x_1, x_1) A_{j2}(x_2, x_2) -  A_{j1}(x_1, x_2) A_{j2}(x_2, x_1)\right), 
\ea
\ee
as well as 
\be
\label{12-mixed}
\ba
Z_X(2,1; \hbar)&=\tr \left(\mA_{j1}^{2} \mA_{j2}\right)-\frac{1}{2} \tr \left(\mA_{j1}^2\right) \tr \, \mA_{j2}+\frac{1}{2} \left(\tr\, \mA_{j1} \right)^2
  \tr\, \mA_{j2}-\tr \, \mA_{j1} \,  \tr  \left(\mA_{j1} \mA_{j2}\right), \\
 Z_X (1,2; \hbar)&=\tr \left(\mA_{j1}\mA_{j2}^{2} \right)-\frac{1}{2} \tr \, \mA_{j1}\,  \tr \left(\mA_{j2}^2\right) +\frac{1}{2}   \tr\, \mA_{j1}\left(\tr\, \mA_{j2} \right)^2
-\tr  \left(\mA_{j1} \mA_{j2}\right)\tr \, \mA_{j2}.
\ea 
  \ee

The generalized spectral determinant encodes the spectral properties of all the operators $\mO_i$ in a single strike. Indeed, one has \cite{cgm},
\be
\label{quot-i}
 {\rm det}\left(1+ \kappa_i \rho_i\right) ={\Xi_X ( {\boldsymbol \kappa}; \hbar) \over \Xi_X \left(\kappa_1, \cdots, \kappa_{i-1},0, \kappa_{i+1}, \cdots, \kappa_{g_\Sigma} ; \hbar\right) }, \qquad 
i=1, \cdots, g_\Sigma. 
 \ee
In addition, 
\be
\label{quot-i0}
{\rm det}\left(1+ \kappa_i \rho_i^{(0)}\right)=\Xi_X(0,\cdots, 0, \kappa_i, 0, \cdots, 0; \hbar), \qquad i=1, \cdots, g_\Sigma, 
\ee
i.e. the generalized spectral determinant specializes to the spectral determinants of the unperturbed operators appearing in the different canonical forms of the mirror curve. 

The standard spectral determinant of a single trace-class operator determines the spectrum of eigenvalues, through its zeros. 
The generalized spectral determinant (\ref{gsd}) vanishes in a codimension one submanifold of the moduli space. It follows from (\ref{quot-i}) that this submanifold 
contains all the information about the spectrum of the operators $\rho_i$ appearing in the quantization of the mirror curve, as a function of the moduli $\kappa_j$, $j\not=i$.

\subsection{Spectral determinants and topological strings}
\label{sc:spec-det}

The main conjectural result of \cite{ghm,cgm} is an explicit formula expressing the generalized spectral determinant and spectral traces in terms of 
of enumerative invariants of the CY $X$. To state this result, we need some basic geometric ingredients. As discussed in the previous section, the CY $X$ has $g_\Sigma$ moduli denoted by $\kappa_i$, $i=1, \cdots, g_\Sigma$. We will 
introduce the associated ``chemical potentials" $\mu_i$ by
\be
\kappa_i =\re^{\mu_i}, \qquad i=1, \cdots, g_{\Sigma}. 
\ee
The CY $X$ also has $r_\Sigma$ mass parameters, $\xi_j$, $j=1, \cdots, r_\Sigma$. Let $n_\Sigma \equiv g_\Sigma+r_\Sigma$. The Batyrev coordinates $z_i$ introduced in (\ref{z-moduli}) can be written as
\be
\label{zmu}
-\log \, z_i= \sum_{j=1}^{g_\Sigma}C_{ij} \mu_j + \sum_{k=1}^{r_\Sigma} \alpha_{ik}\log {\xi_k}, \qquad i=1, \cdots, n_\Sigma.
\ee
The coefficients $C_{ij}$ determine a $n_\Sigma \times g_\Sigma$ matrix which can be read off from the 
toric data of $X$. One can choose the Batyrev coordinates in such a way that, for $i=1, \cdots, g_\Sigma$, the $z_i$'s correspond 
to true moduli, while for $i=g_\Sigma+1, \cdots, g_\Sigma+ r_\Sigma$, they correspond to mass parameters. 
For such a choice, the non-vanishing coefficients 
\be
\label{tmc}
C_{ij}, \quad i,j=1, \cdots, g_\Sigma,
\ee
form an invertible matrix, which agrees (up to an overall sign) with the charge matrix $C_{ij}$ appearing in \cite{kpsw}. 
We also recall that the standard mirror map expresses the K\"ahler moduli $t_i$ of the CY in terms of the Batyrev coordinates $z_i$
\begin{equation}\label{eq:qPeriods}
	- t_i = \log z_i + \tilde{\Pi}_i(\boldsymbol{z}) \ ,\quad i =1\ldots,n_\Sigma\ ,
\end{equation}
where $\tilde{\Pi}_i(\boldsymbol{z})$ is a power series in $z_i$ with finite radius of convergence. Together with \eqref{zmu}, this implies that
\begin{equation}
	t_i = \sum_{j=1}^{g_\Sigma} C_{ij} \mu_j + \sum_{k=1}^{r_\Sigma} \alpha_{ik} \log {\xi_k} + \cO(\re^{-\mu}) \ .
\end{equation}
By using the quantized mirror curve, one can promote the classical mirror map to a {\it quantum mirror map} $t_i(\hbar)$ depending on $\hbar$ \cite{acdkv} (see \cite{huang,hkrs} for examples.)
\begin{equation}
- t_i(\hbar) = \log z_i + \tilde{\Pi}_i(\boldsymbol{z};\hbar) \ ,\quad i =1\ldots,n_\Sigma\ .
\end{equation}
This quantum mirror map will play an important r\^ole in our construction. In addition to the quantum mirror map, we need the following enumerative ingredients. 
First of all, we need the conventional genus $g$ free energies $F_g({\bf t})$ of $X$, $g \ge 0$, in the so-called 
large radius frame (\texttt{LRF}), which encode the information about the Gromov--Witten invariants of $X$. They have the 
structure\footnote{The formula of $F_0(\bf t)$ differs from the one in the topological string literature by the linear term, which usually 
doesn't play a role in non-compact CY models. The addition of this term makes the formulae in the rest of the paper more compact.}
\be
\label{gzp}
\ba
F_0({\bf t})&={1\over 6}\sum_{i,j,k=1}^{n_\Sigma} a_{ijk} t_i t_j t_k  + 4\pi^2 \sum_{i=1}^{n_\Sigma} b_i^{\rm NS} t_i + \sum_{{\bf d}} N_0^{ {\bf d}} \re^{-{\bf d} \cdot {\bf t}},\\
F_1({\bf t})&= \sum_{i=1}^{n_\Sigma} b_i t_i + \sum_{{\bf d}} N_1^{ {\bf d}} \re^{-{\bf d} \cdot {\bf t}}, \\
F_g({\bf t})&= C_g+\sum_{{\bf d}} N_g^{ {\bf d}} \re^{-{\bf d} \cdot {\bf t}}, \qquad g\ge 2. 
\ea
\ee
In these formulae, $ N_g^{ {\bf d}} $ are the Gromov--Witten invariants of $X$ at genus $g$ and multi-degree ${\bf d}$. The coefficients $a_{ijk}$, $b_i$ are cubic and linear 
couplings characterizing the perturbative genus zero and genus one free energies, while 
 $C_g$ is the so-called constant map contribution \cite{bcov}. The constants $b_i^{\rm NS}$, which can be obtained from the 
 refined holomorphic anomaly equation \cite{hk,krewal}, usually appear in the linear term of $F^{\rm NS}(\bf t,\hbar)$ (see below, \eqref{eq:NS-inst}).
The total free energy of the topological string is the formal series, 
\be
\label{tfe}
F^{\rm WS}\left({\bf t}, g_s\right)= \sum_{g\ge 0} g_s^{2g-2} F_g({\bf t})=F^{({\rm p})}({\bf t}, g_s)+ \sum_{g\ge 0} \sum_{\bf d} N_g^{ {\bf d}} \re^{-{\bf d} \cdot {\bf t}} g_s^{2g-2},   
\ee
where
\be
F^{({\rm p})}({\bf t}, g_s)= {1\over 6 g_s^2} \sum_{i,j,k=1}^{n_\Sigma} a_{ijk} t_i t_j t_k + \sum_{i=1}^{n_\Sigma} \(b_i + \frac{4\pi^2}{g_s^2} b_i^{\rm NS}\)t_i + \sum_{g \ge 2}  C_g g_s^{2g-2} 
\ee
and $g_s$ is the topological string coupling constant. 

As found in \cite{gv}, the sum over Gromov--Witten invariants in (\ref{tfe}) 
can be resummed order by order in $\exp(-t_i)$, at all orders in $g_s$. This resummation involves the 
Gopakumar--Vafa (GV) invariants $n^{\bf d}_g$ of $X$, and it has the structure
\be
\label{GVgf}
F^{\rm GV}\left({\bf t}, g_s\right)=\sum_{g\ge 0} \sum_{\bf d} \sum_{w=1}^\infty {1\over w} n_g^{ {\bf d}} \left(2 \sin { w g_s \over 2} \right)^{2g-2} \re^{-w {\bf d} \cdot {\bf t}}.  
\ee
Note that, as formal power series, we have
\be
\label{gv-form}
F^{\rm WS}\left({\bf t}, g_s\right)=F^{({\rm p})}({\bf t}, g_s)+F^{\rm GV}\left({\bf t}, g_s\right). 
\ee
In the case of toric CYs, the Gopakumar--Vafa invariants are special cases of the 
{\it refined BPS invariants} \cite{ikv,ckk,no}. These refined invariants depend on the degrees ${\bf d}$ and on two non-negative 
half-integers or ``spins", $j_L$, $j_R$. We will denote them by $N^{\bf d}_{j_L, j_R}$. We now define the {\it NS free energy} as
\be
\label{NS-j}
F_{\rm NS}({\bf t}, \hbar) = F_{\rm NS}^{\rm pert}({\bf t}, \hbar) + F_{\rm NS}^{\rm inst}({\bf t}, \hbar) \ ,
\ee
where
\be
\label{eq:NS-pert}
F_{\rm NS}^{\rm pert}({\bf t}, \hbar) ={1\over 6 \hbar} \sum_{i,j,k=1}^{n_\Sigma} a_{ijk} t_i t_j t_k +  \( \hbar + \frac{4\pi^2}{\hbar}\) \sum_{i=1}^{n_\Sigma} b^{\rm NS}_i t_i \ ,
\ee
and
\be
\label{eq:NS-inst}
F_{\rm NS}^{\rm inst}({\bf t}, \hbar) = \sum_{j_L, j_R} \sum_{w, {\bf d} } 
N^{{\bf d}}_{j_L, j_R}  \frac{\sin\frac{\hbar w}{2}(2j_L+1)\sin\frac{\hbar w}{2}(2j_R+1)}{2 w^2 \sin^3\frac{\hbar w}{2}} \re^{-w {\bf d}\cdot{\bf  t}} \ . 
\ee
In this equation, the coefficients $a_{ijk}$ are the same ones that appear in (\ref{gzp}). 
By expanding (\ref{NS-j}) in powers of $\hbar$, we find the NS free energies at order $n$, 
\be
\label{ns-expansion}
F^{\rm NS}({\bf t}, \hbar)=\sum_{n=0}^\infty  F^{\rm NS}_n ({\bf t}) \hbar^{2n-1}. 
\ee
The first term 
in this series, $F_0^{\rm NS}({\bf t})$, is equal to $F_0({\bf t})$, the standard genus zero free energy. 

Following \cite{hmmo}, we now define the {\it modified grand potential} of the CY $X$. It is the sum of two functions. The first one is
\be
\label{jm2}
\mathsf{J}^{\rm WKB}_X(\boldsymbol{\mu}, \boldsymbol{\xi}, \hbar)= \sum_{i=1}^{n_\Sigma}{t_i(\hbar) \over 2 \pi}   {\partial F^{\rm NS}({\bf t}(\hbar), \hbar) \over \partial t_i} 
+{\hbar^2 \over 2 \pi} {\partial \over \partial \hbar} \left(  {F^{\rm NS}({\bf t}(\hbar), \hbar) \over \hbar} \right) + {2 \pi \over \hbar} \sum_{i=1}^{n_\Sigma}\(b_i+b_i^{\rm NS}\) t_i(\hbar) + A({\boldsymbol \xi}, \hbar). 
\ee
We note that the function $A({\boldsymbol \xi}, \hbar)$ is only known in a closed form in some simple geometries. The second function is 
the ``worldsheet" modified grand potential, which is obtained from the generating functional (\ref{GVgf}), 
\be
\label{jws}
\mathsf{J}^{\rm WS}_X(\boldsymbol{\mu}, \boldsymbol{\xi}, \hbar)=F^{\rm GV}\left( {2 \pi \over \hbar}{\bf t}(\hbar)+ \pi \ri {\bf B} , {4 \pi^2 \over \hbar} \right).
\ee
It involves a constant integer vector ${\bf B}$ (or ``B-field") which depends on the geometry under consideration. This vector satisfies the following requirement: 
for all ${\bf d}$, $j_L$ and $j_R$ such that $N^{{\bf d}}_{j_L, j_R} $ is non-vanishing, we must have
\be
\label{B-prop}
(-1)^{2j_L + 2 j_R+1}= (-1)^{{\bf B} \cdot {\bf d}}. 
\ee
We note that the characterization above only defines $\bf B$ up to $(2\mathbb{Z})^{n_\Sigma}$. A difference choice of $\bf B$ 
does not change $\mathsf{J}_X^{\rm WS}(\boldsymbol{\mu},\boldsymbol{\xi},\hbar)$. The {\it total, modified grand potential} is the sum of the above two functions, 
\be
\label{jtotal}
\mathsf{J}_{X}(\boldsymbol{\mu}, \boldsymbol{\xi},\hbar) = \mathsf{J}^{\rm WKB}_X (\boldsymbol{\mu}, \boldsymbol{\xi},\hbar)+ \mathsf{J}^{\rm WS}_X 
(\boldsymbol{\mu},  \boldsymbol{\xi} , \hbar), 
\ee
and it was introduced in \cite{hmmo}. Note that if we define
\begin{equation}\label{eq:F-top}
	F_{\rm top}(\mathbf{t},g_s) = \frac{1}{6 g_s^2}\sum_{i,j,k=1}^{n_\Sigma} a_{ijk} t_i t_jt_k + \sum_{i=1}^{n_\Sigma} \(b_i + \frac{4\pi^2}{g_s^2} b_i^{\rm NS}\)t_i + F^{\rm GV}(\mathbf{t},g_s) \ ,
\end{equation}
which differs from $F^{\rm}(\mathbf{t},g_s)$ by the terms proportional to $C_g$, $\mathsf{J}_{X}(\boldsymbol{\mu}, \boldsymbol{\xi},\hbar)$ can be written in a slightly more compact form, i.e.\footnote{Note 
that $F_{\rm top}(\mathbf{t},g_s)$ is $F^{\rm WS}(\mathbf{t},g_s)$ with the contributions from the constant maps removed. 
The latter can be understood as absorbed in $A(\boldsymbol{\xi}, \hbar)$.}
\begin{equation}\label{eq:J}
\begin{aligned}
	J(\boldsymbol{\mu},\boldsymbol{\xi},\hbar) = &\sum_{i=1}^{n_\Sigma}\frac{t_i}{2\pi}\frac{\partial}{\partial t_i} F_{\rm NS}^{\rm inst}(\mathbf{t}(\hbar),\hbar) + \frac{\hbar^2}{2\pi}\frac{\partial}{\partial \hbar}\( \frac{F_{\rm NS}^{\rm inst}(\mathbf{t}(\hbar),\hbar)}{\hbar} \) + A(\boldsymbol{\xi}, \hbar) \\
	&+ \widehat{F}_{\rm top}\( \frac{2\pi}{\hbar}\mathbf{t}(\hbar), \frac{4\pi^2}{\hbar}\) \ ,
\end{aligned}
\end{equation}
where all but the constant $A(\boldsymbol{\xi}, \hbar)$ are hidden in (refined) topological string free energies. 
Here we introduce the notation $\widehat{f}(\mathbf{t})$, meaning that $\mathbf{t}$ is shifted by $\pi \ri \mathbf{B}$ in the terms of order $\exp(-t_i)$ 
(the instanton contributions). In particular,
\be
	\widehat{F}_{\rm top} \( \mathbf{t}, g_s \) = \frac{1}{6 g_s^2}\sum_{i,j,k=1}^{n_\Sigma} a_{ijk} t_i t_jt_k + \sum_{i=1}^{n_\Sigma} \(b_i + \frac{4\pi^2}{g_s^2} b_i^{\rm NS}\)t_i + F^{\rm GV}(\mathbf{t}+\pi\ri\mathbf{B},g_s) \ .
\ee

According to the conjecture in \cite{ghm,cgm}, the 
spectral determinant (\ref{gsd}) is given by 
\be
\label{our-conj}
\Xi_X({\boldsymbol \kappa}; \hbar)= \sum_{ {\bf n} \in \IZ^{g_\Sigma}} \exp \left( \mathsf{J}_{X}(\boldsymbol{\mu}+2 \pi \ri  {\bf n}, \boldsymbol{\xi}, \hbar) \right). 
\ee
The right hand side of (\ref{our-conj}) defines a quantum-deformed (or generalized) Riemann theta function by 
\be
\label{qtf}
\Xi_X({\boldsymbol \kappa}; \hbar)= \exp\left( \mathsf{J}_{X}(\boldsymbol{\mu}, \boldsymbol{\xi}, \hbar) \right) \Theta_X({\boldsymbol \kappa}; \hbar). 
\ee
The r.h.s. of (\ref{our-conj}) can be computed as an expansion around the large radius point of moduli space. In the so-called ``maximally supersymmetric case" $\hbar=2 \pi$, it can be 
written down in closed form in terms of a conventional theta function. There is an equivalent form of the conjecture which gives an integral formula for the fermionic spectral traces: 
\be
\label{multi-Airy}
Z_X({\boldsymbol N}; \hbar)={1\over \left( 2 \pi \ri\right)^{g_\Sigma}} \int_{-\ri \infty}^{\ri \infty} \rd \mu_1 \cdots \int_{-\ri \infty}^{\ri \infty} \rd \mu_{g_\Sigma} \,
 \exp \left\{ \mathsf{J}_X ({\boldsymbol \mu},  {\boldsymbol {\xi}}, \hbar) - \sum_{i=1}^{g_\Sigma} N_i \mu_i \right\}. 
\ee
In practice, the contour integration along the imaginary axis can be deformed to a contour where the integral is convergent. For example, in the genus one case the integration contour is the one defining the Airy function  (see \cite{hmo2,ghm}). 

An important consequence of the representation (\ref{multi-Airy}) is the existence of a 't Hooft-like limit in which one can extract the genus expansion of the conventional 
topological string. The 't Hooft limit is defined by 
\be
\label{thooft-lam}
 \hbar \rightarrow \infty, \qquad N_i \rightarrow \infty, \qquad {N_i \over \hbar} =\lambda_i \, \, \, \, \, \text{fixed}, \quad i=1, \cdots, g_\Sigma. 
 \ee
In this 't Hooft limit, the integral in (\ref{multi-Airy}) can be evaluated in the saddle-point approximation, and in order to have a non-trivial 
result, we have to consider the modified grand potential in the limit 
\be
\label{thooft-mu}
 \hbar \rightarrow \infty, \qquad \mu_i \rightarrow \infty, \qquad {\mu_i \over \hbar} =\zeta_i \, \, \, \, \, \text{fixed}, \quad i=1, \cdots, g_\Sigma. 
 \ee
 In this limit, the quantum mirror map appearing in the modified grand potential becomes trivial. We will assume that the mass parameters $\boldsymbol{\xi}$ scale in such a way that 
\be
\label{mt}
\log \hat \xi_j = {2 \pi \over \hbar} \log  \xi_j, \qquad j=1, \cdots, r_\Sigma,
\ee
%
%
are fixed as $\hbar\rightarrow \infty$. In the regime (\ref{thooft-mu}), the modified grand potential has an asymptotic genus expansion of the form, 
\be
\mJ^{\text{'t Hooft}}_X\left( {\boldsymbol \zeta}, \hat {\boldsymbol \xi}, \hbar \right)= \sum_{g=0}^\infty \mJ^{X}_g \left({\boldsymbol \zeta}, \hat {\boldsymbol \xi}\right) \hbar^{2-2g}, 
\ee 
where
%
\be 
\label{gen-J-as}
\ba
 \mJ^{X}_0 \left( {\boldsymbol \zeta}, \hat {\boldsymbol \xi}\right)&={1\over 16 \pi^4} 
 \widehat F_0 \left( {\bf T}\right) +  A_0 (\hat {\boldsymbol \xi}) \\
 \mJ^{X}_1 \left( {\boldsymbol \zeta}, \hat {\boldsymbol \xi}\right)&=  A_1(\hat {\boldsymbol \xi})+  \widehat F_1 \left( {\bf T}\right), \\
  \mJ^{X}_g \left({\boldsymbol \zeta},\hat  {\boldsymbol \xi}\right)&=  A_g (\hat {\boldsymbol \xi}) +(4 \pi^2)^{2g-2} \left( \widehat F_g \left( {\bf T}\right)-C_g\right), \qquad g\ge 2. 
  \ea
  \ee
In these equations, we have introduced the rescaled K\"ahler parameter
\be
{\bf T}= {2 \pi \over \hbar} {\bf t}. 
\ee
The arguments ${\boldsymbol \zeta}$ and $\hat {\boldsymbol{\xi}}$ of the modified grand potential are related to the rescaled K\"ahler parameters ${\bf T}$ by
\be
\label{tzeta}
T_i -\sum_{j=1}^{r_\Sigma}\alpha_{ij}\log \hat \xi_j= 2 \pi \sum_{j=1}^{g_\Sigma} C_{ij} \zeta_j, \qquad i=1, \cdots, g_\Sigma+ r_\Sigma.
\ee
%
%
%
We have assumed that the function $A\left({\boldsymbol{\xi}}, \hbar\right)$ has the expansion 
\be
A\left({\boldsymbol{\xi}}, \hbar\right)= \sum_{g=0}^\infty A_g(\hat {\boldsymbol \xi}) \hbar^{2-2g}. 
\ee
%
The saddle point of the integral (\ref{multi-Airy}) as $\hbar \rightarrow \infty$ is then given by 
\be
\label{saddle}
\lambda_i =\sum_{j=1}^{n_\Sigma} {C_{ji} \over 8 \pi^3} \left( {\partial \widehat F_0 \over \partial T_j} + 4 \pi^2 b_j^{\rm NS} \right), \qquad i=1, \cdots, g_\Sigma.
\ee
It follows from this equation that the 't Hooft parameters are flat coordinates on the moduli space. The frame defined by these coordinates will be called the 
{\it maximal conifold frame} (\texttt{MCF}). The submanifold in moduli space defined by 
\be
\lambda_i=0, \qquad i=1, \cdots, g_\Sigma,
\ee
has dimension $r_\Sigma$ (the number of mass parameters of the toric CY), and we will call it the {\it maximal conifold locus} (\texttt{MCL}). It is a submanifold of the conifold locus of the CY $X$. 
By evaluating the integral (\ref{multi-Airy}) in the saddle-point approximation, we find that the fermionic spectral traces have the following asymptotic expansion in the 't Hooft limit: 
\be
\label{logz-exp}
\log \, Z({\boldsymbol{N}}; \hbar) \sim \sum_{g\ge 0} \CF_g({\boldsymbol{\lambda}}) \hbar^{2-2g}.
\ee
The leading function in this expansion is given by a Legendre transform, 
\be
\CF_0 ({\boldsymbol{\lambda}})=\mJ^X_0\left({\boldsymbol \zeta}, {\boldsymbol \xi} \right)- {\boldsymbol \lambda}\cdot  {\boldsymbol \zeta}.  
\ee
If we choose the Batyrev coordinates in such a way that the first $g_\Sigma$ correspond to true moduli, and the remaining $r_\Sigma$ correspond to mass parameters, we find 
\be
\label{der-orbi}
{\partial \CF_0 \over \partial \lambda_i}= -\zeta_i =-\sum_{j=1}^{g_\Sigma}{C_{ij}^{-1} \over 2 \pi} \left( T_j  -\sum_{k=1}^{r_\Sigma}\alpha_{jk}\log \xi_k\right), \qquad i=1, \cdots, g_\Sigma,
\ee
%
%
where $C^{-1}$ denotes  the inverse of the truncated matrix \eqref{tmc}. In view of the results of \cite{abk}, the higher order corrections $\CF_g({\boldsymbol{\lambda}})$ can be computed in a very 
simple way: the integral (\ref{multi-Airy}) implements a symplectic transformation from the \texttt{LRF} to the \texttt{MCF}. The functions $\CF_g({\boldsymbol{\lambda}})$ are precisely 
the genus $g$ free energies of the topological string in the \texttt{MCF}.

\subsection{Perturbed $\mathsf{O}_{m,n}$ operators}
\label{sc:Omn}

In order to test the conjectures of \cite{ghm,cgm}, it is very useful to have explicit results on the spectral theory side. In particular, since we have a precise conjecture for the values of the 
fermionic spectral traces of the relevant operators in terms of enumerative invariants, we would like to have independent, analytic computations of these traces. 

In many cases, the operators which appear in the quantization of mirror curves are perturbations of three-term operators of the form 
\be
\label{omn}
\mathsf{O}_{m,n}= \re^{\mx} + \re^\my+ \re^{-m \mx-n \my} \ , \qquad m,n\in \bZ_{>0} \ .
\ee
These operators were introduced and studied in \cite{kasmar}. It turns out that the integral kernel of their inverses $\rho_{m,n}= \mathsf{O}_{m,n}^{-1}$ can be explicitly computed in terms 
of Faddeev's quantum dilogarithm. This makes it possible to calculate the standard traces 
\be
\label{bosmn}
\tr \, \rho_{m,n}^\ell, \qquad \ell=1, 2, \cdots, 
\ee
in terms of integrals over the real line. These integrals can be computed by using the techniques of \cite{garkas}, or by using the 
recursive methods developed in \cite{tw,py,hmo,oz}. Once the ``bosonic" spectral 
traces (\ref{bosmn}) have been computed, the fermionic spectral traces follow by simple combinatorics. Mixed traces, as those 
appearing in (\ref{11-mixed}), (\ref{12-mixed}), can be also obtained in terms of integrals. 

In this section, we will study a four-term operator which is a perturbation of (\ref{omn}). This operator reads, 
\be
\label{pertmn}
\mathsf{O}_{m,n, \xi}= \re^{\mx} + \re^\my+ \re^{-m \mx-n \my}+ \xi \re^{-(1+m) \mx -(n-1) \my}  \ , \qquad m,n\in \bZ_{>0} \ .
\ee
Let us introduce the parameter $\zeta$ by\footnote{Do not confuse the $\zeta$ introduced here with the 't Hooft variables defined in \eqref{thooft-mu}.}
\be
\label{xizeta}
\xi= \re^{2 \pi \mb \zeta}. 
\ee
We will now obtain an explicit expression for the integral kernel of 
\be
\rho_{m,n, \xi}=\mathsf{O}_{m,n,\xi}^{-1}, 
\ee
based on similar derivations in \cite{kasmar} and \cite{kmz}. First, as in \cite{kasmar}, we introduce the Heisenberg 
operators $\mq$ and $\mm$ satisfying the normalized commutation relations
\be
[\mm, \mq]= {1\over 2 \pi \ri}. 
\ee
They are related to the operators $\mx$, $\my$ appearing in (\ref{pertmn}) by 
\begin{equation}\label{eq:xy-pq}
\mathsf{x}\equiv 2\pi\mathsf{b}\frac{(n+1)\mathsf{p}+n\mathsf{q}}{m+n+1},\quad \mathsf{y}\equiv -2\pi\mathsf{b}\frac{m\mathsf{p}+(m+1)\mathsf{q}}{m+n+1}, 
\end{equation}
so that
\be
\label{b-hbar}
\hbar=\frac{2\pi\mathsf{b}^2}{m+n+1}.
\ee
We then have, 
\be
\ba
\re^{-\my/2} \mO_{m,n,\xi} \re^{-\my/2}&= \re^{\mx- \my} + 1+ \re^{-m \mx -(n+1) \my}+ \xi r\re^{-(1+m) \mx -n \my}\\
&= \re^{2 \pi \mb (\mm+ \mq)}+ 1+ \re^{2 \pi \mb \mq} + \xi \re^{-2 \pi \mb \mm}\\
&= \re^{-\pi \mb \mm} \left(  \re^{2 \pi \mb (2\mm+ \mq)}+ \re^{2 \pi \mb \mm}+ \re^{2 \pi \mb \mq}  + \xi \right) \re^{- \pi \mb \mm}.
\ea
\ee
We now use Faddeev's quantum dilogarithm $\fad(x)$ \cite{fk,faddeev-penta,faddeev} (our conventions for this function are as in \cite{kasmar}). 
It satisfies the following identity (a similar identity was already used in \cite{kasmar}):
\be
 \fad(\mathsf{p})\fadi(\mathsf{q})\re^{2\pi\mathsf{b}\mathsf{p}}\fad(\mathsf{q})\fadi(\mathsf{p})
=\re^{2 \pi \mb (2\mm+ \mq)}+ \re^{2 \pi \mb \mm}+ \re^{2 \pi \mb (\mm+\mq)}. 
\ee
Its behaviour under complex conjugation is given by, 
\begin{equation}
	\fadi(z)={1\over \fad(z^*)} \ .
\end{equation}
Let us denote 
\be
\widehat \mO_{m,n,\xi}= \re^{\pi \mb \mm-\my/2} \mO_{m,n,\xi} \re^{\pi \mb \mm-\my/2}. 
\ee
We now recall that Faddeev's quantum dilogarithm satisfies the following difference equation, 
\be
\label{fad-diff}
\frac{\fad(x+c_{\mathsf{b}}+\im \mathsf{b})}{
\fad(x+c_{\mathsf{b}})} 
= \frac{1}{1-q \re^{2 \pi \mathsf{b} x}}, 
\ee
where 
\be
q=\re^{2 \pi \im \mathsf{b}^2}, \qquad c_{\mathsf{b}}=\im { \mathsf{b}+\mathsf{b}^{-1} \over 2}.
\ee
By using this equation, we obtain 
\be
\ba
\fad(\mq) \fadi(\mm) \widehat \mO_{m,n,\xi} \fad(\mm) \fadi(\mq)&= \xi + \re^{2 \pi \mb \mm}= \xi \left( 1+ \re^{2\pi \mb (\mm-\zeta)}\right)  \\
&= \xi {\fad(\mm-\zeta- \ri \mb/2) \over \fad(\mm-\zeta+\ri \mb/2)},
\ea
\ee
By taking the inverse, we find, 
\be
\fad(\mq) \fadi(\mm)  \re^{-\pi \mb \mm+\my/2}\xi \mO^{-1}_{m,n,\xi}  \re^{-\pi \mb \mm+\my/2} \fad(\mm) \fadi(\mq)= {\fad(\mm-\zeta+ \ri \mb/2) \over \fad(\mm-\zeta-\ri \mb/2)}, 
\ee
and we can write, 
\be
\xi \mO^{-1}_{m,n,\xi} = \re^{\pi \mb \mm-\my/2} \fad(\mm) \fadi(\mq) {1\over \fad(\mm-\zeta-\ri \mb/2)} \fad(\mm-\zeta+ \ri \mb/2) \fad(\mq) \fadi(\mm) \re^{\pi \mb \mm-\my/2}.
\ee
We now use the quantum pentagon identity \cite{faddeev-penta}
\be
\fad(\mm-p_0) \fad(\mq) =\fad(\mq) \fad(\mm-p_0 +\mq) 
\fad(\mm-p_0), 
\ee
where $p_0= \zeta-\im \mb /2$, to obtain
\be
\xi \mO^{-1}_{m,n,\xi} = \re^{\pi \mb \mm-\my/2}  {\fad(\mm) \over \fad(\mm-p_0^*)} {1\over 1+ \re^{2 \pi \mb (\mm+ \mq-\zeta)}} {\fad(\mm-p_0) \over \fad(\mm)}\re^{\pi \mb \mm-\my/2}. 
\ee
Let us now introduce the parameters \cite{kasmar}
\be
\label{ac-ops}
a={\mb m\over 2(m+n+1)}, \qquad c= {\mb \over 2(m+n+1)}, \qquad h = a+c-nc, 
\ee
as well as the function
\begin{equation}\label{mypsi}
\mypsi{a}{c}(x)\equiv \frac{\re^{2\pi ax}}{\fad(x-\im(a+c))}. 
\end{equation}
We will also relabel
\be\label{eq:q-relabel}
\mm+ \mq-\zeta \rightarrow \mq. 
\ee
Then, we have
\be
\label{ex-op}
\rho_{m,n, \xi}=\mathsf{O}_{m,n,\xi}^{-1}
=\mypsi{a}{c}^*(\mathsf{p}) \mypsi{nc}{0}(\mathsf{p}-\zeta)\left|\mypsi{a+c}{cn}(\mathsf{q})\right|^2\mypsi{a}{c}(\mathsf{p})\mypsi{nc}{0}^*(\mathsf{p}-\zeta), 
\ee
where we used again the property (\ref{fad-diff}). In particular, its kernel can be written, in the momentum representation
\be
\mm|x)= x|x), 
\ee
as
\be
\label{rhoxy}
(x| \rho_{m,n, \xi}|y)=\rho_{m,n, \xi}(x, y)= {{\overline{f(x)}} f(y) \over 2 \mb \cosh\left( \pi {x-y+\im h \over \mb} \right)}, 
\ee
where
\be
\label{fx}
f(x)= \mypsi{a}{c}(x)\mypsi{nc}{0}^*(x-\zeta)= {\fad(x-\zeta + \ri n c ) \over \fad(x-\im (a+c))} \re^{ 2 \pi (a+ n c) x } \re^{ -2 \pi  c n  \zeta}. 
\ee
It can be easily checked, by using the properties of Faddeev's quantum dilogarithm, that as $\xi \rightarrow 0$ (which corresponds to $\zeta \rightarrow -\infty$), we recover 
the kernel of $\mathsf{O}_{m,n}$ determined in \cite{kasmar}. 

Using the results (\ref{rhoxy}), (\ref{fx}), it is possible to compute explicitly the spectral traces of $\rho_{m,n, \xi}$. Let us consider an example which will be relevant for 
the conjectural relation with the topological string: we consider $m=n=1$, so that we consider a perturbation of the operator associated to local $\IP^2$ \cite{ghm}. Let us focus on the maximally supersymmetric case $\hbar=2 \pi$, which corresponds to $\mb={\sqrt{3}}$. The diagonal integral kernel is given by 
\be
\ba
\rho_{1,1, \xi}(x, x)&= {1\over 2 \mb \cos(\pi/6)} \re^{4 \pi x/\mb- 2 \pi \zeta/\mb} { \fad(x+ \ri \mb/3) \over \fad(x- \ri \mb/3) }  { \fad(x-\zeta+ \ri \mb/6) \over \fad(x-\zeta- \ri \mb/6) } \\
&= {1\over 2 \mb \cos(\pi/6)} { \re^{4 \pi x/\mb- 2 \pi \zeta/\mb}  \over 
(1+ \re^{2 \pi x/  \mb }+ \re^{4 \pi x/\mb })(1+ \re^{2 \pi(x-\zeta)/\mb })}.
\ea
\ee
This can be integrated, to obtain
\be
\label{eq:trRho}
\tr \rho_{1,1, \xi}= \frac{\pi  \left(2-\xi^{1/3} \right)+\sqrt{3} \xi^{1/3}  \log (\xi )}{18 \pi 
   \left(\xi ^{2/3}-\xi^{1/3}+1\right)}. 
   \ee
   Similarly, one finds\footnote{We would like to thank Szabolcs Zakany for adapting the techniques of \cite{oz} to check this result.} 
	\begin{equation}\label{eq:tr11sq}
	\begin{aligned}
		\tr \rho_{1,1,\xi}^2 =& -\frac{6\sqrt{3}(1+\xi)+ \pi(-4+4\xi^{1/3}-13\xi^{2/3}+6\xi)}{108\pi(1+\xi^{1/3})(1-\xi^{1/3}+\xi^{2/3})^2} \\
		&\phantom{}+ \frac{\xi^{2/3}\log \xi}{18\sqrt{3}\pi(1-\xi^{1/3}+\xi^{2/3})^2} + \frac{\xi^{2/3}(\log\xi)^2}{36\pi^2(1+\xi^{1/3})(1-\xi^{1/3}+\xi^{2/3})^2} \ .
	\end{aligned}	
	\end{equation}
These traces are functions of $\xi$ with a branch cut at $\xi=0$. Their limit when $\xi \rightarrow 0$ exists and gives back the traces for the operator $\rho_{1,1}\equiv\rho_{1,1,0}$ calculated in 
for example \cite{kasmar}. The theory for $\xi=1$ is particularly simple and we find 
\be
\label{xi-one-trs}
\tr \rho_{1,1, \xi=1}={1\over 18}, \qquad \tr \rho^2_{1,1, \xi=1}=\frac{7}{216}-\frac{1}{6 \sqrt{3} \pi }.
\ee

\section{The $Y^{3,0}$ Geometry}
\label{sc:Y30-gm}


In this section, we test the conjectural correspondence between spectral theory and topological strings in a genus two example with one mass parameter, namely the $Y^{3,0}$ geometry.

\subsection{Mirror curve and operator content}
\label{sc:Y30}

The generic $Y^{N,q}$ geometry\footnote{To be precise, we are talking about the resolution of the cone over the $Y^{N,q}$ singularity.} 
has been studied in some detail in \cite{bt}. The mirror curve is given by
\be\label{eq:Nqcurve}
a_1 \re^p +a_2 \re^{-p + (N-q) x} + \sum_{i=0}^N b_i \re^{i x}=0 \ ,
\ee
where $p, x$ are variables. Here we are interested in the case $q=0$. In particular, when $N=3$, $q=0$, the mirror curve has the form 
\be
\label{mcurve}
a_1 \re^p + a_2\re^{-p+ 3x} + b_{3} \re^{3x}+ b_{2} \re^{2x} + b_{1} \re^{x} + b_{0} =0.
\ee
The corresponding charge vectors are 
\be
\label{y3-qs}
Q^1=(0,0,1,-2,1,0),\qquad 
Q^2=(0,0,0,1,-2,1),\qquad
Q^3= ( 1,1,-1,0,0,-1),
\ee
and the Batyrev coordinates are 
\be
\label{eq:A2-z}
 z_1= {b_3 b_1 \over b_2^2}, \qquad z_2= {b_0 b_2 \over b_1^2} , \qquad z_3= {a_1 a_2 \over b_3 b_0} \ .
\ee
%

The canonical forms of the mirror curve (\ref{mcurve}) are
\be
\label{op-curves}
\ba
&b_3 \re^x+ \re^y + \re^{-x-y}+  b_0 \re^{-2x}+ b_1\re^{-x} + b_2=0,\\
& b_0 \re^u+ \re^y + \re^{-u-y}+  b_3 \re^{-2u}+ b_2 \re^{-u} +b_1=0, 
\ea
\ee
where we set $a_1=a_2=1$. To obtain the first curve, start from (\ref{mcurve}) and set
\be
p=y+2x. 
\ee
After multiplying the resulting equation by $\re^{-2x}$, we obtain precisely the first curve in (\ref{op-curves}). 
To obtain the second curve, we multiply (\ref{mcurve}) 
by $\re^{-x}$ and we perform the symplectic transformation from $(p,x)$ to $(y, u)$ 
\be
y= 2x-p, \quad u=-x. 
\ee
Note that both curves are identical after exchanging $b_1 \leftrightarrow b_2$ and $b_0 \leftrightarrow b_3$. 
We regard $b_1, b_2$ as moduli of the curve, and $b_3, b_0$ as parameters. We can further set one of $b_0, b_3$ to one by using the remaining $\bC^*$ rescaling freedom, but we refrain from doing it to make the symmetry between the two canonical forms in \eqref{op-curves} apparent.

The operators $\sO_i^{(0)}$, $i=1,2$, obtained by quantization of the canonical forms are
\be
\label{eq:Toda-Os}
\ba
\sO_1^{(0)}&= b_3 \re^\sx+ \re^\sy + \re^{-\sx-\sy}+  b_0 \re^{-2\sx},\\
\sO_2^{(0)}&= b_0 \re^\su+ \re^\sy + \re^{-\su-\sy}+  b_3 \re^{-2\su}. 
\ea
\ee
Both of them can be regarded as perturbations of the operator $\sO_{1,1}$ introduced in (\ref{omn}), up to an overall normalization. In the first case, 
we change the normalization of $\sx$, $\sy$ in such a way that 
\be
\re^\sx \rightarrow b_3^{-2/3} \re^\sx, \qquad \re^\sy \rightarrow b_3^{1/3} \re^\sy, 
\ee
and then we find
\be
 b_3 \re^\sx+ \re^\sy + \re^{-\sx-\sy}+  b_0 \re^{-2\sx}+ b_1 \re^{-\sx} \rightarrow b_3^{1/3} \left( \re^\sx+ \re^\sy +  \re^{-\sx-\sy}+ b_3 b_0 \re^{-2\sx}+ b_1  b_3^{1/3} \re^{-\sx} \right), 
 \ee
 %
Note that, when $b_1=0$, the operator inside the parentheses is precisely the operator $\mO_{1,1,\xi}$ introduced in (\ref{pertmn}), where
\be
\label{eq:xi-z3}
\xi=b_3b_0 = {1\over z_3}. 
\ee
In the second operator the normalization is fixed by requiring
\be
\re^\su \rightarrow b_0^{-2/3} \re^\su, \qquad \re^\sy \rightarrow b_0^{1/3} \re^\sy, 
\ee
and then we find
\be
 b_0 \re^\su+ \re^\sy + \re^{-\su-\sy}+  b_3 \re^{-2\su}+ b_2 \re^{-\su} \rightarrow b_0^{1/3} \left( \re^\su + \re^\sy +  \re^{-\su-\sy}+ b_3 b_0 \re^{-2\su}+ b_2  b_0^{1/3} \re^{-\su} \right), 
 \ee
Note in particular that 
\be
\label{fac-cor}
\ba
\tr \left(\rho_1^{(0)}\right)^N&= b_3^{-N/3} \tr \rho_{1,1,\xi}^N, \\
\tr \left(\rho_2^{(0)}\right)^N&= b_0^{-N/3} \tr \rho_{1,1,\xi}^N.
\ea
\ee
In the following, we will set
\be
\kappa_1=b_2, \qquad \kappa_2= b_1. 
\ee
Note that the truncated matrix appearing in (\ref{tmc}) is in this case the Cartan matrix of $SU(3)$, 
\be
\label{cartan3}
C=\begin{pmatrix} 2 &-1 \\ -1& 2 \end{pmatrix}, 
\ee
with inverse
\be
C^{-1}={1\over 3} \begin{pmatrix} 2 &1 \\ 1& 2 \end{pmatrix}.
\ee

\subsection{'t Hooft expansion of the fermionic traces}

One of the main points of \cite{ghm,cgm}, as we reviewed in section~\ref{sc:spec-det}, is that the topological string amplitudes in the \texttt{MCF} can be obtained from the 't Hooft limit (\ref{thooft-lam}) 
of the fermionic spectral traces (\ref{or-exp}). Therefore, we have the asymptotic expansion given in (\ref{logz-exp}). When the genus of the mirror curve is two, we can write
	\begin{equation}\label{eq:logZ-Fc}
		\log Z(N_1,N_2;\hbar) = \cF(\lambda_1,\lambda_2;\hbar) = \sum_{g=0}^{\infty} \hbar^{2-2g} \cF_{g}(\lambda_1,\lambda_2) \ ,
	\end{equation}
	where $\cF_{g}(\lambda_1,\lambda_2)$ is the topological string free energy at genus $g$ in the \texttt{MCF}, and the 't Hooft parameters $\lambda_i$ are identified with suitable flat coordinates in the \texttt{MCF} as 
	specified in (\ref{saddle}). The genus $g$ topological string free energy can be in turn expanded around the \texttt{MCL} $\lambda_1= \lambda_2=0$, and one finds 
	\begin{equation}\label{eq:Fc-exp}
		\cF_g(\lambda_1,\lambda_2) = \cF^{\rm log}(\lambda_1,\lambda_2) \delta_{g,0} +  \sum_{\substack{i,j=0\\ (i,j)\neq (0,0)}}^\infty \cF_{g;i,j} \lambda_1^i \lambda_2^j \ ,
	\end{equation}
	where $\cF^{\rm log}(\lambda_1,\lambda_2)$ contains $\log \lambda_1, \log \lambda_2$, and the remaining part is a power series in $\lambda_1, \lambda_2$. The numbers $\cF_{g;i,j}$ can be regarded as ``conifold" 
	Gromov-Witten invariants. In order to extract these invariants from the fermionic traces, there are various techniques that we can use. First of all, note that if $N_1=0$ or $N_2=0$, the 
	fermionic traces can be computed from a matrix model with a single set of eigenvalues, as in \cite{mz,kmz}. The matrix integral is given by 
\be
\label{zmn-bis}
Z_{m,n, \xi}(N,\hbar)=\frac{1}{N!}  \int_{\IR^N}  { \rd^N u \over (2 \pi)^N}  \prod_{i=1}^N \re^{-{1\over \mg} V_{m,n, \xi}(u_i,\mg)}  \frac{\prod_{i<j} 4 \sinh \left( {u_i-u_j \over 2} \right)^2}{\prod_{i,j} 2 \cosh \left( {u_i -u_j \over 2} + \ri \pi C_{m,n} \right)}, 
\ee
where 
\be
\mg={m+n+1\over 2 \pi\mb^2}, \qquad C_{m,n}=\frac{m-n+1}{2(m+n+1)}.
\ee
The potential $V_{m,n, \xi}(u,\mg)$ is given by 
\be\label{eq:Vmnxi}
V_{m,n,\xi}(u, \mg)= -\mg \left( {(m+n) \mb^2  \over m+n+1} u - {2 \pi \mb n \zeta \over m+n+1} \right) -2  \mg \log \left| {\fad\left( p+ \ri(a+c) \right) \over \fad(p-\zeta-\ri c n) }\right|, 
\ee
where $u$ is related to $p$ by 
\be
\qquad u={2 \pi \over \mb}p, 
\ee
while $\xi$ is related to $\zeta$ by (\ref{xizeta}). The values of $a$, $c$ are given in (\ref{ac-ops}). Then, due to the correction in (\ref{fac-cor}), we find 
the following formulae for the fermionic traces: 
 \be
 \label{zzpaff}
 \ba
\log  Z(N,0; \hbar)&=\log Z_{1,1,\xi}(N, \hbar) -{\hbar^2  \lambda \over 2 \pi} {1\over 3} \log \widehat b_3, \\
\log  Z(0,N; \hbar)&=\log Z_{1,1,\xi}(N, \hbar) -{\hbar^2  \lambda \over 2 \pi} {1\over 3} \log \widehat b_0,  
  \ea
  \ee
where the hatted parameters are defined as in (\ref{mt}), i.e. 
\be
\widehat b_3= b_3^{2 \pi/\hbar}, \qquad \widehat b_0= b_0^{2 \pi/\hbar}. 
\ee

In practice, the expansion (\ref{eq:Fc-exp}) can be computed as follows. Replacing $\lambda_i$ by $N_i/\hbar$, one gets 
	\begin{equation}
		\log Z(N_1,N_2;\hbar) = \cF^{\rm log}(N_1/\hbar,N_2/\hbar)\hbar^{-2} + \sum_{n\geqslant -1} \hbar^{-n} \sum_{\substack{i,j\geqslant 0,\,(i,j)\neq (0,0)\\ i+j\equiv n\textrm{ mod } 2}}^{i+j\leqslant n+2} \cF_{\tfrac{n-i-j}{2}+1;i,j} N_1^i N_2^j \ .
	\end{equation}
	Note that since $\cF^{\rm log}(\lambda_1,\lambda_2) \sim \log(\lambda_i)\lambda_i^2$, the first term $\cF^{\rm log}(N_1/\hbar,N_2/\hbar)\hbar^{-2}$ is of the order $\cO(\hbar^0)$. We find out that in the power series part of $\log Z(N_1,N_2;\hbar)$, only a finite number of \texttt{MCF} Gromov-Witten invariants contributes at each order $\hbar^{-n}$. 
	Therefore a simple linear algebra calculation allows us to extract the numbers $\cF_{g;i,j}$ from the large $\hbar$ expansion of $\log Z(N_1,N_2; \hbar)$, evaluated with different values of $N_1, N_2$. For instance,
	\begin{equation}\label{eq:cF-von-logZ}
	\begin{aligned}
		\cF_{0;3,0} &= \frac{1}{6}\(-2\log Z(1,0) + \log Z(2,0)\)\big|_{\hbar^{-1}} \ , \\
		\cF_{1;1,0} &= \frac{1}{6}\( 8\log Z(1,0) - \log Z(2,0)\)\big|_{\hbar^{-1}} \ , \\
		\cF_{0;2,1} &= \frac{1}{2}\( \log Z(0,1)+2\log Z(1,0) -2\log Z(1,1) -\log Z(2,0) + \log Z(2,1)\)\big|_{\hbar^{-1}} \ .
	\end{aligned}	
	\end{equation}
	In the above expressions, we suppress ``$;\hbar$'' in $Z(N_1,N_2;\hbar)$ to simplify the notation. We use $|_{\hbar^{-n}}$ to denote the contribution of the power series at the order $\hbar^{-n}$.
	
	To evaluate $Z(N_1,N_2;\hbar)$ directly, we should choose a canonical form of the quantum mirror curve, read off the operators $\sA_{j1}, \sA_{j2}$, and compute the 
	fermionic traces in terms of the spectral traces of $\sA_{j1}, \sA_{j2}$ by \eqref{gtwo-matrix}. We will choose $j=1$, simplify notation by setting $\sA_{1i}= \sA_i$, $i=1,2,$ 
	and put $b_3=1$ for simplicity. The mass parameter $\xi$ is then $b_0$. It is easy to see that
	\begin{equation}
		\sA_{1} = \rho_{1,1,\xi} \ ,\quad \sA_{2} = \rho_{1,1,\xi} \re^{-\sx} \ .
	\end{equation}
	The operator $\rho_{1,1,\xi}$ is an operator of the type $\rho_{m,n,\xi}$ studied in detail in Sec.~\ref{sc:Omn}. Its kernel in the $p$-space (recall the relation between the canonical variables $(x,y)$ and the canonical variables $(p,q)$ in \eqref{eq:xy-pq} as well as the relabeling \eqref{eq:q-relabel}) is given in \eqref{rhoxy}, \eqref{fx} with $m=n=1$.
	 Working carefully in the space of the variable $p$, we find the kernel of the second operator
	\begin{equation}\label{eq:ker-A2}
		A_{2}(x,y) = \re^{-\tfrac{\pi\ri}{9}\sb^2} \re^{-\tfrac{2\pi\sb y}{3}} \rho_{1,1,\xi}\(x,y+\frac{\ri \sb}{3}\) \ .
	\end{equation}
	
	Recall that $\hbar$ is proportional to $\sb$ by \eqref{b-hbar}. To perform the large $\hbar$ expansion of the traces, we need the large $\sb$ expansion of the Faddeev's quantum dilogarithm
	\begin{equation}\label{eq:Phib-expn}
	\begin{aligned}
		\log \Phi_\sb(x+\delta\sb) &= \frac{\sb^2}{2\pi\ri}\sum_{k=0}^\infty \sb^{-k} \sum_{j=0}^{\lfloor k/4 \rfloor} \Li_{2+2j-k}\( -\re^{2\pi\delta} \) \frac{(2\pi)^{k-2j}(-1)^j B_{2j}(1/2)}{(k-4j)!(2j)!} x^{k-4j} \\
		&=-\ri \sb^2 \frac{\Li_2(-\re^{2\pi\delta})}{2\pi} + \ri \sb \log(1+\re^{2\pi\delta}) x + \cO(\sb^{-1}) \ ,
	\end{aligned}	
	\end{equation}
	where $B_{2n}(x)$ are Bernoulli polynomials, and $\delta$ an arbitrary constant. Then, for instance, $\tr\, \sA_1$ has the following form
\be
\tr\,  \sA_1 = \int_{-\infty}^\infty \rd x \, \sA_1(x,x) = \int_{-\infty}^\infty \rd x \, \re^{\alpha \sb^2 + \ri \beta \sb x - \gamma x^2} \sum_{k=0}^\infty \sb^{-k} p_k(x) \ ,
\ee
	where $p_k(x)$ is a polynomial in $x$. To facilitate the calculation of the integral, one can eliminate the linear term $\ri\beta \sb x$ in the exponential by giving an appropriate shift $\delta \sb$ to $x$ before expanding the quantum dilogarithms, such that at each order of $\sb^{-k}$, one has a Gaussian integral or a derivative thereof. The same trick can be applied when the other spectral traces are evaluated.
	
	To simplify the calculation, we take $\xi = 1$. Then we obtain the traces of $\sA_1$, for instance
	\begin{equation}
	\begin{aligned}
		\tr \, \sA_1 &= o(\sb^0) +\frac{4 \pi ^2}{75 \sqrt{3} \sb^4}-\frac{116 \pi ^3}{3375 \sqrt{5} \sb^6}+\frac{452 \pi ^4}{28125 \sqrt{3} \sb^8}+\cO\left(\sb^{-9}\right) \ ,\\
		\tr \, \sA_1^2 &= o(\sb^0) -\frac{8 \pi }{9 \sqrt{15} \sb^2}+\frac{152 \pi ^2}{675 \sb^4}-\frac{12616 \pi ^3}{10125 \sqrt{15} \sb^6}+\frac{1575688 \pi ^4}{2278125
			\sb^8}+\cO\left(\sb^{-9}\right) \ ,
	\end{aligned}
	\end{equation}
	as well as the mixed traces of $\sA_1$ and $\sA_2$
	\begin{equation}
	\begin{aligned}
		\tr \, \sA_1 \sA_2 &= o(\sb^0) -\frac{2 \pi }{5 \sqrt{15} \sb^2}+\frac{26 \pi ^2}{375 \sb^4}-\frac{1468 \pi ^3}{5625 \sqrt{15} \sb^6}+\frac{47242 \pi ^4}{421875
			\sb^8}+\cO\left(\sb^{-9}\right) \ ,\\
		\tr \, \sA_1^2 \sA_2 &= o(\sb^0)-\frac{58 \pi }{225 \sqrt{5} \sb^2}+\frac{1138 \pi ^2}{5625 \sqrt{3} \sb^4}-\frac{17968 \pi ^3}{50625 \sqrt{5} \sb^6}+\frac{10393358 \pi ^4}{18984375
			\sqrt{3} \sb^8}+\cO\left(\sb^{-9}\right) \ .
	\end{aligned}	
	\end{equation}
	Here $o(\sb^0)$ means only terms of the order $\sb^0$. In addition, \eqref{fac-cor} implies that the traces of $\sA_2$ are straightforwardly related to those of $\sA_1$. In fact when $\xi=1$, one can show that
	\begin{equation}
		\tr \, \sA_1^r\sA_2^s = \tr \sA_1^s \sA_2^r \ .
	\end{equation}
	
	Once we have the large $\sb$ expansion of the traces of $\sA_1,\sA_2$, we can easily extract the \texttt{MCF} Gromov-Witten invariants. When $\xi=1$, we get for instance
	\begin{equation}\label{eq:Fc-values}
	\begin{gathered}
		\cF_{0;3,0} = -\frac{16\pi^2}{45\sqrt{15}} \ ,\;\cF_{0;4,0} =  \frac{866 \pi ^4}{30375} \ ,\; \cF_{0;5,0} = -\frac{102848 \pi ^6}{1366875 \sqrt{15}} \ , \\
		\cF_{1;1,0} = \frac{16 \pi ^2}{45 \sqrt{15}} \ ,\; \cF_{1;2,0} = -\frac{146 \pi ^4}{30375} \ ,\; \cF_{1;3,0} = -\frac{9472 \pi ^6}{1366875 \sqrt{15}} \ , \\
		\cF_{0;2,1} = \cF_{0;1,2} = \frac{\pi^2}{\sqrt{15}} \ .
	\end{gathered}	
	\end{equation}
	We will check these numbers against the computation in the topological string theory.
	
Another set of quantities that can be easily computed from operator theory are the derivatives of the genus zero free energy at the \texttt{MCL}. To do 
this computation, we first note that, as shown in \cite{mz}, the linear term in $\lambda$ of the planar free energy of the matrix integral (\ref{zmn-bis}) is given by 
the ``classical" limit (as $\mg\rightarrow 0$) of the potential, evaluated at its minimum. Using the expression of the potential \eqref{eq:Vmnxi} and the expansion of the quantum dilogarithm \eqref{eq:Phib-expn}, the classical limit is found to be
\be
\label{cp}
V^{(0)}_{m,n,\xi}(u)=-{m+n \over 2 \pi} u -{m+n+1\over 2 \pi^2} {\rm Im} \left\{ {\rm Li}_2 \left( -\re^{u + {\pi \ri (m+1) \over m+n+1}} \right) +  
 {\rm Li}_2 \left( -\re^{u-\widehat\zeta + {\pi \ri n \over m+n+1}} \right) \right\} + {n \widehat \zeta \over 2 \pi}, 
 \ee
 where
\be
\widehat \zeta= {2 \pi \zeta \over \mb}. 
\ee
One can check that, for $m=n=1$, the minimum of this potential occurs at
\be
\re^{u_\star}= {1\over 2} \left(1+ {\sqrt{1+ 4 \, \re^{\widehat\zeta}}} \right). 
\ee
If we now use (\ref{zzpaff}), we find that
\begin{equation}
\label{ders-con}
\ba
	\frac{\partial \cF_0}{\partial \lambda_1}\Bigg|_{\lambda_j=0} &= -V^{(0)}_{1,1,\xi} (u_\star) -{1\over 6 \pi} \log \widehat b_3, \\
\frac{\partial \cF_0}{\partial \lambda_2}\Bigg|_{\lambda_j=0} &= -V^{(0)}_{1,1,\xi} (u_\star) -{1\over 6 \pi} \log \widehat b_0. 
\ea
\end{equation}
Thanks to (\ref{der-orbi}), this gives a prediction for the value of the K\"ahler moduli at the \texttt{MCL}, which we will denote by $t_i^{(c)}$, $i=1,2$. In terms of 
\be
\label{z3-bs}
\widehat{z}_3= {1\over \widehat b_3 \widehat b_0}=\re^{-3 \widehat \zeta}, 
\ee
we find that 
\be
\label{eq:tc}
\ba
t_1^{(c)}&=t_2^{(c)}= -2 \log \left[ \frac{1}{2} \left(1+ \sqrt{\frac{4}{\widehat{z}_3^{\,1/3}}+1}\right)\right]\\
&-{3\over  \pi} {\rm Im} \left\{  {\rm Li}_2 \left( -\frac{\re^{2 \pi \ri /3}}{2} \left(1+ \sqrt{\frac{4}{\widehat{z}_3^{\,1/3}}+1}\right) \right)+ 
{\rm Li}_2 \left( -\frac{\re^{ \pi \ri /3} \widehat{z}_3^{\,1/3}}{2} \left(1+ \sqrt{\frac{4}{\widehat{z}_3^{\,1/3}}+1}\right) \right) \right\}.
\ea
\ee
The fact that the value of the periods at the \texttt{MCL} can be computed in terms of classical dilogarithms was first pointed out 
in \cite{rv}, and further developed in \cite{dk}. As explained in \cite{mz,mmrev,cgm}, the conjectural correspondence of \cite{ghm,cgm} allows us to explain 
this number-theoretic property of the periods in many examples. The classical dilogarithm enters as a classical limit of the quantum dilogarithm 
involved in the kernel of the relevant operator. It would be very interesting to prove (\ref{eq:tc}) by using the powerful techniques of \cite{dk}.

	\subsection{Topological string calculations}
In this section, we perform various tests of the general conjectures put forward in \cite{ghm,cgm}, in the $Y^{3,0}$ geometry. In particular, we 
will test the validity of the conjecture \eqref{tzeta}, \eqref{logz-exp}: we will check the prediction for the GW invariants \eqref{eq:Fc-values} in the \texttt{MCF}, as well as the values of 
the K\"{a}hler moduli at the \texttt{MCL} \eqref{eq:tc}. To do this, we review the special geometry of this model in some detail. 

\begin{figure}
		\centering
		\includegraphics[width=0.4\linewidth]{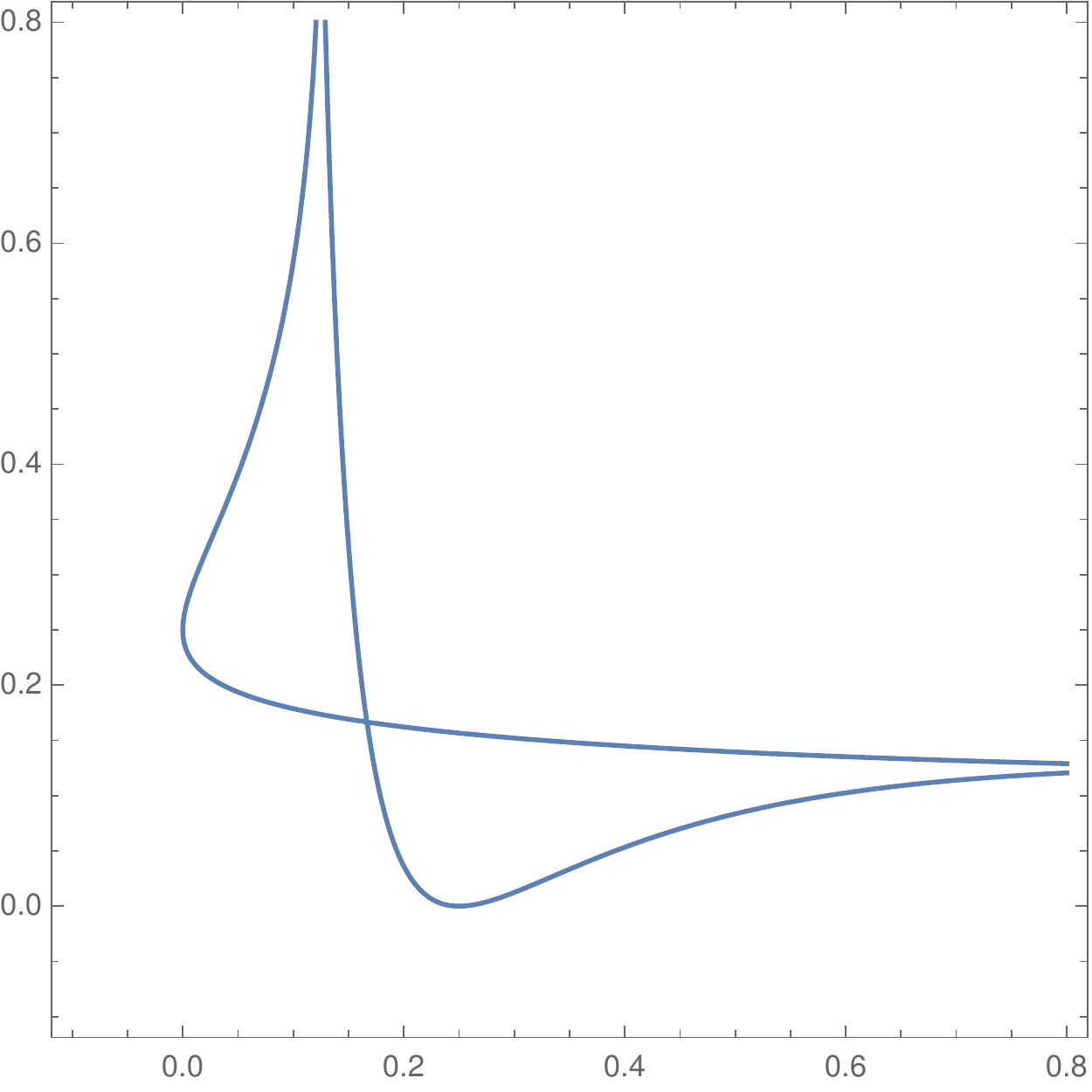}
		\caption{The discriminant locus of the geometry $Y^{3,0}$ with $z_3 = 1$. The two axes are $z_1$ and $z_2$.}\label{fg:A2-disc}
	\end{figure}

From the charge vectors (\ref{y3-qs}), we can write down the Picard-Fuchs operators
	\begin{equation}\label{eq:A2-PF}
	\begin{aligned}
		L_1 &= \theta_{21} (\theta_3-\theta_1) -z_1 \theta_{12}(1+\theta_{12}) \ , \\
		L_2 &= L_1 \: \: (1 \leftrightarrow 2) \ , \\
		L_3 &= \theta_3^2 -z_3(\theta_3-\theta_1)(\theta_3-\theta_2) \ . 
	\end{aligned}
	\end{equation}
	where $\theta_i = z_i \frac{\partial}{\partial_{z_i}}$, and we have defined
	\begin{equation}
		\theta_{12} \equiv 2\theta_2-\theta_1, \: \: \theta_{21} \equiv 2\theta_1-\theta_2 \ .
	\end{equation}	
	Furthermore, using the sum of the charge vectors $Q_1+Q_2+Q_3$, we can get an additional Picard-Fuchs operator
	\begin{equation}
		L_0 = \theta_3^2 - z_1 z_2 z_3 \theta_{12} \theta_{21} \ ,
	\end{equation}
	which helps eliminate the false flat coordinates. Using these Picard-Fuchs operators, we obtain in the \texttt{LRF} the A-periods
	\begin{equation}\label{eq:A2-LRFA}	
	\begin{aligned}
		t_1 =& -\log \left(z_1\right)+\left(-2 z_1+z_2\right)+\left(-3 z_1^2+\frac{3 z_2^2}{2}\right)\\
			&+\left(-\frac{20 z_1^3}{3}+2 z_1^2 z_2-z_1 z_2^2+\frac{10 z_2^3}{3}\right)\\
			&+\left(-\frac{35 z_1^4}{2}+8 z_1^3 z_2-4 z_1 z_2^3+\frac{35 z_2^4}{4}+4 z_1^2 z_2 z_3-2 z_1 z_2^2 z_3\right)\\
			&+\left(-\frac{252 z_1^5}{5}+30 z_1^4 z_2-15 z_1 z_2^4+\frac{126 z_2^5}{5}+24 z_1^3 z_2 z_3-12 z_1 z_2^3 z_3\right)+\cO(z^6),\\
		t_2 =& t_1 \left(z_1 \leftrightarrow z_2\right),\\
		t_\xi =& t_3 = -\log\left(z_3\right),
	\end{aligned}
	\end{equation}
	as well as the B-periods
	\begin{equation}\label{eq:A2-LRFB}
	\begin{aligned}
		\frac{\partial F_0}{\partial t_1}
		=& \log ^2(z_1)+\log (z_1) \log (z_2)+\frac{1}{2} \log ^2(z_2)+\frac{2}{3} \log (z_1) \log
		(z_3)+\frac{1}{3} \log (z_2) \log (z_3)
		\\&+\log (z_1)\( 3 z_1 +\frac{9}{2}  z_1^2 \)+\log (z_2)\( z_1+z_2+\frac{3}{2} z_1^2+\frac{3}{2} z_2^2\)+\log (z_3)\( z_1 +\frac{3}{2}  z_1^2\) 
		\\&+2 z_1+7 z_1^2-z_1 z_2+z_2^2+\cO(z^3),
		\\
		\frac{\partial F_0}{\partial t_2} =& \frac{\partial F_0}{\partial t_1}\left(z_1 \leftrightarrow z_2\right).
	\end{aligned}
	\end{equation}	
	The discriminant of the Picard-Fuchs system is
	\begin{align}
		\Delta =& -729 z_1^4 z_2^4 z_3^3+2187 z_1^4 z_2^4 z_3^2-2187 z_1^4 z_2^4 z_3+729 z_1^4 z_2^4+1215 z_1^3 z_2^3 z_3^2-243 z_1^3 z_2^3 z_3-972 z_1^3 z_2^3\nonumber\\
		&-27 z_1^3 z_2^2 z_3^2+540 z_1^3 z_2^2 z_3+216 z_1^3 z_2^2-27 z_1^2 z_2^3 z_3^2+540 z_1^2 z_2^3 z_3+216 z_1^2 z_2^3-513 z_1^2 z_2^2 z_3\nonumber\\
		&+270 z_1^2 z_2^2+36 z_1^2 z_2 z_3-144 z_1^2 z_2+16 z_1^2+36 z_1 z_2^2 z_3-144 z_1 z_2^2-z_1 z_2 z_3\nonumber\\
		&+68 z_1 z_2-8 z_1+16 z_2^2-8 z_2+1 \ .
	\end{align}
	The zero locus of this discriminant defines the conifold locus. The \texttt{MCL} is the submanifold of the conifold locus where $\lambda_1=\lambda_2=0$. In this case, since we have a mass parameter, 
	this is a one-dimensional complex manifold. If we fix the value of $z_3$, we obtain a point in the \texttt{MCL} which we call a {\it maximal conifold point} (\texttt{MCP}). As in \cite{cgm}, the \texttt{MCP} for a 
	fixed value of $z_3$ corresponds to the point where the components of the conifold locus intersect transversally, see \figref{fg:A2-disc}. 
	
	One ingredient we need for our calculation is the B-field. In this geometry it is given by \cite{hm}
	\begin{equation}
	\label{y30-b}
		\boldsymbol{B} = (0,0,1) \ .
	\end{equation}
This means that in the instanton part of the periods computed above (but not in the log terms) we have to change $z_3 \rightarrow -z_3$. {\it After this change of sign}, 
the parameter $z_3$ should be identified with the parameters in the 
operator as in (\ref{z3-bs}). To simplify the complicated algebraic manipulations, in most of the subsequent analysis we will set $z_3 = 1$, so that $\xi=1$. 
The discriminant locus is plotted in Fig.~\ref{fg:A2-disc}. 
One finds out that the \texttt{MCP} is in that case at
	\begin{equation}
		(z_1,z_2) = (1/6,1/6) \ .
	\end{equation}
We can already test (\ref{eq:tc}) by evaluating numerically the \texttt{LRF} A-period $t_1$ at this point ($t_2$ has the same value as $t_1$ since at the \texttt{MCP} $z_1 = z_2$). One finds, 
	\begin{equation}
		t_1(1/6,1/6,1) = 1.49858\ldots \ ,
	\end{equation}
	which agrees with the prediction of \eqref{eq:tc}. One can perform the same calculation for other values of $\xi$, and the \texttt{LRF} A-periods evaluated at the 
	corresponding \texttt{MCP} also agree with \eqref{eq:tc}. This is a first, highly non-trivial test of the conjectures put forward in \cite{ghm,cgm}.

	Next, let us compute the appropriate periods in the \texttt{MCF}, and then the free energies $\cF_0, \cF_1$, which can be compared with the predictions 
	from the large $\hbar$ expansion of the fermionic traces analyzed in the previous section. Since in those computations $\xi$ is set to 1, we fix the mass parameter $z_3 = 1$ throughout the rest of this section.

	First, we look for good coordinates in the \texttt{MCF}, the coordinates that parametrize the neighborhood of the moduli space near the \texttt{MCP}. The discriminant should be factorized 
	at first order in terms of these coordinates. We find
	\begin{gather}\label{conifoldDiagonals}
	\begin{gathered}
	z_1=\frac{1}{2} \left(3-\sqrt{5}\right) u_1+\frac{1}{2} \left(-3-\sqrt{5}\right) u_2+\frac{1}{6}\ , \\
	z_2=\frac{1}{2} \left(-3-\sqrt{5}\right) u_1+\frac{1}{2} \left(3-\sqrt{5}\right) u_2+\frac{1}{6}\ .
	\end{gathered}
	\end{gather}
	Indeed with the new coordinates $u_1, u_2$, the discriminant is
	\begin{small}
		\begin{equation}
		\Delta = -\frac{1125}{8} u_1 u_2+\left(-\frac{135}{2} \sqrt{5} u_1^3+\frac{405}{2} \sqrt{5} u_1^2 u_2+\frac{405}{2} \sqrt{5} u_1 u_2^2-\frac{135}{2} \sqrt{5}
		u_2^3\right)+\cO(u^4).
		\end{equation}
	\end{small}
	Note that the \texttt{MCF} coordinates $u_1,u_2$ are only defined up to scaling.
	
	Next, we use the Picard-Fuchs equations \eqref{eq:A2-PF} to solve the \texttt{MCF} periods. The two A-periods $\sigma_1, \sigma_2$ should be power series in $u_1, u_2$, and each of them should have leading contributions $u_1$ and $u_2$ respectively. The equations \eqref{eq:A2-PF} allow three linearly independent power series solutions. Two of them have 
	linear leading contributions, while the third one has quadratic leading contributions. The correct combinations are found by the requirement that the 
	A-periods should vanish over the zero locus of the discriminant. In other words, expressed in terms of the \texttt{MCF} A-periods, the discriminant should factorize at every order. We find
	\begin{equation}
	\begin{aligned}
		\sigma_1 =& u_1+\left(\frac{112 u_1^2}{5 \sqrt{5}}-\frac{12 u_1 u_2}{5 \sqrt{5}}+\frac{12 u_2^2}{5 \sqrt{5}}\right)+\left(\frac{8984 u_1^3}{75}-\frac{1212}{25}
		u_1^2 u_2+\frac{588}{25} u_1 u_2^2+\frac{228 u_2^3}{25}\right) \nonumber\\
				&+\left(\frac{3812896 u_1^4}{1125 \sqrt{5}}-\frac{221888 u_1^3 u_2}{125
			\sqrt{5}}+\frac{1824 u_1^2 u_2^2}{5 \sqrt{5}}+\frac{38112 u_1 u_2^3}{125 \sqrt{5}}+\frac{32544 u_2^4}{125 \sqrt{5}}\right)+\cO(u^5)  \ , \\
		\sigma_2 =& \sigma_1 (u_1 \leftrightarrow u_2) \ .
	\end{aligned}
	\end{equation}
	The mirror map follows
	\begin{equation}\label{mirrorMap}
	\begin{aligned}
		u_1 =& \sigma_1+\left(-\frac{112 \sigma_1^2}{5 \sqrt{5}}+\frac{12 \sigma_1 \sigma_2}{5 \sqrt{5}}-\frac{12 \sigma_2^2}{5 \sqrt{5}} \right) + \left(\frac{29912 \sigma_1^3}{375}+\frac{492}{25} \sigma_1^2 \sigma_2-\frac{348}{25} \sigma_1 \sigma_2^2+\frac{1404 \sigma_2^3}{125}\right) \\
			&+\left(-\frac{6903584 \sigma_1^4}{5625 \sqrt{5}}-\frac{843072 \sigma_1^3 \sigma_2}{625
			\sqrt{5}}+\frac{370896 \sigma_1^2 \sigma_2^2}{625 \sqrt{5}}+\frac{18128 \sigma_1 \sigma_2^3}{625 \sqrt{5}}-\frac{117376 \sigma_2^4}{625 \sqrt{5}}\right)+\cO(\sigma^5),\\
		u_2 =& u_1 (\sigma_1 \leftrightarrow \sigma_2),	
	\end{aligned}
	\end{equation}
	
	Similar to the coordinates $u_i$, we have the freedom to rescale the conifold frame A-periods $\sigma_i$. On the other hand, the conifold frame flat coordinates $\lambda_i$ that are identified with the 't Hooft parameters in the large $\hbar$ expansion of the fermionic traces can be expressed in terms of the \texttt{LRF} B-periods according to \eqref{saddle}. The constants $b_j^{\rm NS}$ that appear in \eqref{saddle} are \cite{hm}
	\begin{equation}
		b_1^{\rm NS} = b_2^{\rm NS} = -\frac{1}{6} \ .
	\end{equation}
	By evaluating $\lambda_i$ and $\sigma_i$ at several points in the moduli space between the large radius point (\texttt{LRP}) and the \texttt{MCP}, we find that
	\begin{equation}
		\sigma_i = r_i \lambda_i \ , \quad i=1,2 \ ,
	\end{equation}
	where
	\begin{equation}
		r_1 = r_2 = \frac{\pi^2}{3\sqrt{3}} \ .
	\end{equation}
	This fixes the scaling of the \texttt{MCF} periods.
	
	Next, we wish to compute the two B-periods, which have the form	
	\begin{equation}
		s_i = \lambda_i \log \lambda_i + \textrm{power series }(\lambda_1,\lambda_2) \ ,\quad i=1,2 \ .
	\end{equation}
	The B-periods can be solved by plugging the ansatz into the Picard-Fuchs equation, and also demanding that the two B-periods are the derivatives of a single function, namely 
	the \texttt{MCF} prepotential. Expressed in terms of the \texttt{MCF} A-periods $\lambda_1, \lambda_2$ we find
	\begin{equation}\label{eq:A2-MCFB}
	\begin{aligned}
		s_1=&  \lambda_1 \log (\lambda_1)
		+ \(-\frac{16 \pi ^2 \lambda_1^2}{15 \sqrt{15}}+\frac{2 \pi ^2 \lambda_1 \lambda_2}{\sqrt{15}}+\frac{\pi ^2 \lambda_2^2}{\sqrt{15}}\) \\
		&+ \(\frac{3464 \pi ^4 \lambda_1^3}{30375}-\frac{682 \pi ^4 \lambda_1^2 \lambda_2}{1125}-\frac{106 \pi ^4 \lambda_1 \lambda_2^2}{1125}-\frac{682 \pi ^4 \lambda_2^3}{3375}\)\\
		&+\(-\frac{102848 \pi ^6 \lambda_1^4}{273375 \sqrt{15}}+\frac{131744 \pi ^6 \lambda_1^3 \lambda_2}{50625 \sqrt{15}}-\frac{2624 \pi ^6 \lambda_1^2 \lambda_2^2}{16875 \sqrt{15}}-\frac{5248 \pi ^6 \lambda_1 \lambda_2^3}{50625 \sqrt{15}}+\frac{123808 \pi ^6 \lambda_2^4}{151875 \sqrt{15}}\) + \cO(\lambda_i^5) \ . \\
		s_2 =& s_1 (\lambda_1 \leftrightarrow \lambda_2) \ .
	\end{aligned}
	\end{equation}
	
	The conifold frame periods also satisfy the special geometry relations; in other words
	\begin{equation}
	\label{fdsy}
	\frac{\partial \cF_0}{\partial \lambda_1} = s_1 + \left(\alpha +\frac{1}{2}\right)\lambda_1+\beta \lambda_2-\gamma,\quad \frac{\partial \cF_0}{\partial \lambda_2} = s_2 + \left(\alpha +\frac{1}{2}\right)\lambda_2+\beta \lambda_1 -\gamma \ .
	\end{equation}
	The coefficients $\alpha$, $\beta$ cannot be obtained from this method, but they are determined by 
	the relationship (\ref{der-orbi}) (the value of $\gamma$ is fixed by (\ref{ders-con})). They involve the analytic continuation of the B-periods from the \texttt{LRF} to the \texttt{MCF}. 
	The conjecture of \cite{ghm,cgm} gives a prediction for the values of these coefficients in terms of spectral theory. 
	Such a prediction was tested for the resolved $\IC^3/\IZ_5$ orbifold in \cite{cgm}. It would be interesting to test it in this example as well, 
	as a function of the mass parameter. By integrating the right hand side of (\ref{fdsy}), we get the prepotential $\cF_0$ in the \texttt{MCF}, up to an additive constant:
	\begin{equation}\label{eq:C3Z6-Fc}
	\begin{aligned}
		\cF_0 (\lambda_1, \lambda_2) &= \frac{1}{2}\lambda_1^2 \log(\lambda_1) + \frac{1}{2} \lambda_2^2\log (\lambda_2) + \(\frac{\alpha  \lambda_1^2}{2}
		+\frac{\alpha  \lambda_2^2}{2}+\beta  \lambda_1 \lambda_2\) -\gamma(\lambda_1+\lambda_2)\\
		&+ \(-\frac{16 \pi ^2 \lambda_1^3}{45 \sqrt{15}}+\frac{\pi ^2 \lambda_1^2 \lambda_2}{\sqrt{15}}+\frac{\pi ^2 \lambda_1 \lambda_2^2}{\sqrt{15}}-\frac{16 \pi ^2\lambda_2^3}{45 \sqrt{15}}\)\\
		&+ \(\frac{866 \pi ^4\lambda_1^4}{30375}
		+\frac{866 \pi ^4\lambda_2^4}{30375}+\ldots\)
		+ \(-\frac{102848 \pi ^6 \lambda_1^5}{1366875 \sqrt{15}}
		-\frac{102848 \pi ^6 \lambda_2^5}{1366875 \sqrt{15}}+\ldots\) + \cO(\lambda_i^6) \ .
	\end{aligned}
	\end{equation}
	The coefficients of the prepotential in the second and third lines of (\ref{eq:C3Z6-Fc}) indeed agree with the numbers in \eqref{eq:Fc-values}. 
	We proceed to the genus one standard topological string free energy in the \texttt{MCF}. It has the form,
	\begin{equation}
	\cF_1  = -\frac{1}{12}\log \left(\Delta z_1^a z_2^b \right) - \frac{1}{2} \log \det \(\frac{\partial \lambda_i}{\partial z_j}\).
	\end{equation}
	The two exponents $a$ and $b$ should be identical due to the symmetry of the geometry. Furthermore, we can use the first genus one \texttt{MCF} Gromov-Witten invariant in \eqref{eq:Fc-values} to fix $a=8$. Plugging in the \texttt{MCF} mirror map \eqref{mirrorMap}, we have
	\begin{equation}
	\begin{aligned}
		\cF_1 &=-\frac{1}{12} \log \(\lambda_1 \lambda_2 \) +\(\frac{16 \pi ^2 \lambda_1}{45 \sqrt{15}}+\frac{16 \pi ^2 \lambda_2}{45 \sqrt{15}}\)
		+\(-\frac{146 \pi ^4 \lambda_1^2}{30375}+\frac{533 \pi ^4 \lambda_1 \lambda_2}{3375}-\frac{146 \pi ^4 \lambda_2^2}{30375}\)\\
		&+\(-\frac{9472 \pi ^6 \lambda_1^3}{1366875 \sqrt{15}}-\frac{48256 \pi ^6 \lambda_1^2 \lambda_2}{151875 \sqrt{15}}-\frac{48256 \pi ^6 \lambda_1 \lambda_2^2}{151875 \sqrt{15}}-\frac{9472 \pi ^6 \lambda_2^3}{1366875 \sqrt{15}}\) + \cO(\lambda_i^4) \ .
	\end{aligned}	
	\end{equation}
	Again, the coefficients agree with all the genus one numbers in \eqref{eq:Fc-values} perfectly.

	\subsection{Spectral traces from Airy functions}
	\label{sc:Airy}
	
	In this section we present a check of (\ref{multi-Airy}). On the one hand, since we have the integral kernels of the operators $\sA_1, \sA_2$ 
	associated to the $Y^{3,0}$ geometry, we can directly compute the fermionic traces $Z(N_1,N_2; \hbar)$. In particular, the kernels $A_1(x,y) = \rho_{1,1,\xi}(x,y)$ and 
	$A_2(x,y)$ given in \eqref{eq:ker-A2} are tremendously simplified when $\hbar$ is a rational multiple of $2\pi$. On the other hand, as already pointed in 
	\cite{ghm, cgm}, \eqref{multi-Airy} gives a convenient way to compute  $Z(N_1,N_2; \hbar)$ from the conjectured form of the Fredholm determinant in terms of Airy functions. Let us review the Airy function method and generalize it slightly.

	\begin{figure}
		\centering
		\includegraphics[height=30ex]{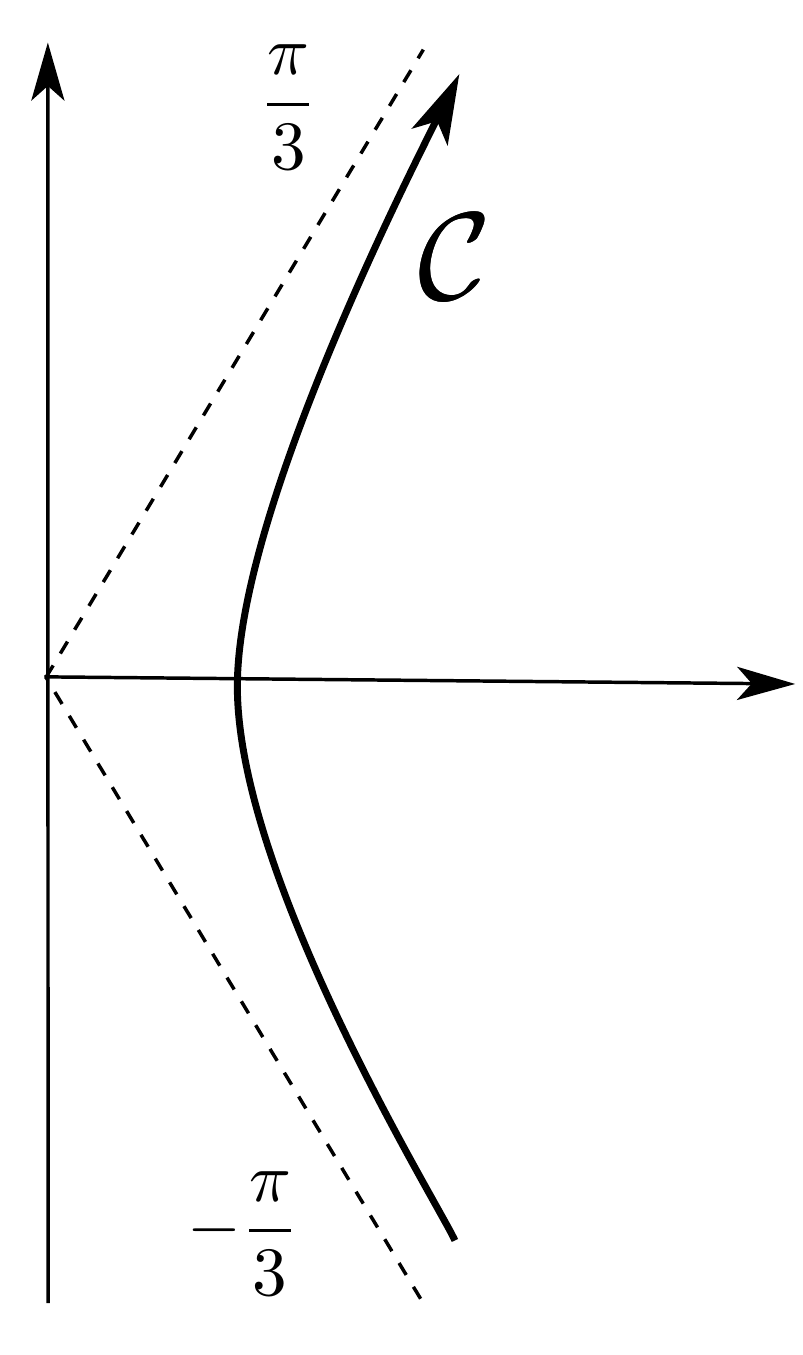}
		\caption{The integration contour $\cC$ used in the computation of $Z(\boldsymbol{N};\boldsymbol{\xi},\hbar)$.}\label{fg:Contour}
	\end{figure}			
	
	We first separate $J(\boldsymbol{\mu};\boldsymbol{\xi},\hbar)$ into the perturbative part $J^{\rm (p)}$, which is a cubic polynomial in $\mu$, and the nonperturbative part $J^{\rm (np)}$, which is a power series in $\re^{-\mu}$. \eqref{multi-Airy} implies that
	\begin{equation}\label{eq:Z-Ai}
		Z(\boldsymbol{N};\boldsymbol{\xi},\hbar) = \int_{\cC} \re^{J^{\rm (p)}+J^{\rm (np)} - \sum_{i=1}^{g_\Sigma} N_i \mu_i} \prod_{i=1}^{g_\Sigma} \frac{\rd \mu_i}{2\pi\ri} \ .
	\end{equation}
	Here the integration path $\cC$ for each $\rd \mu_i$ has been deformed so that it asymptotes to $\re^{\pm \ri\pi/3}\infty$, as seen in Fig.~\ref{fg:Contour}. Let us 
	ignore the nonperturbative contributions for the moment, and let us define
	\begin{equation}
		Z^{\rm (p)}(\boldsymbol{N};\boldsymbol{\xi},\hbar) = \int_{\cC} \re^{J^{\rm (p)} - \sum_{i=1}^{g_\Sigma} N_i \mu_i} \prod_{i=1}^{g_\Sigma} \rd \mu_i \ .
	\end{equation}
	When $g_\Sigma=1$, this is nothing else but an Airy function, since the coefficients of the cubic terms are triple intersection numbers of the geometry and must therefore be real and positive. In the case of $g_\Sigma=2$, we can always find a linear combination of $\mu_{1,2}$
	\begin{equation}
		\nu_i = \sum_{j=1}^{2} K_{ij} \mu_j \ , \quad\quad i = 1,2
	\end{equation}
	such that $J^{\rm (p)}$ has the following form
	\begin{equation}\label{eq:Jp}
		J^{(p)} = \sum_{i=1}^2 \( \frac{C_i}{3}\nu_i^3 + D_i \nu_i^2 + B_i \nu_i\) + \zeta \nu_1\nu_2 + A \ ,
	\end{equation}
	where the coefficient $C_i$ is proportional to $\hbar^{-1}$ and positive, while $D_i, B_i, A$ are functions of the mass parameters $\xi_j$ and $\hbar$ (the coefficient $\zeta$ should not be confused with 
	other occurrences of the same symbol in this paper, in (\ref{thooft-mu}) and (\ref{xizeta})). The Jacobian must not vanish, i.e.
	\begin{equation}
		\(\frac{\partial \mu_i}{\partial \nu_j}\) \neq 0
	\end{equation}
	for the integral measure in \eqref{eq:Z-Ai} to be still meaningful after the coordinate transformation. In addition, we define $M_i$ by
	\begin{equation}
		\sum_{i=1}^2 \mu_i N_i = \sum_{i=1}^2 \nu_i M_i \ .
	\end{equation}
	
	If $\zeta=0$, the quadratic mixing term is absent, and $Z^{\rm (p)}$ splits to a product of Airy functions
	\begin{equation}\label{eq:Zp-Ai}
	\begin{aligned}
		Z^{\rm (p)}(\boldsymbol{N};\boldsymbol{\xi},\hbar) = \(\frac{\partial \mu_i}{\partial \nu_j}\) \re^A &\prod_i (C_i)^{-1/3}  \exp\(D_i C_i^{-1}(M_i - B_i + \tfrac{2}{3}D_i^2 C_i^{-1})\) \\
	&\times\Ai\( (C_i)^{-1/3}(M_i-B_i+D_i^2 C_i^{-1}) \) \ .
	\end{aligned}
	\end{equation}
	This is the scenario discussed in \cite{cgm}. Furthermore, let
	\begin{equation}
		\re^{J^{\rm (np)}} = \sum_{k_i \geqslant 0} P_{\boldsymbol{k}}\(\boldsymbol{\mu};\boldsymbol{\xi},\hbar\) \prod_{i=1}^{n_\Sigma} z_i^{k_i} \ ,
	\end{equation}
	where $P_{\boldsymbol{k}}\(\boldsymbol{\mu};\boldsymbol{\xi},\hbar\)$ is a polynomial in $\mu_i$. Then
	\begin{equation}\label{eq:Znp-Ai}
		Z(\boldsymbol{N}; \boldsymbol{\xi},\hbar ) = \sum_{k_i\geqslant 0} \(\prod_{\ell}\xi_\ell^{-\sum_{i} k_i \alpha_{i\ell}} \) P_{\boldsymbol{k}}\( -\partial_{\boldsymbol{N}};\boldsymbol{\xi},\hbar \)\circ Z^{\rm (p)}\(\boldsymbol{ N} + \boldsymbol{k}\cdot C ;\boldsymbol{\xi}, \hbar\) \ .
	\end{equation}
	Here $P_{\boldsymbol{k}}\( -\partial_{\boldsymbol{N}};\boldsymbol{\xi},\hbar \)$ is the differential operator obtained through replacing variable $\mu_i$ by $-\partial_{N_i}$ in the polynomial $P_{\boldsymbol{k}}\(\boldsymbol{\mu};\boldsymbol{\xi},\hbar\)$, and 
	we use $\circ$ to denote its action on $Z^{\rm (p)}(\{ N_i + \sum_j k_j C_{ji} \} ; \boldsymbol{\xi}, \hbar)$.	
	
	If however the mixing term $\zeta \nu_1\nu_2$ is present,  $Z^{\rm (p)}(\boldsymbol{N};\boldsymbol{\xi},\hbar)$ cannot be split to a product of conventional Airy functions. Instead we need to expand $\exp(\zeta \nu_1\nu_2)$, and replace each term in the expansion by a differential operator that acts on the $Z^{\rm (p)}$ with the mixing term removed, which now can be written as a product of 
	conventional Airy functions, just like what one has done for the terms in $J^{\rm (np)}$. In compact form, one finds
	\begin{equation}\label{eq:Zp-QE}
	\begin{aligned}
	&Z^{\rm (p)}(\boldsymbol{N};\boldsymbol{\xi},\hbar) \\
	= &\(\frac{\partial \mu_i}{\partial \nu_j}\) \re^A \exp\(\zeta\,\partial_{M_1}\partial_{M_2} \)\circ \left(\prod_i (C_i)^{-1/3}\exp\(D_i C_i^{-1}(M_i - B_i + \tfrac{2}{3}D_i^2 C_i^{-1})\)   \right. \\
	& \left. \phantom{\prod_i} \Ai\( (C_i)^{-1/3}(M_i-B_i+ D_i^2 C_i^{-1}) \)  \right)\ .
	\end{aligned}
	\end{equation}
	The formula for computing $Z(\boldsymbol{N};\boldsymbol{\xi},\hbar)$ in terms of derivatives of $Z^{\rm (p)}(\boldsymbol{N};\boldsymbol{\xi},\hbar)$ is the same. 
	In actual computations, both the expansion of $\exp(\zeta \nu_1\nu_2)$ and the expansion of $\re^{J^{\rm(np)}}$ have to be truncated, which greatly constrains the precision 
	of the calculations one can reach. This is the situation one will encounter in Sec.\ \ref{sc:resonance}. We denote the orders of the two truncations by $\deg \zeta$ and $\deg z$, and call them the perturbative degree and the instanton degree respectively.	
	
	Fortunately for the geometry $Y^{3,0}$, when $\xi=1$, which is the only situation that we will consider here, the quadratic mixing term is absent. The perturbative grand potential $J^{\rm (p)}$ is
	\begin{equation}
		J^{\rm (p)}(\boldsymbol{\mu};\hbar) = \frac{1}{12\pi\hbar}\( 8\mu_1^3 -3\mu_1^2\mu_2 - 3 \mu_1\mu_2^2 + 8\mu_2^3\) + \( \frac{\pi}{3\hbar} - \frac{\hbar}{12\pi}\)(\mu_1 + \mu_2) \ ,
	\end{equation}
	and it splits with the following change of variables
	\begin{equation}
		\(\begin{array}{c}
		\mu_1 \\ \mu_2
		\end{array}\) = 
		\begin{pmatrix}
		 1 & \frac{7-3\sqrt{5}}{2} \\
		 \frac{7-3\sqrt{5}}{2} & 1
		\end{pmatrix}\(\begin{array}{c}
		 \nu_1 \\ \nu_2
		\end{array}\) \ .
	\end{equation}
	
	We will consider two particular cases. The first is the maximal supersymmetric case where $\hbar = 2\pi$.  We have already seen that the integral kernel $\rho_{1,1,\xi}$ further simplifies in this case at the end of Sec.~\ref{sc:Omn}. The first fermionic trace $Z(1,0)$ is identified with the trace of $\rho_{1,1,\xi}$, which is given by \eqref{eq:trRho}. When $\xi =1$, one has from (\ref{xi-one-trs}), 
	\begin{equation}\label{eq:A2h2pZ1}
	Z(1,0;\xi = 1, \hbar = 2\pi) ={1\over 18}, 
	\ee
and by using the relation \eqref{gtwo-matrix} between the fermionic traces and the spectral traces, we find in addition, 
	\begin{equation}\label{eq:A2h2pZ2}
		Z(2,0;\xi = 1, \hbar = 2\pi) = \frac{108 - 19\sqrt{3}\pi}{1296\sqrt{3}\pi} \ .
	\end{equation}
	We also use the Airy function method to compute these fermionic traces, including up to seven orders of instanton corrections in the modified grand potential $J(\boldsymbol{\mu};\xi=1,\hbar=2\pi)$. To measure the degree of agreement between the results of the Airy function method and those using the integral kernel, we define the relative matching degree between two numbers $x$ and $y$ to be
	\begin{equation}
		-\log_{10} \Big| \frac{x-y}{y}\Big| \ .
	\end{equation}
	Roughly speaking it gives the number of identical digits between $x$ and $y$. We plot in Figs.~\ref{fg:A2h2p4p3Zs} in blue dots the relative matching degrees between the Airy function results and the analytic results against the order of instanton corrections used in the former method. One finds very good agreement, and it improves consistently when more instanton contributions are included.
	
	\begin{figure}
		\centering
		\subfloat[$Z(1,0)$]{\includegraphics[width=0.4\linewidth]{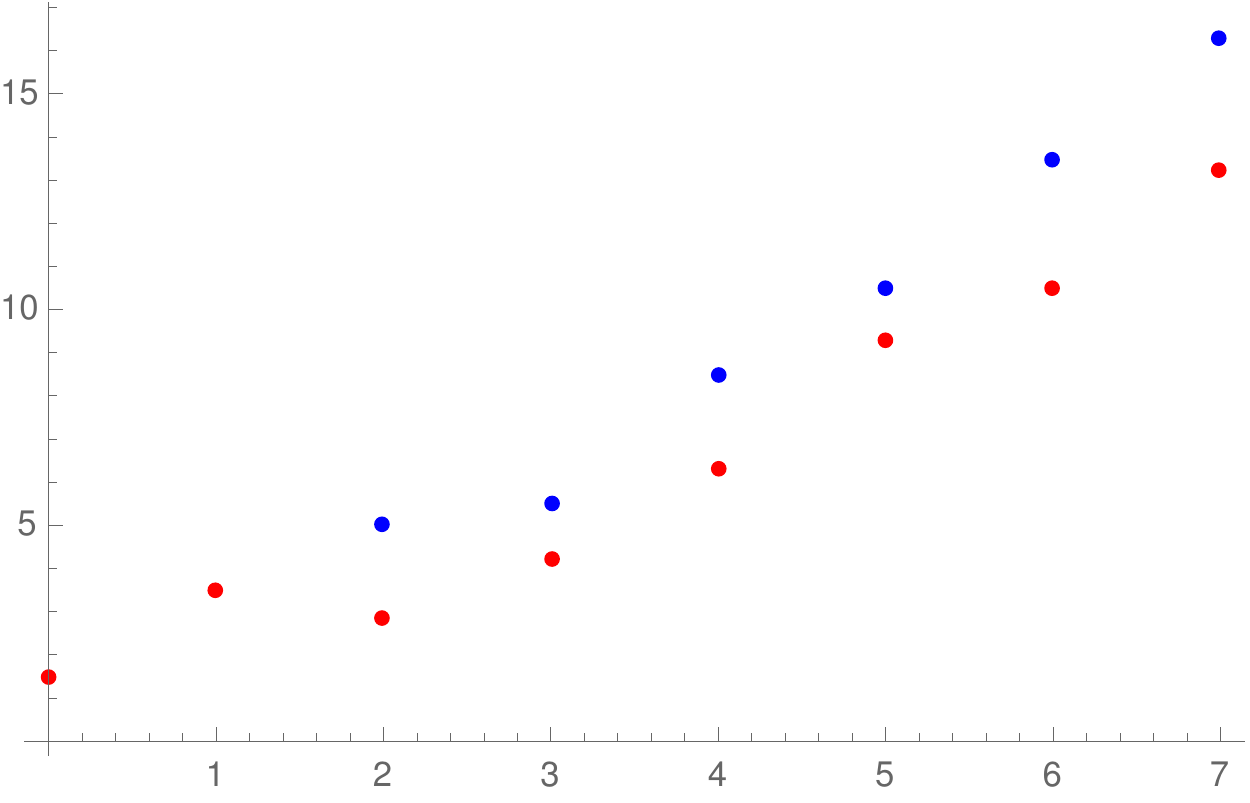}}\hspace{2em}
		\subfloat[$Z(2,0)$]{\includegraphics[width=0.4\linewidth]{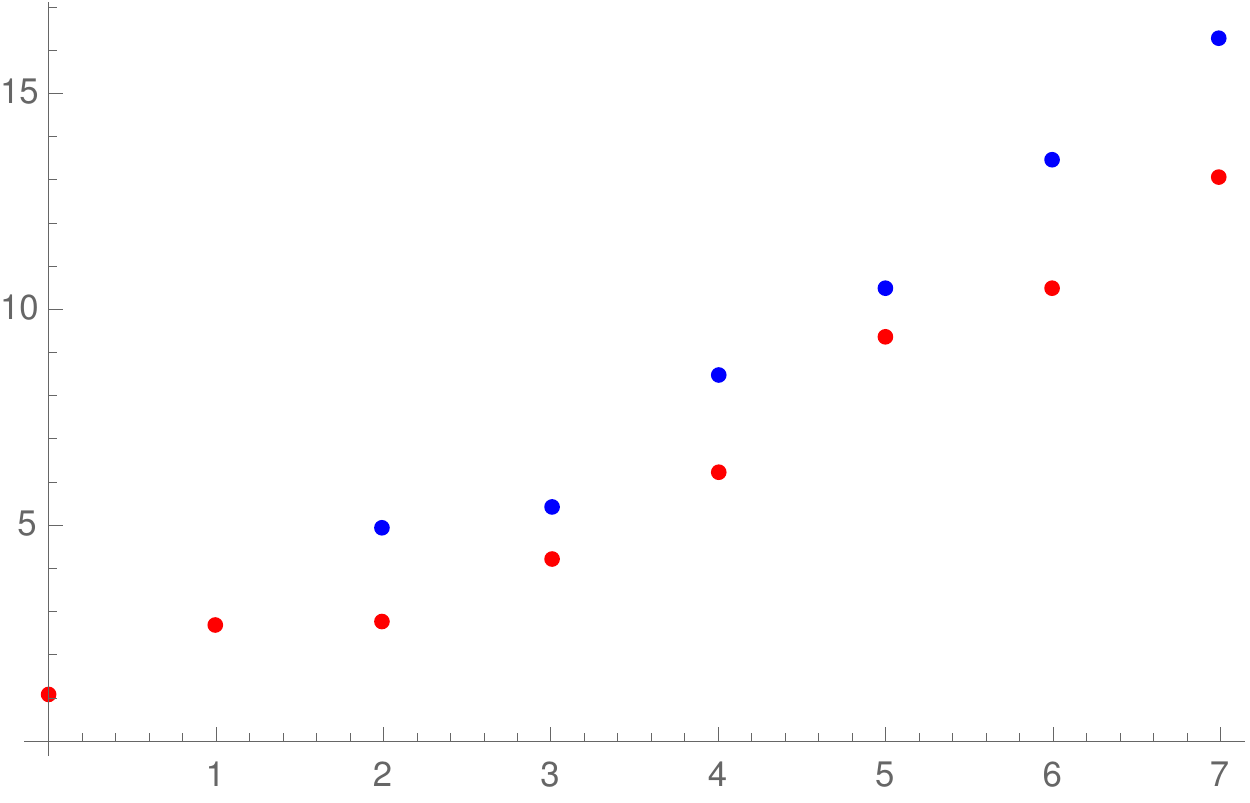}}
		\caption{The relative matching degrees between $Z(1,0)$, $Z(2,0)$ associated to $Y^{3,0}$ geometry ($\xi=1$) computed with the Airy function method and the integral kernel results in \eqref{eq:A2h2pZ1}, \eqref{eq:A2h2pZ2}, plotted against the order of instanton corrections included in the former method. The blue and the red dots correspond to $\hbar = 2\pi$ and $\hbar=4\pi/3$ respectively.}\label{fg:A2h2p4p3Zs}
	\end{figure}
	
	The second case is $\hbar = 4\pi/3$ so that $\sb = \sqrt{2}$. The integral kernel also simplifies. We find 
	\begin{equation}
	\begin{aligned}
		\rho_{1,1,\xi=1}(x,x) =& \frac{1}{2\sb \cos(\pi/6)} \re^{\tfrac{4}{3}\pi\sb x} \frac{\Phi_\sb(x+\tfrac{\ri}{3}\sb^{-1})}{\Phi_\sb (x-\tfrac{\ri}{3} \sb^{-1})} \frac{\Phi_\sb (x+\tfrac{2\ri}{3} \sb^{-1})}{\Phi_\sb(x-\tfrac{2\ri}{3}\sb^{-1})}  \\
		=& \frac{1}{2\sb \cos (\pi/6)} \frac{\re^{\tfrac{4}{3}\pi \sb x}}{1+\sqrt{3} \re^{2\pi\sb^{-1}x} + \re^{4\pi \sb^{-1} x}}  \ ,
	\end{aligned}	
	\end{equation}
	so that the trace is
	\begin{equation}\label{eq:A2h4p3Z1}
		Z(1,0;\xi=1,\hbar = 4\pi/3) = \tr \rho_{1,1,\xi=1} \(\hbar =\frac{4}{3}\pi\) = \frac{\cos\tfrac{\pi}{9} -\sqrt{3}\sin \tfrac{\pi}{9}}{3} \ .
	\end{equation}
	It is also possible to compute the double spectral trace by numerical integration, from which one gets the second fermionic trace, and it reads
	\begin{equation}\label{eq:A2h4p3Z2}
		Z(2,0;\xi = 1, \hbar = 4\pi/3) = 0.003565431804217254350\ldots \ .
	\end{equation}
	Similarly these fermionic traces can be computed by the Airy function method. We give the plots of the relative matching degrees in Figs~\ref{fg:A2h2p4p3Zs} in red dots, with the order of instanton corrections used in the Airy function method increased up to seven. Once again one finds very good agreement between the two results.


	\section{The resolved $\IC^3/\IZ_6$ orbifold}

	In this section we study another genus two example, namely, the 
	total resolution of the orbifold $\IC^3/\IZ_6$, where the action has weights $(4,1,1)$. 
	This is an $A_2$ geometry studied in the first papers on local mirror symmetry \cite{kkv,ckyz}, 
	and it engineers geometrically $SU(3)$ Seiberg--Witten theory. It has also 
	been studied in some detail in \cite{kpsw}. 
	
	\subsection{Mirror curve and operator content}
	
	The charge vectors of the geometry are 
	\be
	\label{cv-c3}
	Q^1=(-2,1,0,0,1,0), \qquad Q^2= (1,-2,1,0,0,0), \qquad Q^3=(0,0,0,1,-2,1). 
	\ee
	We can parametrize the moduli space with the six coefficients $x_i$, $i=0, \cdots, 5$ of the mirror curve subject to three $\bC^*$ scaling relations, or in terms of the Batyrev coordinates
	\be
	\label{eq:C3Z6-z}
	z_1={x_1 x_4 \over x_0^2}, \qquad z_2={x_0  x_2 \over x_1^2}, \qquad z_3= {x_3 x_5 \over x_4^2}. 
	\ee
	It is convenient to set $x_2=x_3=x_5=1$, so that the mirror curve reads
	\be
	\label{m3-can1}
	\re^x+ \re^y + \re^{-4x-y} + x_4 \re^{-2x} + x_0\re^{-x} + x_1=0. 
	\ee
	It follows from the Newton polygon that $x_0$, $x_1$ are true moduli, while $x_4$ is a mass parameter \cite{kpsw}. 
	We rename them to $\kappa_2, \kappa_1$ and $\xi$ respectively. The curve (\ref{m3-can1}) gives then one 
	canonical form of the curve. It is easy to see that the other canonical form is 
	\be
	\label{m3-can2}
	\re^{2x}+\re^y + \re^{-y -2x} + \xi \re^{-x} + \kappa_1\re^x+ \kappa_2=0. 
	\ee
	The operators associated to this geometry are obtained by Weyl quantization of the functions appearing in (\ref{m3-can1}), (\ref{m3-can2}) (this was already discussed in \cite{cgm}):
	\be
	\label{ex2-ops}
	\ba
	\sO_1&=\re^\sx+ \re^\sy + \re^{-4\sx-\sy} + \xi \re^{-2\sx} + \kappa_2\re^{-\sx}, \\
	\sO_2&=\re^{2\sx}+\re^\sy + \re^{-\sy -2\sx} + \xi \re^{-\sx} + \kappa_1\re^\sx.
	\ea
	\ee
It follows that
\be
\label{unpert-c3z6}
\ba
\mO^{(0)}_1&= \re^\sx+ \re^\sy + \re^{-4\sx-\sy} + \xi \re^{-2\sx}, \\
\mO^{(0)}_2&=\re^{2\sx}+\re^\sy + \re^{-\sy -2\sx} + \xi \re^{-\sx}. 
\ea
\ee
 The first operator is a perturbation of the operator $\mO_{4,1}$, while the second operator can be regarded as a perturbation of $\mO_{1,1}$, 
 after rescaling the operator $\mx$ to $\mx' = 2\mx$ (this is of course equivalent to a rescaling of $\hbar$ to $\hbar' = 2\hbar$). The reduced matrix (\ref{tmc}) is again given by (\ref{cartan3}). 
	
In order to calculate the generalized Fredholm determinant, we pick the operators (\ref{ajl}). We choose $j=1$, and we denote $\sA_{1i}= \sA_i$, $i=1,2$. These operators are given by 
\begin{equation}
\sA_1 = \rho_1^{(0)}, \quad \sA_2 = \sA_1 \re^{-\sx} \ .
\end{equation}
Let us note that the operators (\ref{unpert-c3z6}) are not of the form (\ref{pertmn}), and the integral kernel of their inverses is not known for $\xi\not=0$. In order to perform an analytic test of 
the correspondence between spectral theory and topological strings, it is convenient to set $\xi=0$. In this case,
\be
 \sA_1=\rho_{4,1}, 
 \ee
 and the fermionic spectral traces, with $N_1=0$ or $N_2=0$, can be calculated from the matrix model $Z_{m,n}(N, \hbar)= Z_{m,n,\xi=0}(N,
 \hbar)$ studied in \cite{mz}. We then find,
 \be\label{eq:Z-c3z6}
 \ba
 Z(N,0; \hbar )&= Z_{4,1}(N; \hbar), \\
 Z(0,N; \hbar)&= Z_{1,1}(N; 2\hbar). 
 \ea
 \ee
 The 't Hooft expansion of these matrix integrals, perturbatively in $\lambda$, has been obtained in \cite{mz}. To obtain mixed traces, we need the integral kernel of $\sA_2$. 
We first need to convert the variables $(x,y)$ to $(p,q)$ by \eqref{eq:xy-pq}, which specifies in this case to 
	\begin{equation}
		x = 2\pi\sb \frac{2p+q}{6} \ ,\quad y = -2\pi\sb\frac{4p+5q}{6}. 
	\end{equation}	
One finds, 
\begin{equation}
A_2(p,p') = (p| \rho_{4,1}\re^{-\sx} |p') = \re^{-\tfrac{2\pi}{3}\sb p}\re^{-\tfrac{\ri\pi}{18}\sb^2} \rho_{4,1}\(p,p'+\frac{\ri\pi}{6}\) \ .
\end{equation}
%
In addition, one can use the classical potentials $V^{(0)}_{4,1}(u)$, $V^{(0)}_{1,1}(u)$ associated to $\sO_{4,1}, \sO_{1,1}$ to obtain the first derivatives of 
$\CF_0(\lambda_1, \lambda_2)$ when $\lambda_1=\lambda_2=0$ \cite{mz}. We note that these potentials can be obtained from the general perturbed potential (\ref{cp}) 
by sending $\widehat{\zeta}$ to $-\infty$, which corresponds to $\xi\rightarrow 0$. The evaluation of $V^{(0)}_{m,m}(u)$ at its minimum $u_\star$ gives \cite{mz}
\begin{equation}\label{eq:Vmn-min}
V^{(0)}_{m,n}(u_\star) = -\frac{m}{2\pi}\log \chi_m -\frac{m+n+1}{2\pi^2} \imag \Li_2(-q^{m+1}\chi_m) \ ,
\end{equation}
where
\begin{equation}
\chi_m(q) = \frac{q^m-q^{-m}}{q-q^{-1}} \ , \quad q = \exp \(\frac{\pi}{m+n+1}\) \ .
\end{equation}
We finally obtain, 
\begin{equation}
\label{mcpp-c3}
\begin{aligned}
\frac{\partial \cF_0}{\partial \lambda_1}\Big|_{\lambda_j = 0}  &= -V^{(0)}_{4,1}(u_\star)= \frac{3}{\pi^2}\imag \Li_2\( - \re^{\tfrac{\pi\ri}{3}}\)\ ,\\
\frac{\partial \cF_0}{\partial \lambda_2}\Big|_{\lambda_j = 0}  &= -2 V^{(0)}_{1,1}(u_\star) = \frac{3}{\pi^2}\imag \Li_2\(-\re^{\tfrac{2\pi\ri}{3}}\)\ .
\end{aligned}
\end{equation}
The factor of $2$ in the second equation is due to the rescaling of $\hbar$. As in the previous example, 
this calculation can be related to the evaluation of the \texttt{LRF} A-periods at the 
corresponding conifold point, and we will test it in the next section. 

\subsection{Topological string calculations}	
		We wish to perform the same tests as in the previous sections, namely, the \texttt{LRF} A-periods at the \texttt{MCP}, as well as the \texttt{MCF} free energies. First of all, we know that \cite{fhm}
	\begin{equation}
		\mathbf{B} = (0,0,0) \ ,
	\end{equation}
	so no additional signs need to be given to the Batyrev coordinates when comparing the topological string results with the operator analysis results.
	
	From the charge vectors (\ref{cv-c3}) we can write down the Picard-Fuchs operators
	\begin{equation}
	\begin{aligned}
		L_1 &= (2\theta_2 - \theta_1)(2\theta_3-\theta_1) - z_1 (2\theta_1-\theta_2+1)(2\theta_1-\theta_2) \ ,\\
		L_2 &= -\theta_2 (2\theta_1-\theta_2)-z_2(2\theta_2-\theta_1+1)(2\theta_2-\theta_1) \ ,\\
		L_3 &= \theta_3^2 - z_3(2\theta_3-\theta_1 + 1)(2\theta_3-\theta_1) \ .
	\end{aligned}
	\end{equation}
	which annihilate the periods. Here $\theta_i \equiv z_i \partial_{z_i}$. Among all the A-periods, the one associated to $z_3$ is special. Since $z_3$ is a mass parameter, the corresponding flat coordinate $Q_3 = \exp(-t_3) 
	$ is a rational function of $z_3$. In fact \cite{kpsw}
	\begin{equation}
	\label{t3mm}
		z_3 = \frac{Q_3}{(1+Q_3)^2} \ .
	\end{equation}
This relation is valid in any reference frame of the moduli space. The other two A-periods in the \texttt{LRF} are 
\begin{equation}\label{eq:C3Z6-LRFA}
\begin{aligned}
t_1 =& -\log \left(z_1\right)+\left(-2 z_1+z_2+z_3\right)+\left(-3 z_1^2+\frac{3 z_2^2}{2}+\frac{3 z_3^2}{2}\right)+\cO(z^3)\ , \\
t_2 =& -\log \left(z_2\right)+\left(z_1-2 z_2\right)+\left(\frac{3 z_1^2}{2}-3 z_2^2\right)+\cO(z^3)\ .
\end{aligned}
\end{equation}
The two B-periods are
\begin{equation}\label{eq:C3Z6-LRFB}
	\begin{aligned}
		\frac{\partial F_0}{\partial t_1} =& \frac{2}{3} \log ^2 \left(z_1\right)+ \frac{2}{3} \log \left(z_1\right) \log \left(z_2\right)+\frac{2}{3} \log ^2\left(z_2\right) +\frac{2}{3}
		\log \left(z_1\right) \log \left(z_3\right)+\frac{1}{3} \log \left(z_2\right) \log \left(z_3\right)
		\\
		&+2 \log (z_1)
		z_1+2 \log (z_2) z_2 +\log (z_3) \(z_1 -\frac{2}{3}z_3\) +2 z_1+\cO(z^2),
		\\
		\frac{\partial F_0}{\partial t_2} =& \frac{1}{3} \log ^2 
		 \left(z_1\right)+\frac{4}{3} \log \left(z_1\right) \log \left(z_2\right)+\frac{4}{3} \log ^2\left(z_2\right)+\frac{1}{3}
		\log \left(z_1\right) \log \left(z_3\right)+\frac{2}{3} \log \left(z_2\right) \log \left(z_3\right)
		\\
		&+2 \log \left(z_1\right)
		z_2+4 \log \left(z_2\right) z_2+\log \left(z_3\right) \(z_2-\frac{1}{3}  z_3 \) +2 z_2 +\cO(z^2).
	\end{aligned}
\end{equation}
We also have, from the results in \cite{kpsw} (see also \cite{fhm}), 
	\begin{equation}
		b_i^{\rm NS} = -b_i = -\frac{1}{6} \ ,\quad i =1,2\ .
	\end{equation}
The discriminant of the Picard-Fuchs system reads in this case
	\begin{equation}
	\begin{aligned}
		\Delta =& 729 z_1^4 \left(1-4 z_3\right){}^2 z_2^4+108 z_1^3 \left(9 z_2-2\right) \left(4 z_3-1\right) z_2^2+\left(1-4 z_2\right){}^2\\
		&-4 z_1 \left(36 z_2^2-17
		z_2+2\right)+2 z_1^2 \left(108 \left(4 z_3+1\right) z_2^3-27 \left(28 z_3-5\right) z_2^2\right.\\
		&\left. + 72 \left(4 z_3-1\right) z_2-32 z_3+8\right)
	\end{aligned}
	\end{equation}	
and it defines the conifold locus. 
	
We want to compare now the topological string free energies in the appropriate \texttt{MCF} to the 't Hooft expansion of the fermionic spectral traces. Since these have been computed for $\xi=0$, 
the scaling limit for the mass parameter in the 't Hooft regime is no longer the one 
presented in (\ref{mt}). In this example, as in the local $\IF_2$ model studied in \cite{mz}, we want to keep $\xi=0$ fixed in this limit. This corresponds to $Q_3=-1$, therefore the value of $t_3$ is fixed to $\pm \pi \ri$. However, 
this means that the scaled variable $T_3$ goes to zero as $\hbar\rightarrow \infty$. In other words, when computing the topological string free energy to compare with the 't Hooft limit of the spectral 
traces, we have to set $T_3=0$, which by (\ref{t3mm}) means setting 
\be
z_3 = {1\over 4}.
\ee
The conifold locus corresponding to this value of $z_3$ is plotted in \figref{fg:C3Z6-disc}. The \texttt{MCP} occurs at the point of  
transversal intersection of the two branches of the locus, which takes place at
\begin{equation}
(z_1, z_2) = \(\frac{4}{27}, \frac{1}{4} \)\ .
\end{equation}
\begin{figure}
\centering
\includegraphics[width=0.4\linewidth]{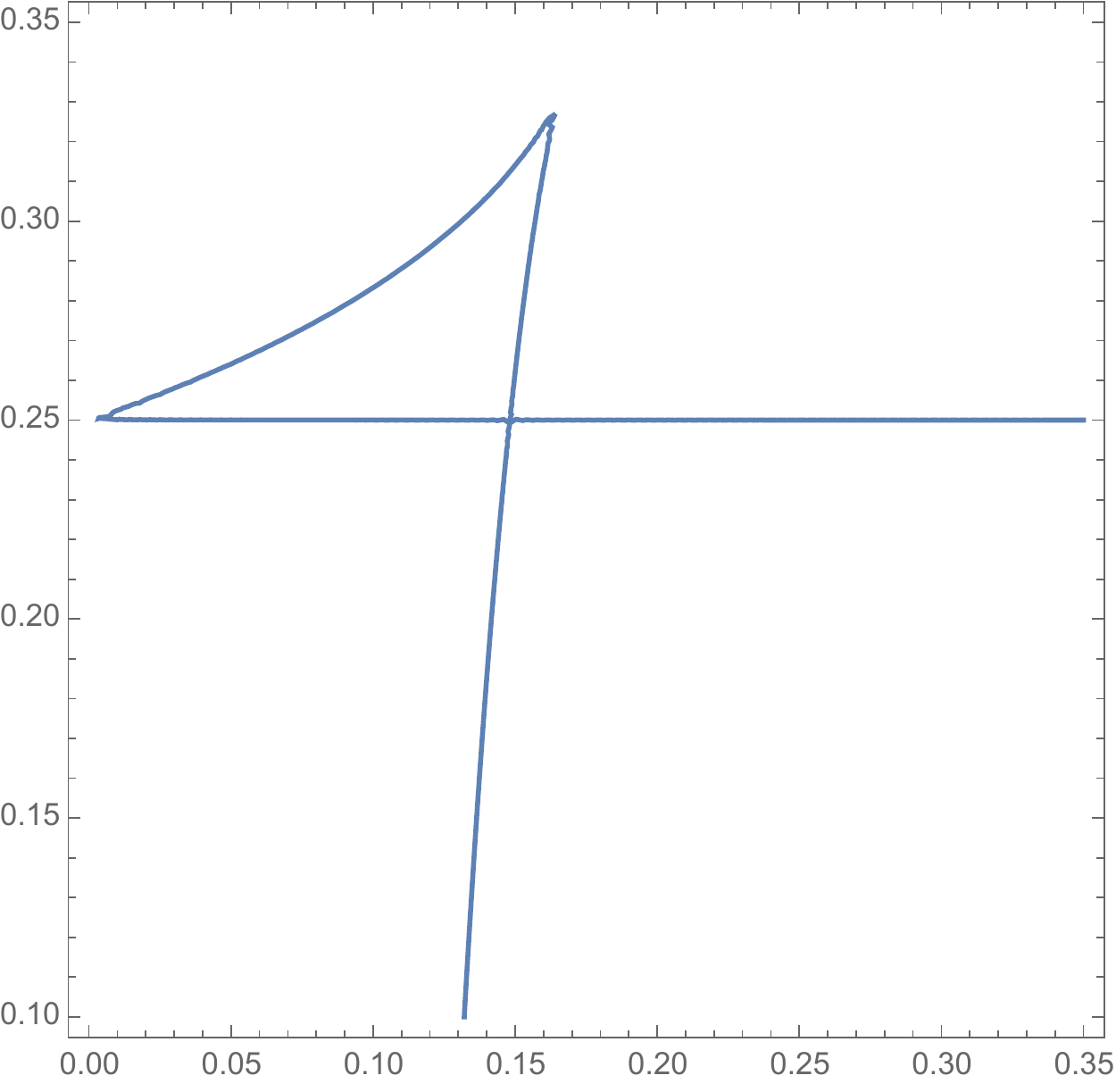}
\caption{The discriminant locus of the resolution of local $\bC^3/\bZ_6$ with $z_3 = 1/4$. The two axes are $z_1$ and $z_2$.}\label{fg:C3Z6-disc}
\end{figure}
This can be tested numerically, by verifying that the conifold coordinates given in (\ref{saddle}) vanish at that point. 

We can now test that the prepotential, when expanded around the \texttt{MCP}, 
reproduces the expansion of the fermionic spectral traces. To compute the prepotential, we use special geometry. 
We choose the coordinates near the \texttt{MCP} to be $\rho_1, \rho_2$, which are related to $z_1, z_2$ by
	\begin{equation}
	\begin{aligned}
		z_1 = \rho_1 + \frac{4}{27} \ , \quad z_2 = \rho_2 + \frac{1}{4} \ .
	\end{aligned}	
	\end{equation}
	The conifold A-periods $\sigma_1, \sigma_2$ are solved from the Picard-Fuchs equation, from which the mirror maps follow
	\begin{equation}
	\begin{aligned}
		\rho_1 &= \left(\frac{2 \sigma_1}{27}+\frac{\sigma_2}{2}\right)+\left(\frac{\sigma_1^2}{6}+\frac{33 \sigma_2^2}{32}\right)+\cO(\sigma_i^3) \ ,
		\\
		\rho_2 &= \frac{\sigma_1}{2}+\left(\frac{3 \sigma_1^2}{8}+\frac{27 \sigma_1 \sigma_2}{32}\right)+\cO(\sigma_i^3) \ .
	\end{aligned}
	\end{equation}
	They satisfy the condition that the discriminant factorizes when expressed in terms of $\sigma_1, \sigma_2$. To compare these conifold A-periods with those that are identified with the 't Hooft parameters $\lambda_1, \lambda_2$, which appear in the large $\hbar$ expansion of the fermionic traces, we evaluate both $\sigma_i$ and $\lambda_i$ at several points between the \texttt{LRP} and the \texttt{MCP}. We find 	
	\begin{equation}
		\sigma_i = r_i \lambda_i \ ,\quad i=1,2\ ,
	\end{equation}
	\begin{equation}
		r_1 = -\frac{4\pi^2}{3\sqrt{3}} \ ,\quad r_2 =-\frac{16\pi^2}{9\sqrt{3}} \ .
	\end{equation}
	Then we solve for the conifold B-periods $s_1, s_2$ by using again the Picard-Fuchs equations:
	\begin{equation}
	\begin{aligned}
		s_1 =& \lambda_1 \log (\lambda_1)
		+\(-\frac{7 \pi ^2 \lambda_1^2}{6 \sqrt{3}}+\sqrt{3} \pi ^2 \lambda_1 \lambda_2+\frac{\pi ^2 \lambda_2^2}{2\sqrt{3}}\) \\
		&+\(\frac{41 \pi ^4 \lambda_1^3}{54}-\frac{49}{12} \pi ^4 \lambda_1^2 \lambda_2-\frac{1}{2} \pi ^4 \lambda_1 \lambda_2^2-\frac{11 \pi ^4 \lambda_2^3}{108}\) + \cO(\lambda_i^4) \ ,\\
		s_2 =&\lambda_2 \log (\lambda_2)
		+\(\frac{1}{2} \sqrt{3} \pi^2 \lambda_1^2+\frac{\pi ^2 \lambda_1 \lambda_2}{\sqrt{3}}-\frac{\pi ^2 \lambda_2^2}{6\sqrt{3}}\) \\
		&+ \(-\frac{49}{36} \pi ^4 \lambda_1^3-\frac{1}{2} \pi ^4 \lambda_1^2 \lambda_2-\frac{11}{36} \pi^4 \lambda_1 \lambda_2^2+\frac{\pi ^4 \lambda_2^3}{486}\) + \cO(\lambda_i^4) \ .
	\end{aligned}
	\end{equation}
	The conifold periods satisfy the special geometry relation
	\begin{equation}
	\label{sg-dp}
		\frac{\partial \cF_0}{\partial \lambda_i} = \gamma_i + \sum_{j=1}^2 \alpha_{ij} \lambda_j + s_i \ ,\quad i=1,2\ ,
	\end{equation}
	where $\gamma_i, \alpha_{ij}$ are constants. Integrating the right hand side, we obtain the conifold prepotential
	\begin{equation}
	\label{foc3z6}
	\begin{aligned}
		\cF_0 =& \frac{1}{2} \lambda_1^2 \log (\lambda_1)+\frac{1}{2} \lambda_2^2 \log (\lambda_2)+q(\lambda_1, \lambda_2)
		+\(-\frac{7 \pi ^2 \lambda_1^3}{18 \sqrt{3}}+\frac{1}{2} \sqrt{3} \pi ^2 \lambda_1^2 \lambda_2+\frac{\pi ^2 \lambda_1 \lambda_2^2}{2 \sqrt{3}}-\frac{\pi ^2 \lambda_2^3}{18 \sqrt{3}}\)\\
		&+\(\frac{41 \pi ^4 \lambda_1^4}{216}-\frac{49}{36} \pi ^4 \lambda_1^3 \lambda_2-\frac{1}{4} \pi^4 \lambda_1^2 \lambda_2^2-\frac{11}{108} \pi ^4 \lambda_1 \lambda_2^3+\frac{\pi ^4 \lambda_2^4}{1944}\) + \cO(\lambda_i^5)\ ,
	\end{aligned}	
	\end{equation}	
where $q(\lambda_1,\lambda_2)$ is a quadratic polynomial involving $\gamma_i, \alpha_{ij}$ in (\ref{sg-dp}). The coefficients of the prepotential can be 
compared with the predictions from the large $\hbar$ expansion of the fermionic traces, 
and we find a complete agreement: when $\lambda_2=0$, the expansion agrees precisely with the expansion obtained in \cite{mz} for the matrix model $Z_{4,1}(N, \hbar)$. When $\lambda_1=0$, 
the expansion agrees with that obtained for the matrix model $Z_{1,1}(N, \hbar)$. However, due to the rescaling of $\hbar$ in (\ref{eq:Z-c3z6}), in order to compare with the numbers presented in \cite{mz}, 
we have to rescale $\lambda_2 \rightarrow 2 \lambda_2$ and multiply the result by $1/4$. The crossed terms appearing in (\ref{foc3z6}) can be also checked against 
the 't Hooft expansion of the mixed traces, similarly to what we did in the case of the $Y^{3,0}$ geometry. Finally, we note that, with the scaling we are using, the equation (\ref{der-orbi}) simplifies to
\begin{equation}\label{eq:C3Z6-con-B}
\frac{\partial \cF_0}{\partial \lambda_i} = -\sum_{j=1}^2 \frac{C_{ij}^{-1}}{2\pi} T_j  \ ,\quad i =1,2\ .
\end{equation}
When comparing to (\ref{sg-dp}), we conclude that the coefficients $\gamma_i$ can be obtained by evaluating the \texttt{LRF} A-periods at the \texttt{MCP}. On the other hand, spectral theory gives 
the prediction (\ref{mcpp-c3}) for their values. A numerical evaluation confirms indeed the validity of the prediction. (\ref{eq:C3Z6-con-B}) can be used to fix the values of the coefficients $\alpha_{ij}$ in (\ref{sg-dp}). 

As explained in \cite{mz}, with the scaling we are using for the mass parameters, a test of (\ref{logz-exp}) at next-to-leading order in $\hbar^{-1}$ involves both $\CF_1$ and the 
derivatives of $\CF_0$ w.r.t. the mass parameter. It would be interesting to perform such a test in this example.

	\subsection{Spectral traces from Airy functions}
	
	We now test the fermionic traces computed by the Airy function method against the direct integration of the 
	integral kernel of the operators. To simplify the calculation, we set $\xi=0$, in which case 
	the operators $\sA_1, \sA_2$ reduce to $\rho_{4,1}$ and $\rho_{1,1}$ (the latter after the rescaling of $\hbar$), and we 
	can use the formulae (111) and (112) in \cite{kasmar} to compute the first two traces. The calculations are further simplified if we choose special values of 
	$\hbar$, where the integral kernel is reduced to hyperbolic functions. For $\hbar = 2\pi$, the results of the first few traces are
	\begin{equation}
	\begin{aligned}
	Z(1,0;\xi=0,\hbar=2\pi) &= \frac{1}{6\sqrt{3}} \ , \\
		Z(0,1;\xi=0,\hbar=2\pi) &= \frac{1}{36} \ , \\
		Z(2,0;\xi=0,\hbar=2\pi) &= \frac{1}{432} \ . \\
		Z(0,2;\xi=0,\hbar=2\pi) &=\frac{5}{324} - \frac{1}{12\sqrt{3}\pi} \ , \\
	\end{aligned}	
	\end{equation}
	The results for $Z(0, N; \xi=0, \hbar=2 \pi)$ can be also read from the results in \cite{oz} for the operator $\rho_{1,1}$ and $\hbar=4 \pi$. 
	Equally simple is the case when $\hbar =\pi$, and we get
	\begin{equation}
	\begin{aligned}
	Z(1,0;\xi=0,\hbar=\pi) &= \frac{1}{2\sqrt{3}} \ ,\\
		Z(0,1;\xi=0,\hbar=\pi) & = \frac{1}{9} \ ,\\
		Z(2,0;\xi=0,\hbar=\pi) & = \frac{1}{36},\\
		Z(0,2;\xi=0,\hbar=\pi) & =  -\frac{1}{81} + \frac{1}{12 \sqrt{3} \pi} \ . 
		\end{aligned}	
	\end{equation}	
	
	\begin{figure}
		\centering
		\subfloat[$Z(1,0)$]{\includegraphics[width=0.4\linewidth]{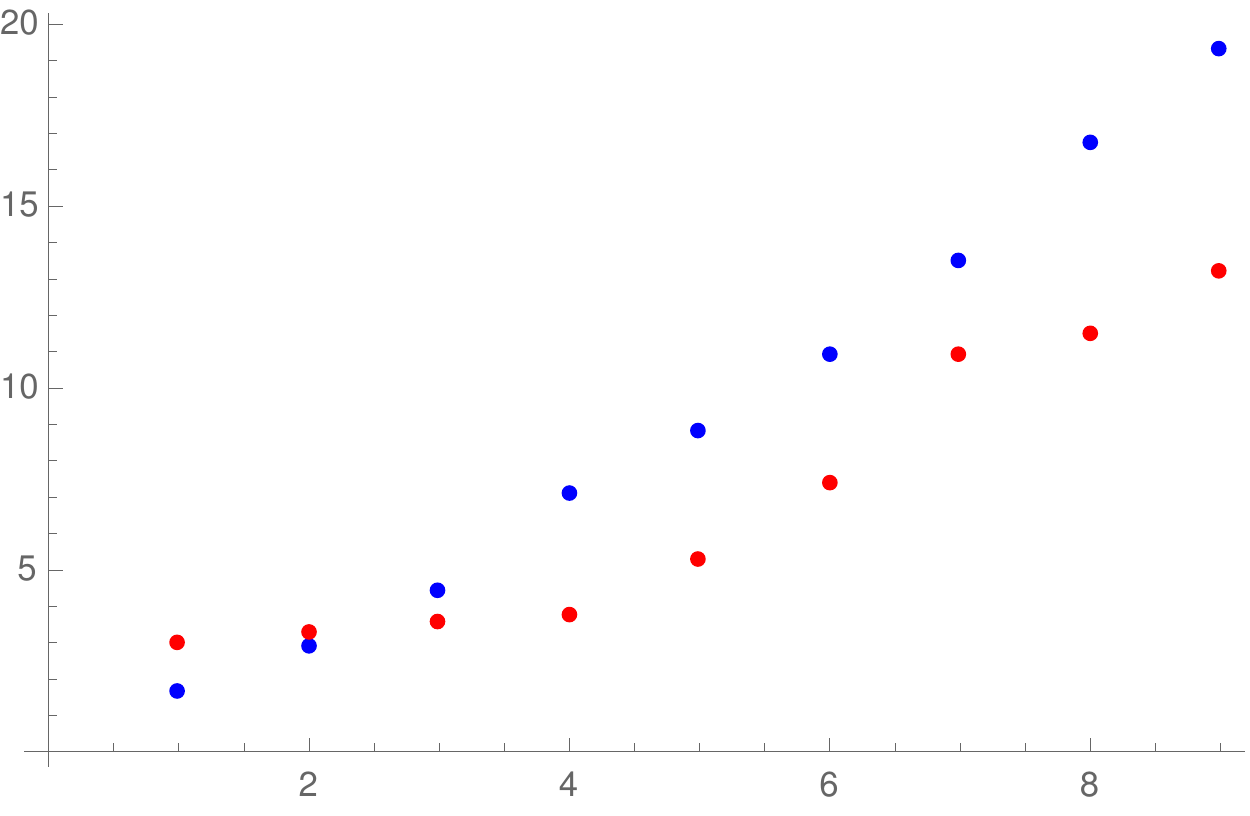}}\hspace{2em}\subfloat[$Z(0,1)$]{\includegraphics[width=0.4\linewidth]{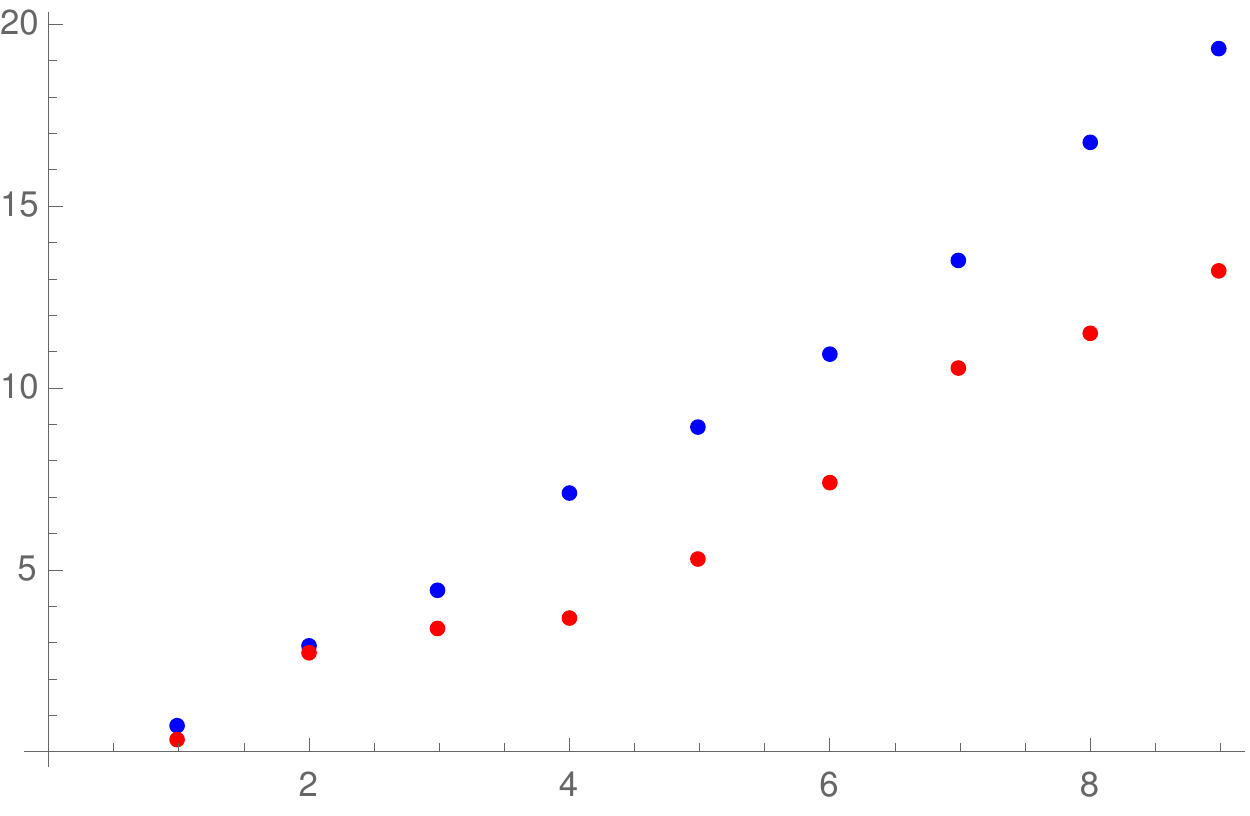}} \\
		\subfloat[$Z(2,0)$]{\includegraphics[width=0.4\linewidth]{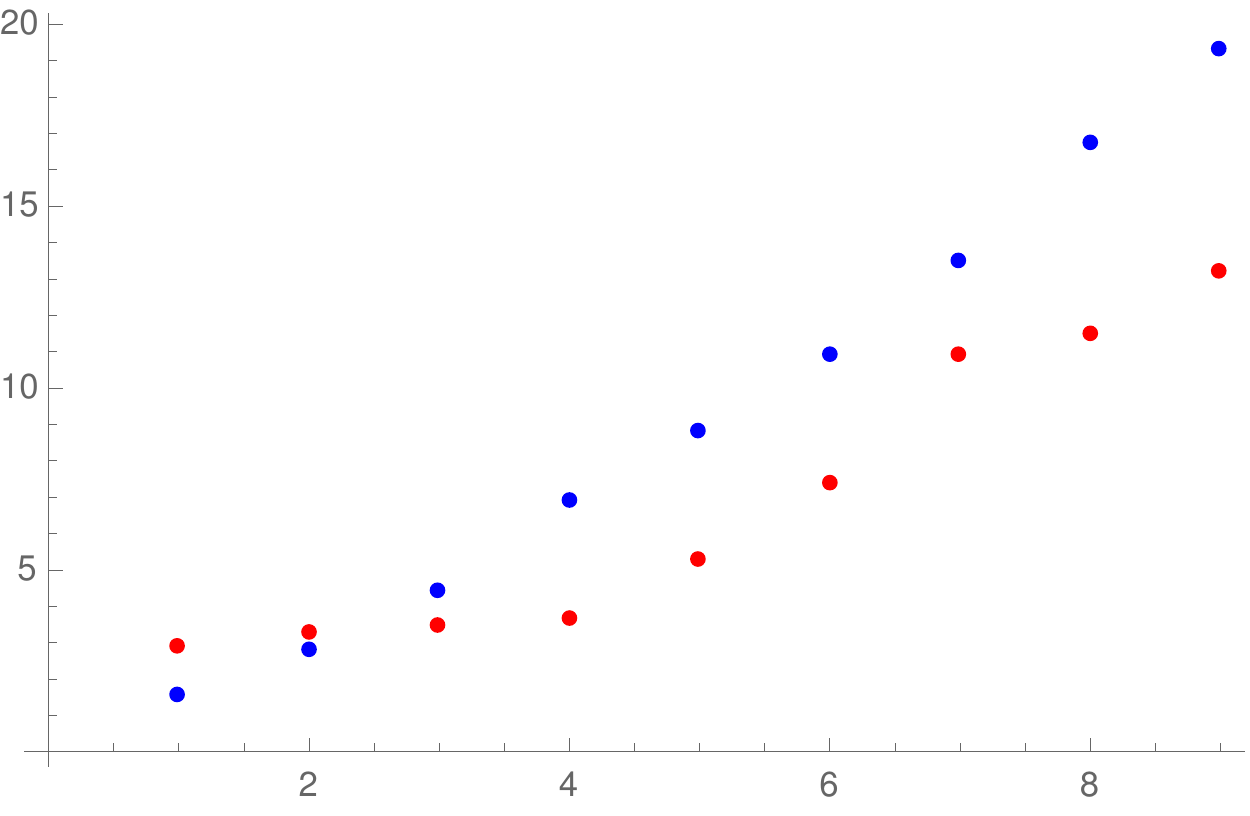}}\hspace{2em}\subfloat[$Z(0,2)$]{\includegraphics[width=0.4\linewidth]{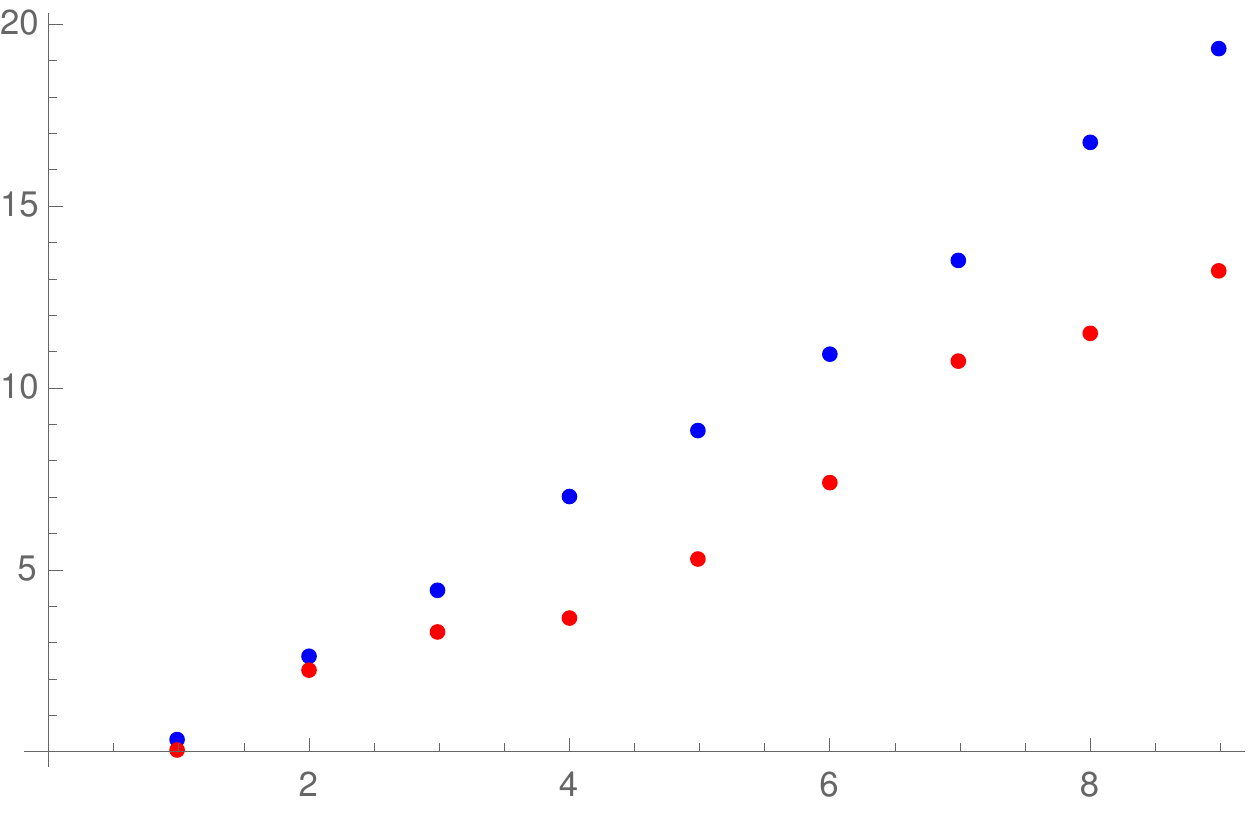}}
		\caption{The relative matching degrees between the four fermionic traces $Z(1,0)$, $Z(0,1)$, $Z(2,0)$, $Z(0,2)$ associated to resolved $\bC^3/\bZ_6$ orbifold ($\xi=0$) computed by the Airy function method and those by integrating the kernel, plotted against the order of instanton corrections used in the former method. The blue and red dots correspond to $\hbar=2\pi, \pi$ respectively.}\label{fg:C3Z6-h2p1p-Zs}
	\end{figure}	
	
	The calculation with the Airy function technique is relatively easy, as the quadratic mixing term is absent in the modified grand potential. The perturbative part of the grand potential is
	\begin{equation}
	\begin{aligned}
		J^{\rm (p)}(\boldsymbol{\mu};\hbar) =& \frac{1}{6\pi\hbar}\(4\mu_1^3 - 6\mu_1^2\mu_2 + 3 \mu_1 \mu_2^2 +4\mu_2^3\) \\
		&+ \( \frac{\pi}{3\hbar} - \frac{\hbar}{12\pi} \)(\mu_1+\mu_2) -\frac{(\log Q_3)^2}{8\pi\hbar} \mu_1  - \frac{(\log Q_3)^3}{36\pi\hbar} \ .
	\end{aligned}
	\end{equation}
	We choose the following change of variables
	\begin{equation}
		\(\begin{array}{c}
		 \mu_1 \\ \mu_2
		\end{array}\) = \begin{pmatrix}
		1 & \frac{1}{2} \\
		0 & 1
		\end{pmatrix} \(\begin{array}{c}
		 \nu_1 \\ \nu_2 
		\end{array}\) \ ,
	\end{equation}
	after which the modified grand potential becomes
	\begin{equation}
	\begin{aligned}
		J^{\rm (p)}(\boldsymbol{\mu};\hbar) = \frac{1}{12\pi\hbar}\( 8\nu_1^3 + 9\nu_2^3 \) +\(  \frac{\pi}{6\hbar} -\frac{\hbar}{24\pi} \)(2\nu_1 + 3\nu_2) -\frac{(\log Q_3)^2}{16\pi\hbar}(2\nu_1+\nu_2) -\frac{(\log Q_3)^3}{36\pi\hbar} \ .
	\end{aligned}
	\end{equation}
	Note that corresponding to $\xi =0$ in the operators, here we should plug in $Q_3=-1$, as we have discussed in the previous section. We compute the same four fermionic traces with both $\hbar = 2\pi$ and $\hbar = \pi$, with instanton corrections included up to order 9. The relative matching degrees with the kernel results are plotted in Fig.~\ref{fg:C3Z6-h2p1p-Zs}. Yet again we see impressive agreement.

	\section{Resonances and quantum spectral curves}
	\label{sc:resonance}
	
	

	As we mentioned in the introduction, the operators arising from quantum mirror curves depend on the mass parameters of the geometry. 
	One example is the operator (\ref{pertmn}) with $m=n=1$, 
	\begin{equation}
	\label{ex-op-1}
		\sO_{1,1,\xi} = \re^{\sx} + \re^{\sy} + \re^{-\sx - \sy} + \xi \re^{-2\sx} \ .
	\end{equation}
	This operator occurs in the analysis of the quantum mirror curve for the $Y^{3,0}$ geometry considered in Sec.~\ref{sc:Y30}. 
	When $\xi>0$, the inverse of this operator is trace class, by an argument presented in \cite{kasmar}. Another example is the operator associated to the local $\IF_2$ geometry, 
	\begin{equation}
		\label{ex-op-2}
		\sO_{\IF_2,\xi} =\re^{\sx} + \re^{\sy} + \re^{-2\sx-\sy} + \xi \re^{-\sx} \ ,
	\end{equation}
	which is a perturbation of the operator $\mO_{2,1}$ (although different from (\ref{pertmn})). The inverse operator is trace class when $\xi >-2$, as it can be seen by using the relation 
	to the operator associated to local $\IF_0$ \cite{kmz,gkmr}. 
	
	What happens when the values of the parameters are negative? A first hint of their behavior can be obtained by considering 
	the region of the phase space $(x,y)$ in which their classical counterparts are bounded by $\re^E$ (see for example \cite{mmrev} for more details on this type of argument). For the operator 
	$\sO_{1,1,\xi}$, we note that the value of the function
	\be
		\cO_{1,1,\xi}(x,y) =\re^x + \re^y +\re^{-x-y} +\xi \re^{-2x} \ ,\quad (x,y)\in \bR^2
	\ee
	is unbounded from below along the directions
	\be
		x < y < -2x \ , \quad x \rightarrow -\infty \ ,
	\ee
	as shown on the left in \figref{fg:regions}. This suggests that the inverse operator $\rho_{2,1,\xi<0}$ is not compact. Similarly, the value of the function
	\be
		\cO_{\IF_2,\xi}(x,y) =\re^x + \re^y  +\re^{-2x-y} +\xi \re^{-x} \ ,\quad (x,y)\in\bR^2,
	\ee
	is unbounded from below along the direction
	\be
		y = -x \ , \quad x\rightarrow -\infty \ ,
	\ee
	when $\xi<-2$, as shown on the right in \figref{fg:regions}. This suggests that $\rho_{\IF_2,\xi<-2}= \sO_{\IF_2,\xi}^{-1}$ is not compact. The situation arising here is very similar to what 
	happens in the quartic oscillator discussed in the introduction, since for 
	negative coupling, the classical counterpart of the Hamiltonian (\ref{qosc}) is also unbounded from below. 
	
\begin{figure}
\centering
\includegraphics[width=0.35\linewidth]{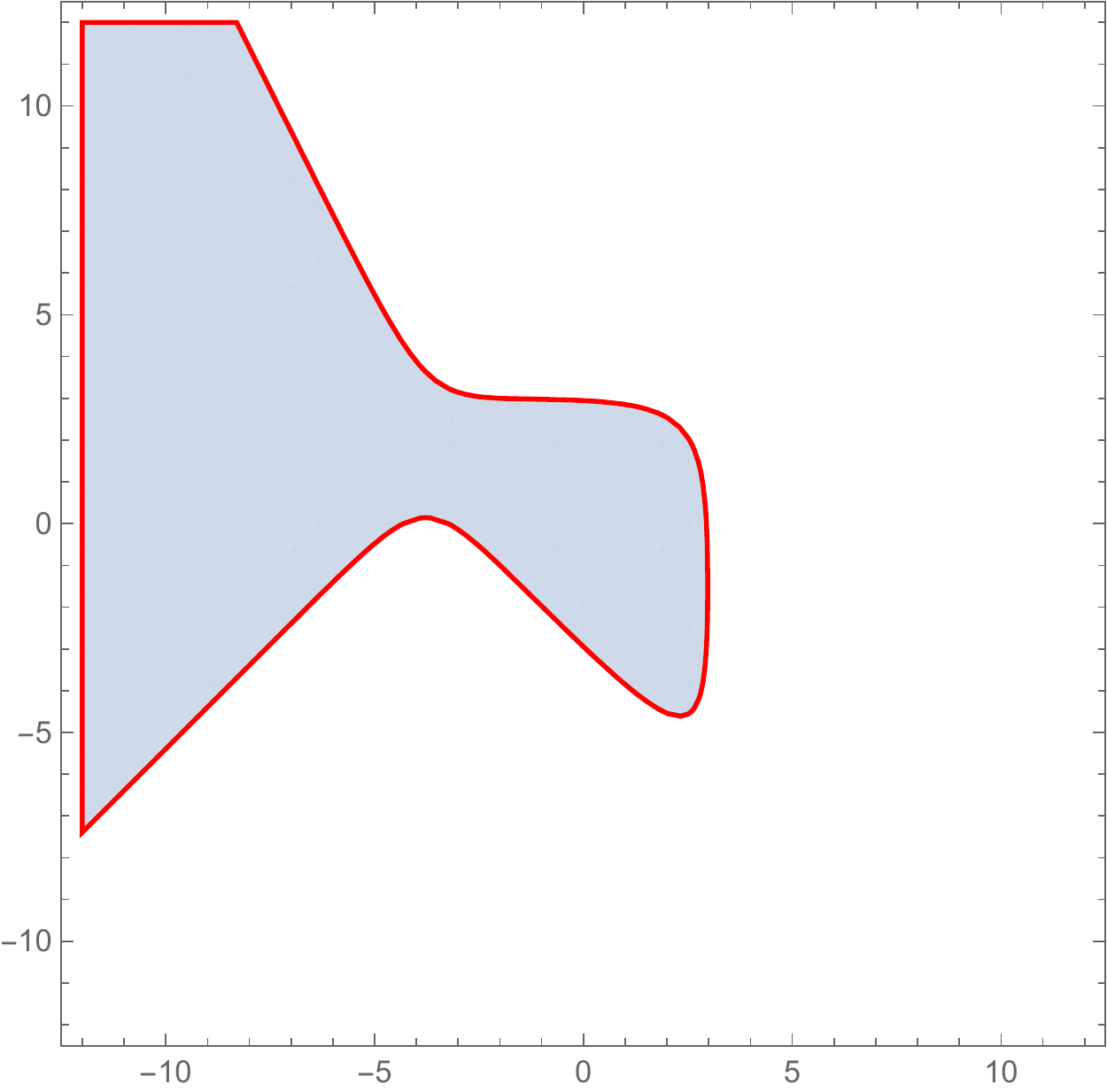}\hspace{3em}
\includegraphics[width=0.35\linewidth]{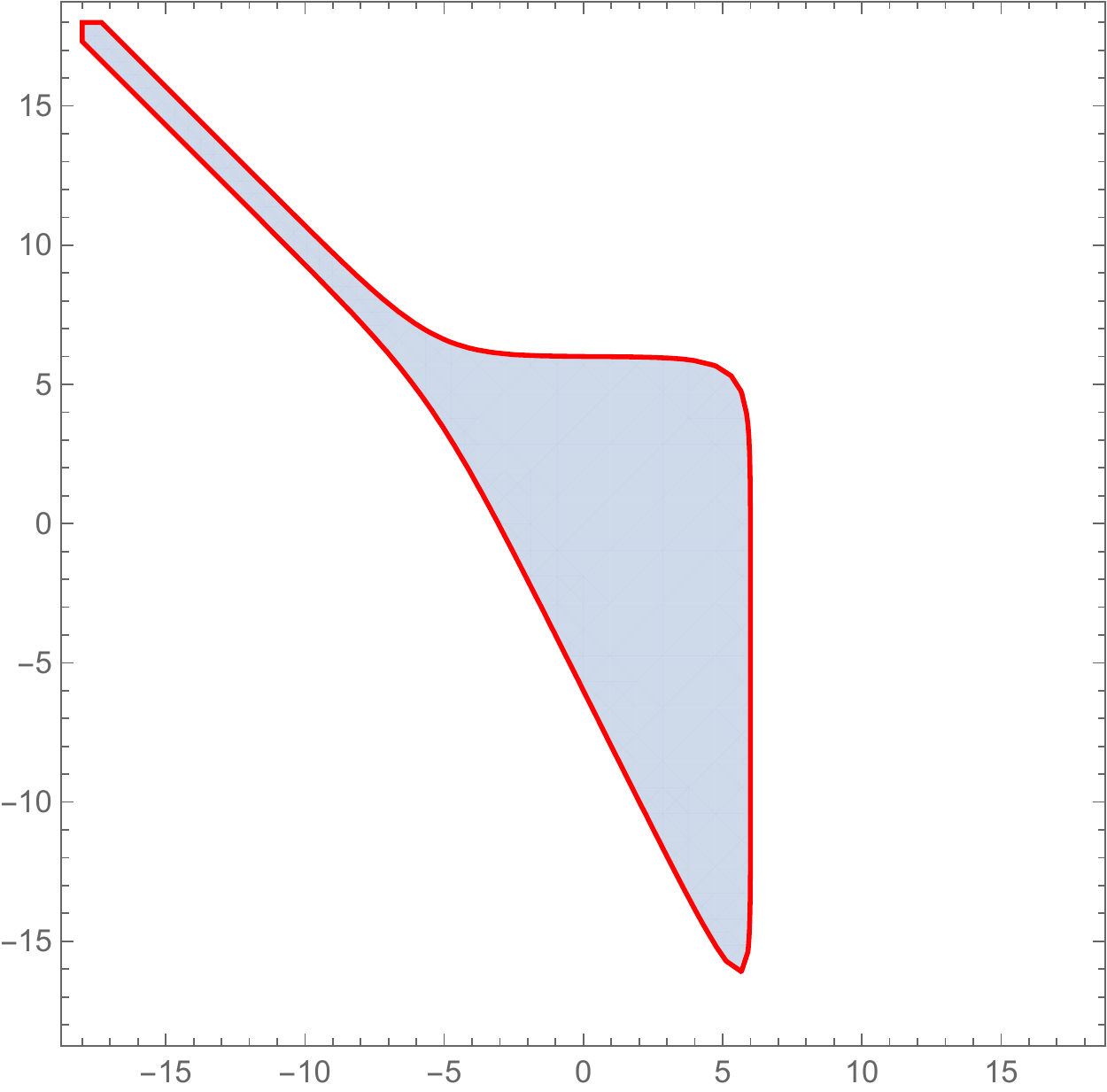}
\caption{The regions in the $(x,y)$ plane where $\cO_{1,1,\xi}(x,y)\le \re^E$, with $E=3$ and $\xi=-1/100$ (left) and where $\cO_{\IF_2,\xi}(x,y)\le \re^E$, with $E=6$ and $\xi=-5/2$ (right).}\label{fg:regions}
\end{figure}	
	More generally, we can consider the operators (\ref{ex-op-1}) and (\ref{ex-op-2}) when the parameter $\xi$ takes arbitrary complex values. It turns out that the values of the spectral traces 
	can be analytically continued to complex values for the mass parameter, in such a way that the conjectural formula (\ref{multi-Airy}) remains true. This is very satisfying, since 
	after all the mass parameters in topological string theory are naturally complex numbers. As in the case of the quartic oscillator with $g<0$, we will see that, when $\xi$ is negative and 
	the operators are non-compact, we can define a natural notion of resonance for the operators $\rho_{1,1,\xi<0}$, $\rho_{\IF_2,\xi<-2}$, i.e. we will find an infinite, discrete set of complex eigenvalues. 
	The spectral traces will be then given by sums over this resonant spectrum.


	To see this in some detail, let us first look at the spectral traces of $\rho_{\IF_2,\xi}$ and $\rho_{1,1,\xi}$ when $\xi>0$. These traces can be computed analytically, 
	with the help of the Faddeev's quantum dilogarithm, especially when $\hbar = 2\pi$. 
	For $\rho_{1,1,\xi}$, the first two traces have already been written down in \eqref{eq:trRho} and \eqref{eq:tr11sq}. For convenience, we repeat the results here:
	\begin{equation}\label{eq:tr-11}
	\begin{aligned}
		\tr \rho_{1,1,\xi}(\hbar = 2\pi) =& \frac{\pi(2-\xi^{1/3})+\sqrt{3}\xi^{1/3}\log \xi}{18\pi(1-\xi^{1/3} + \xi^{2/3})} \ ,\\
		\tr \rho_{1,1,\xi}^2(\hbar = 2\pi)  =& -\frac{6\sqrt{3}(1+\xi)+ \pi(-4+4\xi^{1/3}-13\xi^{2/3}+6\xi)}{108\pi(1+\xi^{1/3})(1-\xi^{1/3}+\xi^{2/3})^2} \\
		&\phantom{}+ \frac{\xi^{2/3}\log \xi}{18\sqrt{3}\pi(1-\xi^{1/3}+\xi^{2/3})^2} + \frac{\xi^{2/3}(\log\xi)^2}{36\pi^2(1+\xi^{1/3})(1-\xi^{1/3}+\xi^{2/3})^2} \ .
	\end{aligned}	
	\end{equation}
	For $\rho_{\IF_2,\xi}$, we quote results from \cite{kmz,gkmr}:
	\begin{equation}\label{eq:tr-21}
	\begin{aligned}
		\tr \rho_{\IF_2,\xi}(\hbar=2\pi) &= \frac{1}{4\pi}\frac{\cosh^{-1}(\xi/2)}{\sqrt{\xi-2}} \ ,\\
		\tr \rho_{\IF_2,\xi}^2(\hbar=2\pi) &= \frac{1}{16\pi^2} \left[ \( 2\frac{\cosh^{-1}(\xi/2)}{\sqrt{\xi^2-4}} + 1 \)^2 - 1 -\frac{\pi^2}{\xi+2} \right] \ .
	\end{aligned}	
	\end{equation}
One notices immediately that these functions of $\xi$ can be analytically continued to the whole complex plane 
with appropriate branch cuts. The traces of $\rho_{1,1,\xi}$ are multivalued functions with a branch cut starting at $\xi=0$, while the traces of  $\rho_{\bF_2,\xi}$ 
are multivalued functions of $\xi$ with a branch cut starting at $\xi=-2$. This is true for all the traces of the two operators: the traces of $\rho_{1,1,\xi}(\hbar=2\pi)$ 
are rational functions of $\xi^{1/3}, \log \xi$, while 
the traces of $\rho_{\bF_2,\xi}(\hbar=2\pi)$ are always rational functions of $\cosh^{-1}(\xi/2), \sqrt{\xi-2}, \sqrt{\xi+2}$ 
(the apparent branch point at $\xi=2$ is always spurious). We now recall that the fermionic spectral traces can be 
recovered from the standard traces, by the standard formula 
	\begin{equation}
		Z(N) = {\sum_{\{m_\ell\}}}' \prod_{\ell}\frac{(-1)^{(\ell-1)m_\ell} (\tr\rho^\ell)^{m_\ell}}{m_\ell ! \ell^{m_\ell}} \ ,
	\end{equation}
	with the prime meaning we sum over all the partitions $\{m_\ell\}$ that satisfy
	\begin{equation}
		\sum_\ell \ell m_\ell = N \ .
	\end{equation}
We can use this result to define the fermionic spectral traces as functions (with branch cuts) on the complex plane of the mass parameter $\xi$. 
This allows in turn to define an analytic continuation of the Fredholm determinants by using 
the fermionic spectral traces,  
\be
\label{spec-cont}
\Xi_{\bF_2, \xi}(\kappa)=1+ \sum_{N=1}^\infty Z_{\IF_2,\xi}(N; \hbar) \kappa^N, \qquad \Xi_{1,1, \xi}(\kappa)=1+ \sum_{N=1}^\infty Z_{1,1, \xi} (N; \hbar) \kappa^N.
\ee
For general complex values of $\xi$, the operators are no longer of trace class, so the above formulae define just formal power series in $\kappa$. In the trace class case, 
the fermionic spectral traces decrease rapidly as $N$ increases, in such a way that the spectral determinants are entire functions of $\kappa$. We do not have a proof in spectral theory that this is still 
the case for the analytically continued fermionic traces. In order to make progress on this issue, we turn to topological string theory. 
	
In topological string theory, the mass parameters are naturally complex. Therefore, the integral on the 
r.h.s. of the Airy function formula (\ref{multi-Airy}) can be computed for arbitrary complex mass parameters, in terms 
	of topological string data. The resulting expression inherits the same branch cut structure discussed above, due to the form of the mirror map for the mass parameters.  
	We can now ask whether the conjectural formula (\ref{multi-Airy}) remains true for arbitrary values of $\xi$, i.e. we can ask whether the analytically continued 
	spectral traces can be still computed in terms of topological string theory. We plot in Figs.~\ref{fg:Zl-F2} and respectively in Figs.~\ref{fg:Zl12-A2} the relative matching degrees 
	between the first two spectral traces $\tr\rho, \tr\rho^2$ computed by the Airy function method, and the results from the integral kernel, for the operators $\rho_{\bF_2,\xi=-3}$ and $\rho_{1,1,\xi=-1}$ with $\hbar =2\pi$\footnote{In calculating 
	these values, we have made a consistent choice of sign for the imaginary part of the multivalued functions.}. We find very good 
	agreement\footnote{When applying the Airy function method in the case of $\rho_{1,1,\xi}$, the quadratic mixing term $\zeta\nu_1\nu_2$ does not disappear as long as $\xi\neq 1$. Therefore, 
	as discussed in Sec.~\ref{sc:Airy}, we have to perform both the perturbative expansion and the non-perturbative expansion. 
	In Figs.~\ref{fg:Zl12-A2}, the relative matching degree for $\rho_{1,1,\xi=-1}$ is plotted against not only the instanton degree 
	but also the perturbative degree. The double expansion truncation greatly constrains the numerical precision we can reach, and in the end, as one finds in the plot, we get at most approximately $7\sim 8$ matching digits.}.	
	\begin{figure}
		\centering
		\subfloat[$Z_1$]{\includegraphics[width=0.35\linewidth]{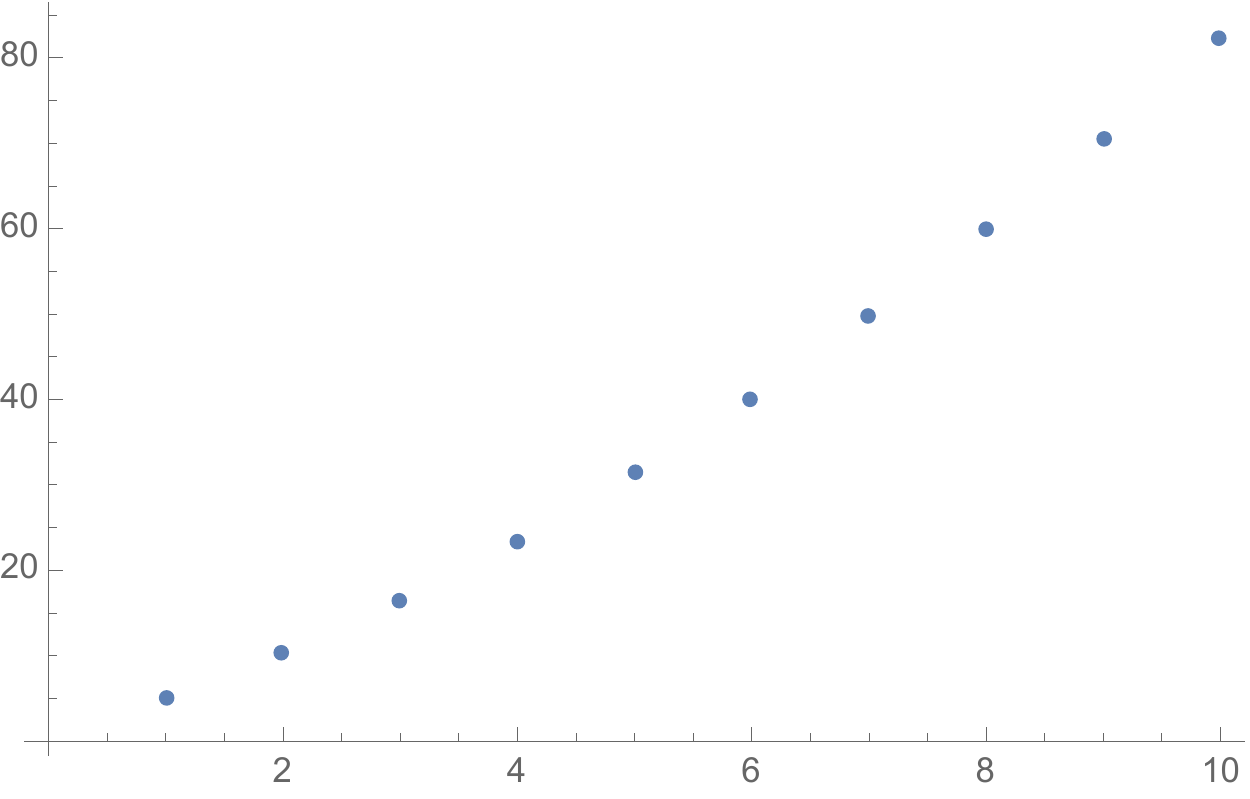}}\hspace{1.2em}
		\subfloat[$Z_2$]{\includegraphics[width=0.35\linewidth]{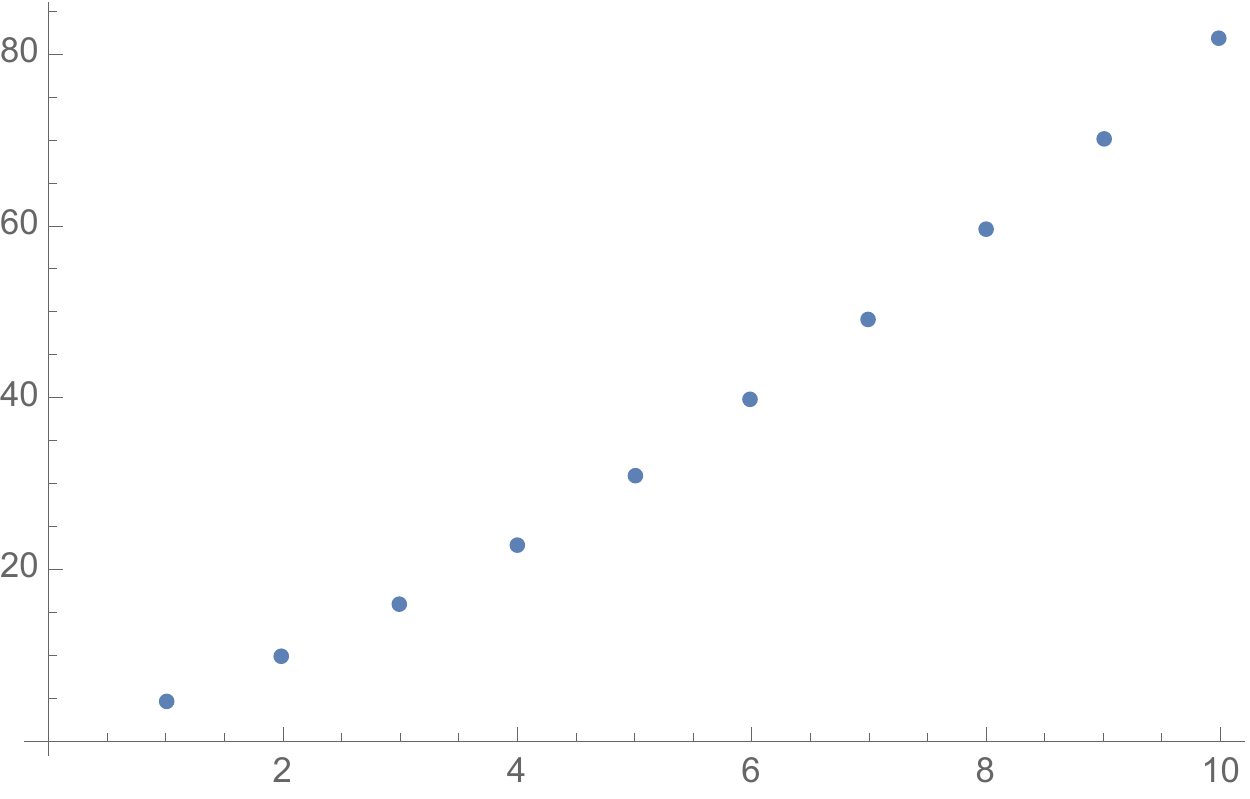}}
		\caption{The relative matching degrees between $\tr\rho_{\bF_2,\xi}, \tr\rho^2_{\bF_2,\xi}$ with $\xi=-3$ and $\hbar=2\pi$ computed by the Airy function method and the analytic spectral traces in \eqref{eq:tr-21}, plotted against the order of instanton corrections included in the calculation of the former.}\label{fg:Zl-F2}
	\end{figure}

	\begin{figure}
		\centering
		\subfloat[perturbative degree]{\includegraphics[width=0.35\linewidth]{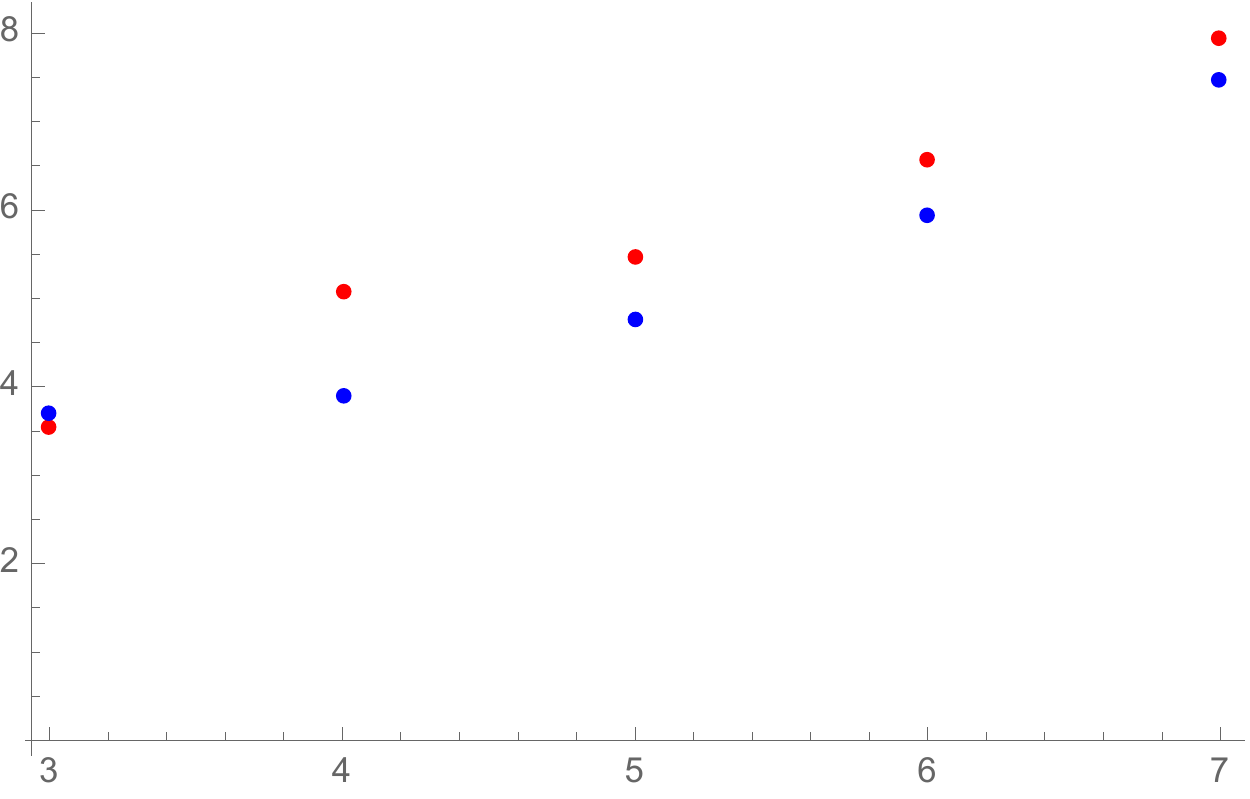}}\hspace{1.2em}
		\subfloat[instanton degree]{\includegraphics[width=0.35\linewidth]{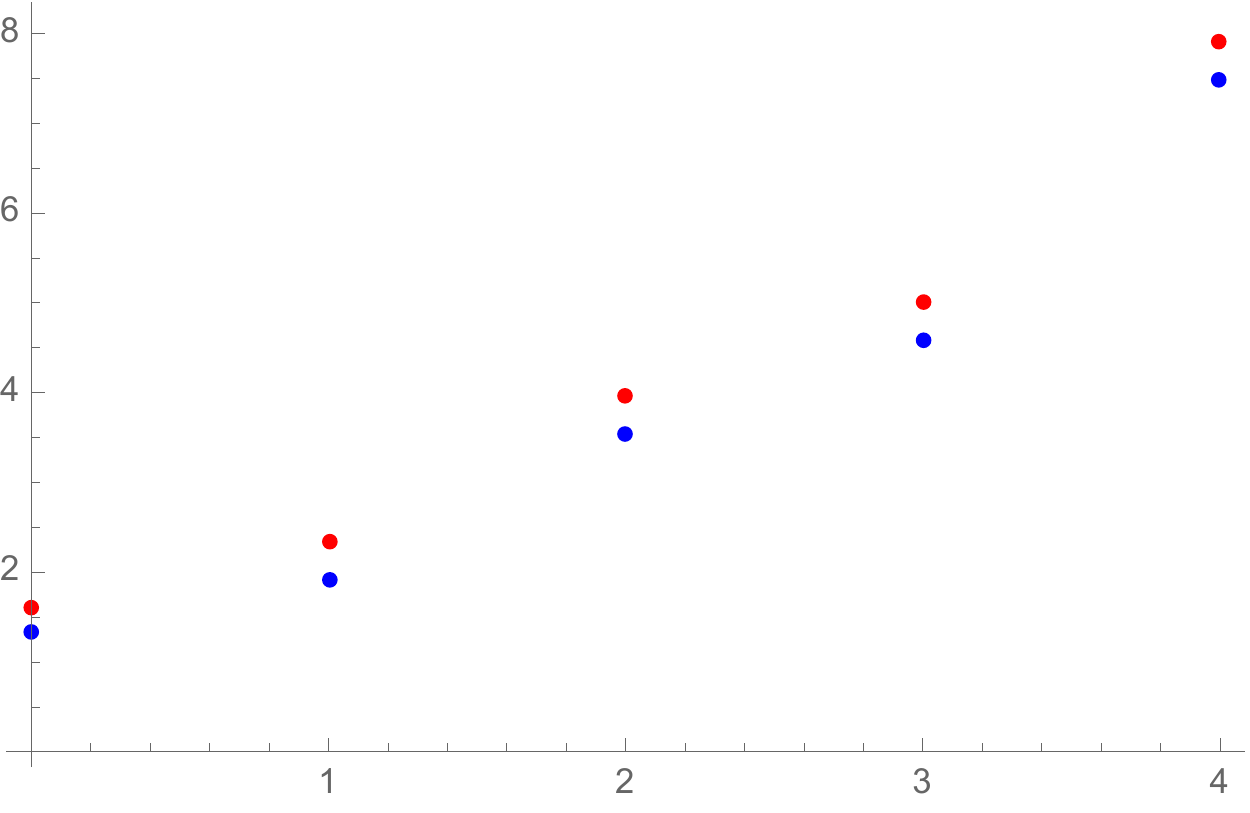}}
		\caption{The relative matching degrees between $\tr\rho_{1,1,\xi}$ (red points), $\tr\rho_{1,1,\xi}^2$ (blue points) with $\xi=-1$ and $\hbar=2\pi$ computed by the Airy function method and the spectral traces in \eqref{eq:tr-11}, plotted (a) against the perturbative degree, with fixed instanton degree 4, and (b) against the instanton degree, with fixed perturbative degree 7, used in the former method.}\label{fg:Zl12-A2}
	\end{figure}

We conclude that, at least in these examples (and others we have tested) the spectral traces are still given by the r.h.s. of (\ref{multi-Airy}). 
It is natural to conjecture that the formula (\ref{multi-Airy}) remains true for generic values of $\xi$. Assuming this is the case, we can deduce some properties of the analytically 
continued spectral traces, in particular their growth rate. We will do this analysis for both geometries simultaneously, eliminating the corresponding labels. 
When $N$ is large, $Z(N)$ is dominated by the perturbative component $Z^{\rm (p)}(N)$. 
Invoking the Airy function formula \eqref{eq:Zp-Ai}, roughly speaking we have
	\begin{equation}\label{eq:ZN-growth}
		|Z(N)| \sim \big|\re^{ D\cdot C^{-1} N}\big| \Ai(C^{-1/3} N)  , \qquad N\gg 1. 
	\end{equation}
	The coefficients $C=C(\hbar)$ and $D=D(\xi,\hbar)$ have been defined in the perturbative part of the modified grand potential $J^{\rm (p)}$ in \eqref{eq:Jp}. Of the two, 
	$C(\hbar)$  is proportional to the classical triple intersection number in the CY, and for any positive real value of $\hbar$ it is always a positive constant. By using the asymptotic bound of the Airy function, we have
	\begin{equation}
		|Z(N)| \lesssim \big| \re^{D\cdot C^{-1} N}\big|  \frac{\re^{-\tfrac{2}{3} C^{-1/2} N^{3/2}}}{2\sqrt{\pi} C^{-1/12} N^{1/4}} \ .
	\end{equation}
	The growth of $Z(N)$ is dominated by $\re^{-\tfrac{2}{3} C^{-1/2} N^{3/2}}$, therefore the corresponding Fredholm determinant is an entire function, regardless of the value of the mass $\xi$. The coefficient 
	$D(\xi,\hbar)$ is a function of $\xi$, but it only affects the growth behavior at subleading order.  
	We conclude that the functions (\ref{spec-cont}), defined for generic $\xi$ as formal power series, are entire. 
	
	In fact, by using topological string theory, we can be more precise about the structure of the Fredholm determinant. A useful theorem in \cite{simon-paper} shows that the entire function $\Xi_\rho(\kappa)$ has the infinite product form
	\begin{equation}\label{eq:Xi-product}
		\Xi_\rho(\kappa) = \prod_{j=1}^\infty (1-\kappa z_j^{-1}) \ ,
	\end{equation}
	if the following three conditions are satisfied:
	\begin{enumerate}
		\item $\Xi_\rho(0) = 1$ \ ;
		\item $\sum_{j=1}^\infty |z_j|^{-1} < \infty$ \ ;
		\item for any $\epsilon>0$, $|\Xi_\rho(\kappa)| \leqslant C(\epsilon) \exp(\epsilon |\kappa|)$ \ .
	\end{enumerate}
	Let us now use the information provided by topological string theory to see if these assumptions are still valid for arbitrary complex mass parameters. The first condition is satisfied by construction (the Airy function method is normalized by $Z(0) = \Xi_\rho(0) =1$). The second condition roughly speaking says the first coefficient $Z(1)$ is finite, and we have shown it by actual computation for $\rho_{\bF_2,\xi}$ and $\rho_{1,1,\xi}$. The last condition puts a stricter bound on the growth of the coefficients. 
	So we continue our estimate of $Z(N)$ from \eqref{eq:ZN-growth}. Using the fact that asymptotically
	\begin{equation}
		n^{1/2} > \alpha \log n + \beta \ ,\forall \alpha \in \bR_+, \, \beta\in \bR \ ,
	\end{equation}
	and that
	\begin{equation}
		\log n! < n \log n \ ,
	\end{equation}
	we have
	\begin{equation}
	\begin{aligned}
		|\Xi_\rho(\kappa)| \leqslant \sum_{N=0}^\infty \big|Z(N)\big| |\kappa|^N 
		&\lesssim \sum_N (\big|\re^{D\cdot C^{-1}}\big| |\kappa|)^N \re^{-\tfrac{2}{3} C^{-1/2} N ( \tfrac{3}{2}C^{1/2}\log N + \beta)}\\
		&\lesssim \sum_N \frac{1}{N!}\(\re^{|D\cdot C^{-1} |-\tfrac{2}{3} C^{-1/2}\beta}|\kappa|\)^N \\
		&= \exp \(\re^{|D\cdot C^{-1}| -\tfrac{2}{3} C^{-1/2}\beta}|\kappa|  \) \ .
	\end{aligned}
	\end{equation}
	Since we can tune the constant $\beta$ for the factor $\re^{|D\cdot C^{-1}| -\tfrac{2}{3} C^{-1/2}\beta}$ to be as (positively) small as possible, 
	the Fredholm determinant grows in $\kappa$ more slowly than any exponential (this is in agreement with the $\exp ((\log \kappa)^3)$ behavior noted in (109) and (110) of \cite{mmrev}). 
	The last condition of the theorem is also satisfied. Note that this argument does not depend on the value of the mass parameter $\xi$, and so the infinite product form of the Fredholm determinant is 
	valid even when the mass parameters are complex.
	
	We conclude that, even for complex values of the mass parameter, the Fredholm determinants of the operators $\rho_{\bF_2,\xi}$, $\rho_{1,1,\xi}$ have an infinite discrete set of complex 
	zeros on the complex $\kappa$ plane. These zeros can be used to define the sets,
		\begin{equation}
		\label{sets}
	\begin{aligned}
		\Delta_{\bF_2,\xi} =& \left\{ -z_j^{-1} \; :\; \Xi_{\bF_2,\xi}(z_j) = 0 \right\} \ , \\
		\Delta_{1,1,\xi} =& \left\{ -z_j^{-1} \; :\; \Xi_{1,1,\xi}(z_j) = 0 \right\} \ .
	\end{aligned}	
	\end{equation}
	As a concrete example, we list in Tab.~\ref{tb:F2-nxi} the first few entries (sorted by decreasing real components) of the set $\Delta_{\bF_2,\xi}$ for $\xi=-3$ and $\hbar = 2\pi$. When up to order nine instanton corrections are used in the computation of the Fredholm determinant, these entries have up to 70 stabilized digits.
	
When the value of $\xi$ is such that the 
corresponding operators are of trace class, the sets (\ref{sets}) are simply the spectra of the operators. 
When the operators are not compact, we regard these values as {\it resonances}, or resonant eigenvalues, of 
the operators. They are indeed natural analytic continuations of the eigenvalues occurring in the trace class case. In addition, 
one can verify that the resonant eigenvalues obtained in this way (like those listed in Tab.~\ref{tb:F2-nxi}) agree with the solutions of the 
quantization conditions of \cite{ghm,hm,fhm} for complex values of $\xi$. It would be very interesting to see if these resonances can be obtained directly from the operators, by using generalizations of the 
complex dilatation techniques that are so powerful for Schr\"odinger operators. 

	\begin{table}
		\centering
		\begin{tabular}{*{3}{>{$}l<{$}}}\toprule
			n & \real \lambda & \imag \lambda \\\midrule
			0 & 0.08232509967018131347\ldots & -0.02921973315814942282\ldots\\
			1 & 0.01825649964473641110\ldots & -0.00352353844067812079\ldots\\
			2 & 0.00618641344745375484\ldots & -0.00092900112363282145\ldots\\\bottomrule
		\end{tabular}
		\caption{The resonant eigenvalues $\lambda\in \Delta_{\bF_2,\xi}$ when $\xi=-3$ and $\hbar =2\pi$.}\label{tb:F2-nxi}
	\end{table}

	\section{Quantum mirror curves and cluster integrable systems}
	\label{sc:cis}

	We defined in Sec.~\ref{sc:mc} the quantum mirror curve as an operator obtained by quantizing the mirror curve to a toric CY threefold $X$:
	\begin{equation}
		\sW_{X,i} = \sO_i + \kappa_i = \sO_i^{(0)} + \sum_{j} \kappa_j\sP_{ij} \ .
	\end{equation}
	This can be regarded as a tool to encode the operators $\sO_i$ or $\sO^{(0)}_i$, which are usually the three-term operators $\sO_{m,n}$ or their perturbations. There is another spectral problem that can be 
	associated to a toric CY manifold. Goncharov and Kenyon \cite{gk} assign to every 2d convex Newton polygon $\cN$ an integrable system called {\it cluster integrable system}. The number of Poisson commuting 
	Hamiltonians $H_i(q_k,p_k)$ is the same as the number $g_\Sigma$ of inner integral vertices in $\cN$. If we choose $\cN$ to be the 2d support of the toric fan of a toric CY threefold $X$, it turns out that the spectral curve of 
	the classical integrable system coincides with the mirror curve of topological string theory on $X$. The $g_\Sigma$ true moduli are identified with the 
	conserved Hamiltonians, while the $r_\Sigma$ mass parameters are identified with the Casimir parameters, up to signs. 
	
	As an example, let us revisit the $Y^{N,0}$ geometry. By setting $q=0$ in the mirror curve for the general $Y^{N,q}$ geometry in \eqref{eq:Nqcurve}, we get the mirror curve to the $Y^{N,0}$ geometry\footnote{In contrast to \eqref{eq:Nqcurve}, we use the symbol $y$ in place of $p$ since the latter has been reserved as the momentum variable of the underlying integrable system.}
	\begin{equation}\label{eq:cs-An}
		a_1 \re^y + a_2 \re^{Nx - y} + \sum_{i=0}^{N} b_i \re^{i x}  = 0 \ .
	\end{equation}
	The mirror curve has genus $g_\Sigma=N-1$, corresponding to the true moduli $b_i, i =1,\ldots,N-1$. The other coefficients $a_1, a_2, b_0, b_N$ are mass parameters, while the $\bC^*$ actions reduce their degrees-of-freedom to one. The associated cluster integrable system is nothing else but the relativistic, periodic $\widehat{A}_{N-1}$ Toda lattice \cite{efs,marshakov,fm}. The spectral curve is \eqref{eq:cs-An} with the following identification
	\begin{equation}\label{eq:is-prm-An}
		a_1 = a_2 = R^N \ , \quad b_i = (-1)^{N-i} H_{N-i}\ , \quad i = 0,1,\ldots, N \ ,
	\end{equation}
	where $H_0= H_{N} = 1$ are trivial. Indeed the true moduli are identified with the non-trivial Hamiltonians $H_i, i=1,\ldots,N-1$, while the 
	mass parameter is identified with the only Casimir parameter $R$. Let us now follow the convention in Sec.~\ref{sc:Y30-gm}, 
	rename the true moduli $b_i$ by $\kappa_{N-i}$ for $i=1,\ldots,N-1$, the mass parameter $b_0$ by $\xi$, and set $a_1,a_2,b_N$ to 1. 
	After proper scaling, the identification \eqref{eq:is-prm-An} can be written as
	\begin{equation}\label{eq:kappa-H}
		\xi = (-1)^N R^{-N} \ ,\quad \kappa_i = (-1)^i R^{-i} H_i \ ,\quad i=1,\ldots,N-1 \ .
	\end{equation}
	Furthermore, the Batyrev coordinates $z_i$ can also be expressed in terms of $H_i$ and $R$, and they read
	\begin{equation}\label{eq:z-H}
	\begin{aligned}
		z_i &= \frac{H_{i-1}H_{i+1}}{H_i^2} \ ,\quad i=1,\ldots,N-1 \ ,\\
		z_N &= (-1)^N R^{2N} \ .
	\end{aligned}	
	\end{equation}
	
	It is straightforward to quantize the classical cluster integrable system. The dynamic variables $p_k,q_k$ are promoted to operators subject to the usual commutation relation
	\begin{equation}
		[ \sq_i, \sp_j ] = \ri\hbar \delta_{i,j} \ .
	\end{equation}
	The $g_\Sigma$ Hamiltonian functions $H_i(p_k,q_k)$ are accordingly promoted to $g_\Sigma$ mutually commuting Hamiltonian operators
	\begin{equation}
		[ \sH_i(\sp_k,\sq_k), \sH_j(\sp_k,\sq_k)] = 0 \ .
	\end{equation}
Solving the quantum cluster integrable system means finding the spectra and eigenfunctions of these Hamiltonians. In \cite{hm,fhm} it was conjectured that the spectra 
can be obtained from exact quantization conditions, akin to those obtained in \cite{ns} for the Toda lattice, and involving topological string data. It turns out that, when $g_\Sigma=1$, the 
single Hamiltonian of the cluster integrable system is precisely the (single) operator $\mO$ obtained from the mirror curve. The quantization condition of \cite{hm,fhm} reduces in that case to the one 
proposed in \cite{wzh}, which has been checked to be equivalent to the vanishing of the Fredholm determinant of \cite{ghm}. 
	
	When $g_\Sigma>1$, the two spectral problems associated to a toric CY involve two very different sets of operators. 
	In particular, when mass parameters/Casimir parameters are present, sometimes the Hamiltonian operators $H_i$ and the operators $\sO^{(0)}_i$ 
	have real positive spectrum in different regions of the parameter space. This is the most conspicuous in the example of the $Y^{N,0}$ geometry/$\widehat{A}_{N-1}$ Toda lattice when 
	$N$ is odd. In the integrable system interpretation, the Casimir $R$ has to be positive for the Hamiltonians $\sH_i$ to have a positive real spectrum, 
	while in the operator interpretation, the mass parameter $\xi$ should be non-negative for the operators $\sO^{(0)}_1$ in \eqref{eq:Toda-Os} to have a positive 
	spectrum bounded from below. The dictionary \eqref{eq:kappa-H} dictates that the two scenarios are mutually exclusive.

	In spite of these differences, it turns out that, even in the higher genus case, both spectral problems are closely related. In particular, 
	the spectra of the Hamiltonians of the cluster integrable system are located in the zero locus of the 
	generalized Fredholm determinant $\Xi_X(\boldsymbol{\kappa};\boldsymbol{\xi},\hbar)$. Let us give a heuristic argument for this. 
	First, we set the Casimirs $R_i$ to positive values, as required by the integrable system interpretation. 
	As we have seen in Sec.~\ref{sc:resonance}, even if the corresponding mass parameters are negative, we can analytically continue the conjectural Fredholm determinant $\Xi_X(\boldsymbol{\kappa};\boldsymbol{\xi},\hbar)$ 
	constructed from topological string amplitudes into this region of the parameter space, and interpret it as the Fredholm determinant in terms of the resonant eigenvalues. Let us now write the quantum mirror curve as
		\begin{equation}
		\sW_{X,i}= \sO^{(0)}_i \( 1+\sum_{j=1}^{g_\Sigma} \kappa_j \sA_{ij} \). 
	\end{equation}
If the generalized spectral determinant vanishes, i.e. if
\be
\det \(1+\sum_{i=1}^{g_\Sigma}\kappa_j \sA_{ij}\) = 0, 
\ee
and assuming that $\sO^{(0)}_i$ has a trivial kernel, there should be a state $|\Psi\rangle$ annihilated by the quantum mirror curve: $\sW_{X,i} |\Psi\rangle = 0$. 
We can interpret this condition as a quantum Baxter equation for the cluster integrable system, arising from the quantization of the spectral curve. The values of the quantum 
Hamiltonians must be such that this equation is satisfied. Therefore, the vanishing of the generalized spectral determinant gives one constraint for the spectra of the quantum integrable system. 
In order to find the precise spectra, one needs $g_\Sigma-1$ additional 
conditions. 
	
	In  \cite{fhm} it was noted, in a genus two example, that one can impose an additional condition to determine the spectrum, 
	involving the generalized Fredholm determinant but in terms of ``rotated" variables. This observation 
	has been extended in \cite{swh}, where it has been proposed to use the vanishing of $g_\Sigma-1$ additional functions, obtained from the generalized 
	Fredholm determinant of \cite{cgm} but with different choices of the B-field. 
	
	In some cases one can consider much simpler conditions, namely, that the eigenvalues of the Hamiltonians should all be real and positive. We will now use the example of the $Y^{3,0}$ 
	geometry/relativistic $\widehat{A}_2$ Toda lattice to demonstrate that, in genus two, the spectrum of the cluster integrable system can be 
	sometimes obtained from the vanishing equation of the Fredholm determinant, together with the positive reality condition.
	
	We use the dictionary between the geometric moduli and the quantities of the integrable system in \eqref{eq:is-prm-An} specified to $N=3$, and the $\mathbf{B}$-field given by (\ref{y30-b}). 
	To simplify the calculation, we consider $\hbar=2\pi$, and choose $\xi=\re^{\ri\pi}=-1$ or equivalently $R=1$ in order to compare with the results in \cite{hm}. Then the computation of the coefficients $Z(\boldsymbol{N};\boldsymbol{\xi},\hbar)$ is just a continuation of that at the end of Sec.~\ref{sc:resonance}. Limited by computer capacities, we can reach at most the truncation at the perturbative degree 7 and the instanton degree 4, which as one finds in the precision plots Figs.~\ref{fg:Zl12-A2} limits the number of accurate digits of the results of $Z(\boldsymbol{N};\boldsymbol{\xi},\hbar)$ to about seven or eight.

	\begin{table}
		\centering
		\begin{tabular}{*{5}{>{$}c<{$}}}\toprule
			(n_1,n_2) &\real \Xi & \imag \Xi & {\rm slope}\,\real \Xi=0 & {\rm slope}\,\imag \Xi=0 \\\midrule
			(0,0) & -3.70 \times 10^{-8} & -1.24\times 10^{-7} & -2.846040 & -2.846038 \\
			(1,0) & 6.76\times 10^{-6} & -4.46\times 10^{-6} &-6.920725 & -6.920736 \\\bottomrule
		\end{tabular}
		\caption{Values of $\real\Xi(H_1,H_2)$ and $\imag\Xi(H_1,H_2)$ of relativistic $\widehat{A}_2$ Toda lattice ($\hbar=2\pi,R=1$) 
		when the eigenvalues of $\sH_1,\sH_2$ are plugged in, as well as the slopes of the two curves $\real\Xi(H_1,H_2)=0$ and $\imag\Xi(H_1,H_2)=0$ 
		at the points corresponding to the eigen-energies on the $(H_1,H_2)$ plane.
		}\label{tb:TodaA2-Xi}
	\end{table}	
	
	\begin{figure}
		\centering
		\subfloat[$(n_1,n_2)=(0,0)$]{\includegraphics[width=0.4\linewidth]{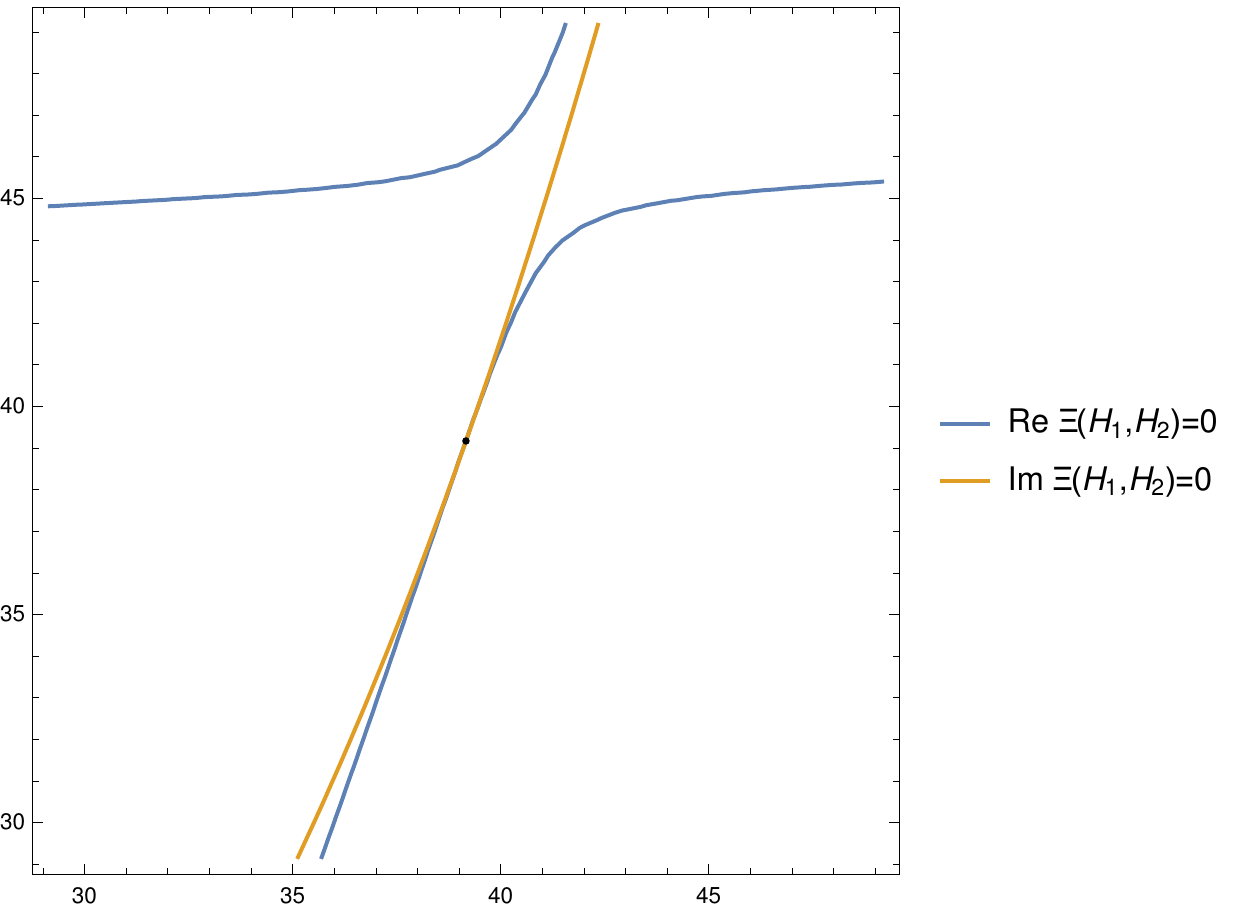}}\hspace{1em}
		\subfloat[$(n_1,n_2)=(1,0)$]{\includegraphics[width=0.4\linewidth]{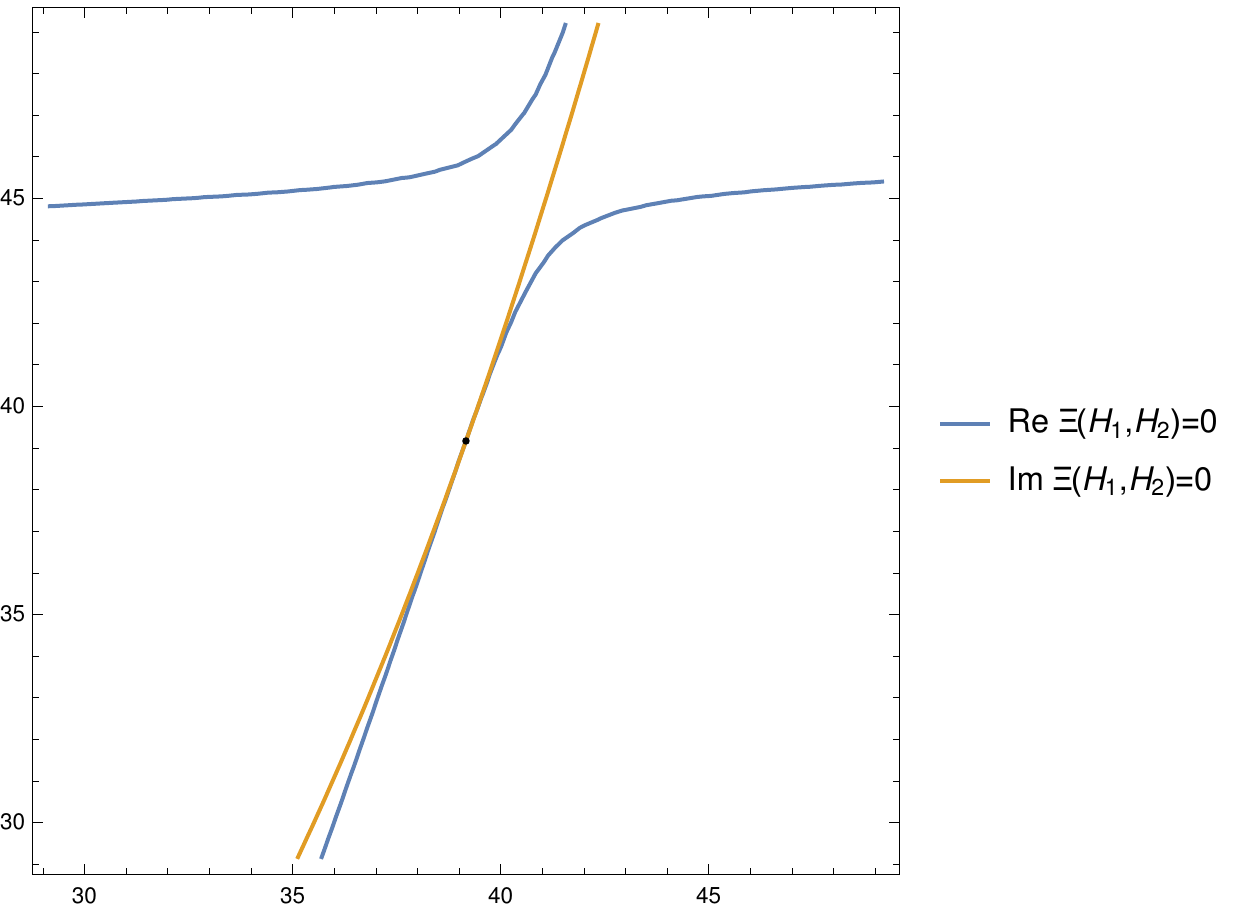}}
		\caption{The two curves $\real\Xi(H_1,H_2)=0$ and $\imag\Xi(H_1,H_2) = 0$ on the $\bR^2$ plane parametrized by $H_1,H_2$ of relativistic $\widehat{A}_2$ Toda lattice ($\hbar=2\pi, R=1$), zoomed in around the points (black) corresponding to the eigen-energies at levels $(0,0)$ and $(1,0)$.}\label{fg:TodaA2-tangency}
	\end{figure}

	We computed $Z(N_1 ,N_2; \xi=\re^{\ri\pi}, \hbar=2\pi)$ up to $N_1+N_2 = 10$.
	 Since these traces have complex values, the vanishing equation of the Fredholm determinant together with 
	 the positive reality conditions of $H_1, H_2$ is equivalent to two equations, the vanishing of both the real part and the 
	 imaginary part of the Fredholm determinant. To check that these two equations are enough to reproduce the spectrum of the $\widehat{A}_2$ Toda lattice in \cite{hm}, we construct the real and the imaginary determinants
	\begin{equation}
	\begin{aligned}
		\real \Xi(H_1,H_2) \equiv \real\Xi(H_1,H_2;-1,2\pi) = \sum_{N_1,N_2} \real Z(N_1,N_2;-1,2\pi)(-H_1)^{N_1} H_2^{N_2}  \ ,\\
		\imag \Xi(H_1,H_2) \equiv \imag\Xi(H_1,H_2;-1,2\pi) = \sum_{N_1,N_2} \imag Z(N_1,N_2;-1,2\pi)(-H_1)^{N_1} H_2^{N_2}  \ .
	\end{aligned}
	\end{equation}
	We expect both of them to vanish when the eigenvalues of $\sH_1,\sH_2$ are plugged in. Indeed when the eigenvalues at levels $(n_1,n_2) = (0,0), (1,0)$, which are computed in \cite{hm} and reproduced in the last rows of Tabs.~\ref{tb:TodaA2-spec}, are plugged in, the absolute values of $\real \Xi(H_1,H_2)$ and $\imag \Xi(H_1,H_2)$, as seen in Tab.~\ref{tb:TodaA2-Xi}, are within the error margins of the Fredholm determinant estimated from the precisions of the fermionic traces, thus it is consistent with the previous statement. We also plot in Figs.~\ref{fg:TodaA2-tangency} the curves $\real \Xi(H_1,H_2)=0$ and $\imag \Xi(H_1,H_2) = 0$ in the real plane of $(H_1,H_2)$, and find that the points corresponding to the eigen-energies at $(n_1,n_2) = (0,0), (1,0)$ lie on both curves.
	
	In practice, computing the spectrum of $\widehat{A}_2$ Toda lattice from the double vanishing equations of $\real\Xi(H_1,H_2)$ and $\imag \Xi(H_1,H_2)$ may not be preferable, as the two curves $\real \Xi(H_1,H_2)=0$ and $\imag \Xi(H_1,H_2) = 0$ intersect tangentially at the points of eigen-energies, as seen in Figs.~\ref{fg:TodaA2-tangency}. This is confirmed by the numerical calculation of the slopes of the two curves at the points of eigen-energies, as seen in Tab.~\ref{tb:TodaA2-Xi}. Here we compute the slope $k_f(H_1,H_2)$ of a curve $f(H_1,H_2) = 0$ by
	\begin{equation}
		k_f(H_1,H_2) \equiv \frac{\rd H_2(H_1;f)}{\rd H_1} = -\frac{\partial_{H_1}f(H_1,H_2)}{\partial_{H_2} f(H_1,H_2)} \ .
	\end{equation}
	Therefore starting from the Fredholm determinant, we can use any two of the following three equations to compute the spectrum of $\widehat{A}_2$ quantum Toda lattice
	\begin{equation}\label{eq:trieq-A2}
	\begin{gathered}
		\real \Xi(H_1,H_2) = 0 \ , \\
		\imag \Xi(H_1,H_2) = 0 \ , \\
		k_{\real \Xi}(H_1,H_2) = k_{\imag \Xi}(H_1,H_2) \ .
	\end{gathered}	
	\end{equation}
	As seen in Tab.~\ref{tb:TodaA2-spec}, all three combinations give solutions consistent with the eigen-energies given in \cite{hm}, while the two combinations including the identical slope equation provide solutions with better precision.
	
	\begin{table}
		\centering
		\begin{tabular}{*{2}{>{$}c<{$}} *{2}{>{$}l<{$}}}\toprule
			(n_1,n_2)&& H_1 & H_2 \\\midrule
			\multirow{4}{5ex}{(0,0)}	&\real\Xi=0, \imag \Xi=0  & 39.1604\ldots        & 39.1468\ldots \\
			&\real\Xi=0, k_{\real \Xi} = k_{\imag \Xi} 	             & 39.167844\ldots     & 39.167912\ldots \\
			&\imag \Xi=0, k_{\real \Xi} = k_{\imag \Xi}	           & 39.167834\ldots     & 39.167873\ldots \\\cmidrule{2-4}
			&\textrm{numerical}                  		                         & 39.16781907\ldots & 39.16781907\ldots \\\midrule
			\multirow{4}{5ex}{(1,0)}	&\real\Xi=\imag \Xi=0    & 61.82\ldots         & 151.41\ldots \\
			&\real\Xi=0, k_{\real \Xi} = k_{\imag \Xi}  	       & 61.9680\ldots      & 152.3712 \ldots \\
			&\imag \Xi=0, k_{\real \Xi} = k_{\imag \Xi}  	      & 61.9700\ldots      & 152.3858\ldots \\\cmidrule{2-4}
			&\textrm{numerical}                                                & 61.966419\ldots   & 152.359672\ldots \\\bottomrule			
		\end{tabular}
		\caption{The eigen-energies of $\sH_1, \sH_2$ of relativistic quantum $\widehat{A}_2$ Toda lattice ($\hbar=2\pi, R=1$) computed from the Fredholm determinant, using any two of the combinations of \eqref{eq:trieq-A2}. 
			The numerical results are quoted from \cite{hm}.
		}\label{tb:TodaA2-spec}
	\end{table}

	\section{Conclusions and discussion}
	
	In this paper we continued the program started in \cite{ghm,cgm}, and we studied the spectral theory of trace-class operators associated to mirror curves of toric CY threefolds. 
	We focused on two geometries with genus two mirror curves and nontrivial mass parameters: the $Y^{3,0}$ geometry, and the resolved $\bC^3/\bZ_6$ orbifold. 
	We tested various conjectures put forward in \cite{ghm, cgm}. In particular, we checked the expression (\ref{multi-Airy}) for the fermionic spectral traces in various cases, and we 
	verified that their 't Hooft expansion reproduces the free energies of the standard topological string in the so-called maximal conifold frame. We were also able to write down the 
	integral kernel of a large class of perturbations of the three-termed operator $\sO_{m,n}$ studied in \cite{kasmar}, and this allowed us to perform precision tests of the conjectures in the case of the $Y^{3,0}$ geometry. 
	
	We studied certain perturbations of the $\sO_{m,n}$ operators in which the mass parameters no longer satisfy the relevant 
	positivity conditions. As a consequence, the resulting operators are not in general of trace class. 
	Using the correspondence with topological string theory, we obtained an analytic continuation of the Fredholm determinant. This in turn allowed us to define 
	an infinite set of complex valued ``resonant'' eigenvalues of the operators, in analogy to what happens in ordinary quantum mechanics. 
	
	We also explored the connection between the spectral theory of quantum mirror curves, and the cluster integrable systems arising from toric CY threefolds. We showed that, in some 
	cases, when the spectral curve is of genus two, the spectrum of the integrable system can be obtained from the generalized Fredholm determinant, after imposing positive reality conditions for the spectra.
	
	There are many important questions remaining. For example, in our discussion of the resonant eigenvalues associated to mirror curves, we used an analytic continuation of the 
	spectral traces and the Fredholm determinant. This is equivalent to a continuous deformation of the operator in the complex plane, which is one of the key ideas underlying the rigorous 
	definition of resonances for Schr\"odinger operators (see for example \cite{reed-simon}). 
	It would be interesting to put our definition of resonances in quantum spectral curves on a rigorous footing. More pragmatically, one would like to 
	calculate the resonant eigenvalues directly from the operator, as in the method of complex dilatation in ordinary quantum mechanics.  
	
	Another important open problem is to find the eigenfunctions of the quantum mirror curves. A general strategy to address this issue, extending the ideas of \cite{ghm}, 
	has been recently put forward in \cite{mz-open}\footnote{See \cite{kashani-poor, sciarappa} for other attempts to obtain the eigenfunctions.}. This has led to exact formulae for the 
	eigenfunctions in the maximally supersymmetric case of local $\IF_0$. However, much more work is needed in order to have concrete results for general geometries, and in particular 
	for the higher genus case. 
	
	Of course, the key puzzle is the origin of the conjectured form of the Fredholm determinant in terms of topological string free energies. Recently, this has been achieved when the CY is 
	local $\IF_0$, in a certain 4d limit \cite{bgt}, by showing that both sides of the conjectural formula (\ref{our-conj}) are 
	solutions to a certain Painlev\'{e} equation. It would be interesting to generalize this approach to the 5d case. In particular, the ubiquitous presence of mass parameters in toric CY threefolds indicates a certain 
	hierarchy of Painlev\'{e} difference equations probably related to \cite{sakai}.
	
	
	Finally, there are still some mysteries concerning the relationship between the Fredholm theory of quantum mirror curves and the quantization conditions for 
	cluster integrable systems. The compatibility between both approaches, explored in detail in \cite{wzh,swh}, requires an intriguing relationship 
	between the refined and the conventional topological free energies, as well as a set of B-fields providing additional vanishing functions on moduli space. This set has been found in many examples in a 
	rather {\it ad hoc} manner. It would be interesting to give a first-principle explanation for the 
	existence of this set, and to understand 
	better the relationships between the free energies. Work in this direction will appear in \cite{gg}.

	\section*{Acknowledgments}
	We would like to thank Gian--Michele Graf, Alba Grassi, Yasuyuki Hatsuda, Rinat Kashaev, 
	Amir-Kian Kashani-Poor, Thorsten Schimannek and Szabolcs Zakany for valuable discussions. 
	SC and MM are supported in part by the Fonds National Suisse, 
subsidies 200021-156995 and 200020-141329, and by the NCCR 51NF40-141869 ``The Mathematics of Physics" (SwissMAP). JG is supported by the grant ANR-13-BS05-0001.


\end{document}